\documentclass[preprint,showpacs,amsmath,amssymb,prc]{revtex4}
\usepackage{graphicx}
\usepackage{dcolumn}
\usepackage{bm}

\begin{document}
    
\title{Measurement of the Neutron Lifetime by Counting \\
Trapped Protons in a Cold Neutron Beam}
    
\author{J.~S.~Nico, M.~S.~Dewey, and D.~M.~Gilliam}
\affiliation{National Institute of Standards and 
Technology, Gaithersburg, MD 20899}

\author{F.~E.~Wietfeldt}
\affiliation{Tulane University, New Orleans, LA 70118}

\author{X.~Fei and W.~M.~Snow}
\affiliation{Indiana University and Indiana University Cyclotron Facility, Bloomington, IN 47408}

\author{G.~L.~Greene} \affiliation{University of Tennessee/Oak Ridge
National Laboratory, Knoxville, TN 37996}

\author{J.~ Pauwels, R.~Eykens, A.~Lamberty, and J.~Van~Gestel} \affiliation{European Commission, Joint Research Centre,\\
Institute for Reference Materials and Measurements, 2440 Geel, Belgium}

\author{R.~D.~Scott} \affiliation{Scottish Universities Research and  Reactor Centre,
East Kilbride, G75 0QU, UK}

\date{\today}
\begin{abstract}
A measurement of the neutron lifetime $\tau_{n}$ performed by the absolute counting of in-beam neutrons and their decay protons has been completed.  Protons confined in a quasi-Penning trap were accelerated onto a silicon detector held at a high potential and counted with nearly unit efficiency.  The neutrons were counted by a device with an efficiency inversely proportional to neutron velocity, which cancels the dwell time of the neutron beam in the trap.  The result  is $\tau_{n} = (886.6\pm1.2{\rm [stat]}\pm3.2{\rm [sys]})$~s, which is the most precise measurement of the lifetime using an in-beam method.  The systematic uncertainty is dominated by neutron counting, in particular the mass of the deposit and the $^{6}$Li({\it{n,t}}) cross section. The measurement technique and apparatus, data analysis, and investigation of systematic uncertainties are discussed in detail.
\end{abstract}

\maketitle
    
\section{Introduction}\label{sec:intro}

Precision measurements of the beta decay of a free neutron address fundamental questions in particle physics, astrophysics, and cosmology.  The decay can be described by the transformation of a $d$ quark into an $u$ quark through the emission of a virtual $W$ boson that decays into an electron and an antineutrino.  As the simplest semi-leptonic decay, the study of neutron decay tests the charged current sector of the Standard Model.  Improving the precision of the neutron lifetime is fundamental to testing the validity of the theory.

There are three distinct experimental strategies for measuring the neutron lifetime.  One can confine neutrons in material ``bottles'' or magnetic fields and measure the number of neutrons remaining as a function of time~\cite{MAM89,MAM93,ARZ00,PAU89}.  The number of neutrons in the bottle, $N(t)$, is measured and fit to the exponential decay function $N(t)=N(0)e^{-t/\tau_{n}}$ to extract $\tau_{n}$.  The second approach measures simultaneously both the rate of neutron decays ${dN/dt}$ and the average number of neutrons $N$ in a well-defined volume of a neutron beam~\cite{SPI88}.  The neutron lifetime is determined from the differential form of the radioactive decay law ${dN/dt}=-N/{\tau_{n}}$.  A third approach proposes to measure the lifetime using ultracold neutrons that are magnetically confined in superfluid $^{4}$He~\cite{HUF01}.  The decay electrons are registered via scintillations in the helium thus allowing one to directly fit for the exponential decay of the trapped neutrons.  Accurate measurements using each of these independent methods are important for establishing the reliability of the results for $\tau_{n}$.

\begin{figure}
\includegraphics[width=6in]{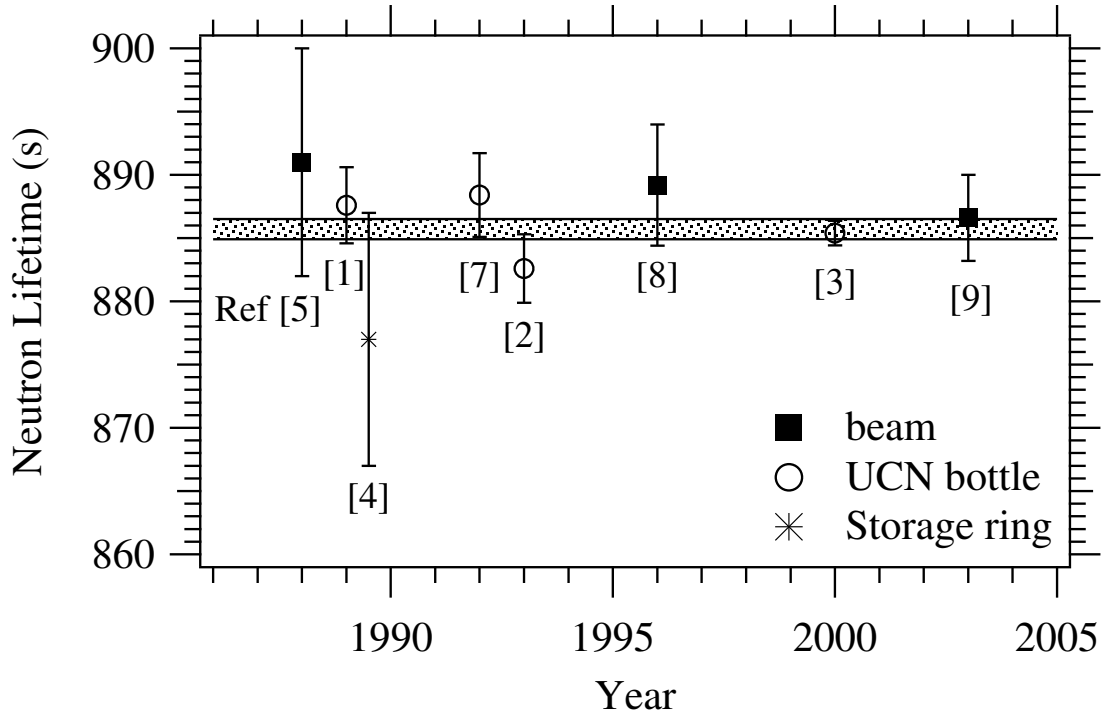}
\caption{\label{fig:history}A comparison of recent neutron lifetime measurements. The shaded band is $\pm 1$ standard deviation of the weighted average including this result.}
\end{figure}

Figure~\ref{fig:history} shows a summary of recent  measurements.  Seven of the  experiments~\cite{SPI88,MAM89,NEZ92,MAM93,BYR96,ARZ00, DEW03} contribute to a current neutron lifetime world average of $\tau_{n} = (885.7\pm0.8)$~s~\cite{EID04}. While the agreement among the results is very good, the four more precise measurements utilize ultracold neutrons that are confined to a bottle or gravitational trap.  The two experiments using a beam of cold neutrons do not have much statistical influence.  Given the different set of systematic problems that the two classes of experiments encounter, a precision measurement of the lifetime using a cold neutron beam not only reduces the overall uncertainty of $\tau_{n}$, but it is also an important independent check on the robustness of the central value.

Accurate determination of the parameters that describe neutron decay  can provide important information regarding the completeness of the three-family picture of the Standard Model through a test of the unitarity of the Cabibbo-Kobayashi-Maskawa (CKM) matrix. Neutron decay can be used to determine the CKM matrix element $|V_{\mathrm{ud}}|$ with high precision in a fashion that is  free of some of the theoretical uncertainties present in $0^{+} \to 0^{+}$ nuclear decays, which are still used for the most precise determination of $|V_{\mathrm{ud}}|$. Neutron decay is the system that offers the best prospect of a significant improvement in the direct determination of $|V_{\mathrm{ud}}|$. Such a measurement can be used to test whether the weak interaction in the charged-current sector is purely V-A (as in the Standard Model) or has right-handed components. Neutron decay also dictates the time scale for Big Bang Nucleosynthesis, and its lifetime remains the most uncertain nuclear parameter in cosmological models that predict the cosmic $^4$He abundance.

This paper describes a measurement of the neutron lifetime by counting beam neutrons and trapped protons~\cite{DEW03}. It presents a refined analysis of the data and treatment of the systematic effects.
The remainder of this section gives additional motivation and background for this measurement. Section~\ref{sec:apparatus} discusses the principles behind the experimental technique as well as details of the apparatus. The method of analysis is presented in Section~\ref{sec:analysis}. In Section~\ref{sec:systematics} we consider the treatment of the systematic effects in detail. They are divided into two main classes: systematics that affect the neutron counting efficiency and those that affect the proton counting efficiency. Lastly, the new result is given in Section~\ref{sec:results}.

\subsection{Neutron Lifetime and the Standard Model}

Most of the constraints on physics beyond the Standard Model come from high energy collider experiments. However, precision measurements in neutron and nuclear beta decay can test certain Standard Model extensions in the charged-current sector with comparable or superior sensitivity to colliders for certain types of Standard Model extensions~\cite{HER01}. As the simplest nuclear beta decay, the free neutron provides a particularly attractive laboratory for the study of the charged-current sector of the weak interaction. Because free neutron decay is unencumbered by the many nucleon effects present in all other nuclear decays, measurements of the parameters that describe neutron decay can be related to the fundamental weak couplings in a more straightforward fashion. 

In the Standard Model, beta decay of the free neutron is a mixed vector/axial-vector current process characterized to a good approximation by two coupling strengths: $g_{v}$ and $g_{a}$, the vector and axial-vector coupling coefficients.  The probability distribution for neutron beta decay can be written as~\cite{JAC57}:
\begin{eqnarray}\label{eq:decaydist}
dW &\propto& (g_{v}^{2} + 3 g_{a}^{2}) F(E_{e})\nonumber\\&&\left[1+a\frac{\Vec{p}_{e}\cdot\Vec{p}_{\nu}} {E_{e}E_{\nu}}+ \Vec{\sigma_{n}}
\cdot \left( A\frac{\Vec{p}_{e}} {E_{e}}+ B\frac{\Vec{p}_{\nu}} {E_{\nu}} \right)\right],
\end{eqnarray}
where $F(E_e)$ is the beta electron energy spectrum; $\Vec{p_e}$, $\Vec{p_{\nu}}$; $E_e$ and $E_{\nu}$ are the momenta and kinetic energies of the decay electron and antineutrino; $a$, $A$, and
$B$ are angular correlation coefficients; and $\Vec{\sigma_n}$ is the initial spin of the decaying neutron. In addition, one obtains an expression for the neutron lifetime
\begin{equation}
\tau_n = \frac{2 \pi^3
\hbar^7}{m_e^5c^4}\frac{1}{f(1+\delta_R)(g_v^2+3g_a^2)}
\end{equation}
and can define the coupling constant ratio $\lambda = {g_{a}}/{g_{v}}$. Here $f(1+\delta_R)=1.71489 \pm 0.00002$ is a theoretically calculated phase space factor including radiative corrections~\cite{TOW95}.  The parameter $\lambda$ can be extracted from measurement of either $a$, $A$, or $B$, and thus, with a measurement of neutron lifetime $\tau_{n}$, $g_{v}$ and $g_{a}$ can be determined uniquely.

A strong motivation for more accurate measurements of neutron decay parameters arises from the results of nuclear beta decay experiments. The most precise way of determining $g_v$ to date has
been from superallowed $0^+ \rightarrow 0^+$ nuclear $\beta$ decays between isobaric analog states.  The current result, $g_{v} = (1.41517  \pm 0.00046) \times 10^{-62}~\text{J m}^{3}$, can be related to the  
$|V_{\mathrm{ud}}|$ matrix element of the CKM matrix element via $g_{v}^{2} \propto V_{\mathrm{ud}}^{2} G_{F}^2$,  where $G_F$ is known precisely from muon decay.  This yields a value of $| V_{\mathrm{ud}} |  = 0.9740 \pm 0.0005$, with the uncertainty dominated by theoretical corrections.  This value can be used to test the unitarity condition of  the CKM matrix ($|V_{\mathrm{ud}}|^2 + |V_{\mathrm{us}}|^2 +|V_{\mathrm{ub}}|^2 = 1$), with the values of $|V_{\mathrm{us}}|$ and  $|V_{\mathrm{ub}}|$ taken from the current recommendations of the  
Particle Data Group~\cite{EID04}.  Using these values, the unitarity  sum is $\sum_i | {V_{\mathrm{ui}} |}^2 = 0.9969 \pm 0.0015$, a value which  differs from unitarity by 2.1 standard deviations. 

Compared with nuclear beta decay, neutron beta decay offers a theoretically cleaner environment for extracting $g_{v}$ due to the absence of other nucleons, although some radiative corrections are common to both systems.  Combining the world-average values of $\tau_{n}$ (including the value reported in this paper) and $\lambda$, one can extract a value for $g_{v}$, yielding $g_{v} = (1.4153 \pm 0.0027) \times 10^{-62}~\text{J m}^{3}$. Using this value of $g_{v}$, one can apply the same unitarity test,
giving a sum value of $\sum_i | {V_{\mathrm{ui}} |}^2 = 0.9971 \pm 0.0039$, 0.75 standard deviations below unity.  This result agrees with both the nuclear result and unity.  If neutron measurements are to
address definitively a possible incompatibility with the Standard Model, as suggested by the nuclear beta-decay results, both the neutron lifetime and $\lambda$ must be determined with higher precision.

A precision determination of $| V_{\mathrm{ud}} |$ should be seen in the context of the overall effort in high energy physics at beauty and charm factories to determine with high precision all the parameters of the CKM matrix. With the recently approved CLEO-c project, for example, it should be possible in the next few years to measure the CKM matrix element $V_{\mathrm{cd}}$ to 1~\% accuracy if lattice gauge theory calculations of the required form factors can improve sufficiently to match the expected precision of the data~\cite{SHI03}. If successful, this would make possible another independent check of CKM unitarity using the first column: 
$|V_{\mathrm{ud}}|^2+|V_{\mathrm{cd}}|^2+|V_{\mathrm{td}}|^2=1$. Thus, a precision measurement of $| V_{\mathrm{ud}} |$ can be used for two separate checks of CKM unitarity.

Furthermore, assuming unitarity of the upper row of the CKM matrix and the Wolfenstein parametrization, a precision determination of $| V_{\mathrm{ud}} |$ can be used to infer the Wolfenstein parameter $\lambda_W=V_{\mathrm{ud}}$, which is needed for the tests of the unitarity triangles at B factories.   Among the phenomena that have been considered as possible causes for violation of CKM unitarity include right-handed currents~\cite{HOL77}, supersymmetry~\cite{KUR02}, exotic fermions~\cite{LAN88,MAA90}, and additional Z bosons~\cite{LAN92,MAR87} among others. One notes that while the sum above is dominated by $|V_{\mathrm{ud}}|$, the contribution of $|V_{\mathrm{us}}|$ is significant and there remains a question of the reliability of the currently accepted value and its uncertainty. There has been recent discussion regarding the value of $|V_{\mathrm{us}}|$ from kaon decay based on new results and evaluations of kaon semileptonic decays rates. If one were use the value of $|V_{\mathrm{us}}|$ from some recent evaluations~\cite{FRA04}, the discrepancy with unitarity disappears. Efforts are now under way to clarify this situation using kaon decay data from several collaborations~\cite{BAL03}. There are also renewed theoretical efforts to attempt to extract $| V_{\mathrm{us}} |$ from hyperon decay~\cite{CAB03}.

\subsection{Neutron Lifetime and Nucleosynthesis}

The neutron lifetime also influences the predictions of the theory of Big Bang Nucleosynthesis (BBN) for the primordial helium abundance in the universe and the number of different types of light neutrinos $N_{\nu}$.  Since a large fraction of the uncertainty in the BBN prediction for the primordial $^4$He/H abundance ratio comes from the uncertainty of the neutron lifetime~\cite{LOP99, BUR99}, improved neutron lifetime measurements are useful for sharpening the BBN prediction.  With the recent high-precision determination of the cosmic baryon density reported by the WMAP measurement of the microwave background~\cite{SPE03}, the BBN prediction for the $^4$He abundance is higher than the  value inferred from observation~\cite{CYB03}. However, systematic uncertainties in the astronomical determinations of the $^4$He/H ratio are still believed to dominate the difference between theory and observation.  Furthermore, comparisons of BBN helium abundance calculations to observation using the number of known light neutrinos ($N_{\nu}=3$) are consistent with the value derived from $Z$ decay~\cite{EID04}.

\section{Experimental Method and Apparatus}\label{sec:apparatus}
\subsection{The ``In-Beam'' Technique}

The measurement presented here requires accurate counting of neutrons and neutron decay products (protons) from a cold neutron beam.  Such an in-beam lifetime measurement must overcome several technical difficulties.  These include the accurate measurement of the relatively low number of neutron decay events in the presence of background, accurate measurement of the decay detection volume, and accurate measurement of the mean number of neutrons within the decay detection volume.  Each of these difficulties is directly addressed in this experiment in a manner similar to that of previous experiments utilizing the in-beam technique~\cite{WIL89,BYR90,BYR96,SNO00}. 

\begin{figure}
\includegraphics[width=6in]{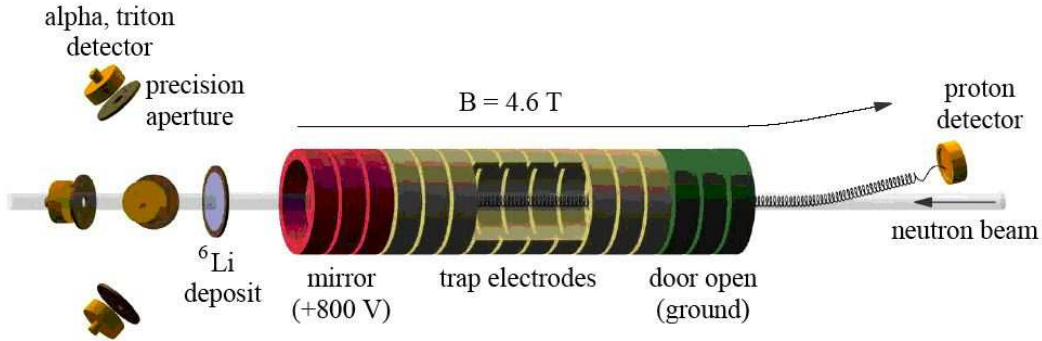}
\caption{\label{fig:schematic}An illustration of the experimental method for measuring the lifetime by counting neutrons and trapped protons.}
\end{figure}

An illustration of the experimental method is shown in Fig.~\ref{fig:schematic}.  The technique of trapping protons to increase the signal-to-background was first proposed by Byrne {\it et al.} and is described in detail elsewhere~\cite{BYR89,WIL89}. A trapping region of length $L$ intercepts
the entire neutron beam.  Within the volume of this region, neutron decay is observed by detecting decay protons with an efficiency $\epsilon_{p}$.  The neutron beam is characterized by a velocity dependent fluence rate $I(v)$.  The mean number of neutrons in the trap at any time is given by
\begin{equation}
N_{n}=L \int_{A} da I(v) \frac{1}{v}
\label{eq:mean_n},
\end{equation}
where $A$ is the trap cross-sectional area having non zero fluence. Thus, the rate at which decay events are detected, $\dot{N_{p}}$, is
\begin{equation}
    \dot{N_{p}}=\tau^{-1}\epsilon_{p}L\int_{A}da\,I(v)\frac{1}{v}
\label{eq:mean_p}.
\end{equation}

After leaving the trap, the neutron beam passes through a detector whose efficiency for detecting a low energy neutron is proportional to $1/v$.  Following the usual convention in thermal neutron physics, we define the efficiency for the neutron detector, $\epsilon_{o}$, as the ratio of the reaction product rate to the neutron rate incident on a ${^6}$LiF deposit for neutrons with a velocity $v_{o} = 2200$~m/s. The corresponding efficiency for neutrons of other velocities is $\epsilon_{o}v_{o}/v$.  Therefore, the total charged particle count rate, denoted $\dot N_{\alpha+t}$ to indicate the neutron capture reaction products, is
\begin{equation}
    \dot N_{\alpha+t}=\epsilon_{o}v_{o}\int_{A}da\,I(v)\frac{1}{v}
\label{eq:mean_alpha}.
\end{equation}

The integrals in Eq.~(\ref{eq:mean_p}) and Eq.~(\ref{eq:mean_alpha}) are identical.  The velocity dependence of the neutron detector efficiency compensates for the fact that the faster neutrons in the beam spend less time in the decay volume.  This cancelation is exact given two assumptions: (i) the neutron absorption efficiency in the ${^6}$LiF target is exactly proportional to $1/v$ and (ii) the neutron beam intensity and its velocity dependence do not change between the trap and the target. The deviation from the $1/v$ law in the $^{6}$Li(n,t)$^{4}$He cross section has been shown to be less than 0.01~\%~\cite{BER61} and changes in the neutron beam due to decay-in-flight and residual gas interaction are less than 0.001~\%. The cancelation allows this technique to make full use of the broad neutron energy spectrum from the reactor cold source.  Thus, we obtain an expression for the neutron lifetime $\tau_{n}$ in terms of measurable quantities
\begin{equation}
    \tau_{n}=\frac{L}{\dot{N_{p}}}
    \frac{\dot N_{\alpha+t}}{\epsilon_{o}}\frac{\epsilon_{p}}{v_{o}}
\label{eq:mean_tau}.
\end{equation}
The challenge of the in-beam technique is to measure the quantities $\dot N_{\alpha+t}$, $\dot{N_{p}}$, $L$, $\epsilon_{p}$, and $\epsilon_{o}$ accurately.

\subsection{Neutron Beamline}

The experiment was performed using cold neutrons at the National Institute of Standards and Technology Center for Neutron Research (NCNR).  The NCNR operates a 20 MW, heavy-water-moderated, research reactor that provides fission neutrons moderated to thermal energies by the D$_{2}$O primary reactor coolant.  Cold neutrons were produced by a cold neutron moderator situated adjacent to the reactor core.  It consisted of a spherical shell of
liquid hydrogen maintained at a temperature of 20~K.  Neutrons emerged from the cold source in a pseudo-Maxwellian distribution with an effective temperature of 40~K. The slower average velocity of  cold neutrons increased the number of neutrons that decay in the fiducial volume of the proton trap.

Neutron guides coated with $^{58}$Ni efficiently transported the cold neutrons approximately 68~m from the cold source to the experimental area at the end of neutron guide 6 (NG-6)~\cite{HUF03} in the NCNR Guide Hall.  The average thermal-equivalent neutron fluence rate was measured to be $1.4\times10^{9}$~$\rm~cm^{-2}~s^{-1}$ at the local guide shutter at the NG-6 end station.  Immediately after exiting the guide shutter, the neutron beam passed through a beam filter of single-crystal bismuth cooled to liquid nitrogen temperature.  The filter attenuated fast neutrons and gamma rays originating from the reactor core that may contribute to the background signal.  Cooling the filter elements to 77~K significantly increased the transmission of cold neutrons though the filter by reducing the scattering from phonons in the solid.

After the neutron beam exited the filter, it was collimated by two $^{6}$LiF apertures, which are almost totally absorbing for low energy neutrons.  The diameter of the first aperture (C1) was varied as a systematic check on the effect of the beam diameter on the measured lifetime.  The second one (C2) had a diameter of 8.4~mm and was not changed during the measurement.  In between these two beam-defining apertures were several $^{6}$LiF beam scrapers, which remove scattered and highly divergent neutrons. The scrapers were mounted inside a 4.9~m He-filled flight tube wrapped with $^{10}$B-loaded rubber.  After passing through C2, the beam entered a 1~m section of pre-guide and then entered the vacuum system through the silicon window of a 7~mm diameter quartz guide tube.  After passing through the trap, the beam traveled 83~cm to the neutron counter. It exited the vacuum system through a silicon window and was stopped in a $^{6}$LiF beam dump. 

\subsubsection{Vacuum System}\label{sec:VacuumSystem}

The vacuum system consisted of three main sections: the proton detector, the bore of the superconducting solenoid (where the proton trap resided), and the neutron detector.   Rough vacuum was achieved by an oil-free turbo pump, and ultra high vacuum (UHV) was maintained by two ion pumps. All parts of the system that could withstand typical UHV bake-out temperatures were routinely baked after every exposure of the vacuum system to air. The solenoid bore is the most notable exception to that procedure. The bore could be isolated from both the proton detector end and neutron detector end by gate valves, thus allowing access to either end without the necessity of warming-up and venting the inner bore. The pressure in the system measured at the ion pumps was typically  $10^{-9}$~mbar. It  was reasonably assumed that the pressure at the trap was significantly below that value due to the cryopumping of the solenoid bore.

\subsection{Proton Counting}

The detection of protons was accomplished through the use of a silicon detector and a proton trap, which consists of a 4.6~T magnetic field along the beam axis and an annular electrostatic trap composed of 16 electrodes segmented along the beam direction.  In trapping mode these
electrodes impose a potential well over a volume of the neutron beam of depth approximately $+800$~eV, which is well above the maximum proton kinetic energy of $751$~eV, and confine the protons axially.  Since the protons from neutron decay have a maximum cyclotron radius that is less than $1$~mm in the 4.6~T field, the decay protons are radially confined as well.  The protons from neutron decay are therefore trapped with unit efficiency except at the ends of the trap, where potential gradients affect the efficiency.  After a trapping time of order $10$~ms, the trapped protons are ejected from the trap, guided adiabatically along the magnetic field lines that bend protons out of the neutron beam, and accelerated onto a detector held at a high negative potential.

\subsubsection{The Proton Trap}\label{sec:Penningtrap}

The ideal proton trap for this experiment would consist of a perfectly uniform axial magnetic field and an axial electrostatic square well potential whose height on both ends exceeds the maximum kinetic energy of neutron decay protons (751~eV).  In this case the length $L$ of the trap would be well-defined, and all protons created within this length would be trapped with 100~\% efficiency. One could determine $\tau_n$ by applying Eq.~(\ref{eq:mean_tau}) to the data from a single trap length. An exact square well potential cannot be realized in this experiment. There is a region near each end of the trap, which we collectively refer to as the {\em end region}, where the electrostatic potential is above ground but less than the maximum applied voltage. Protons created in the central, grounded region are always trapped, but those created in the end region are trapped with less than 100\% efficiency. For this reason the trap is segmented into 16 electrodes and we vary the trap length.  The electrode structure is assembled in a manner that allows accurate determination of the segment repeat distance.  The lengths of the individual electrodes, and therefore the changes in the length of the trap, must be accurately known.

\begin{figure}
\includegraphics[width=6in]{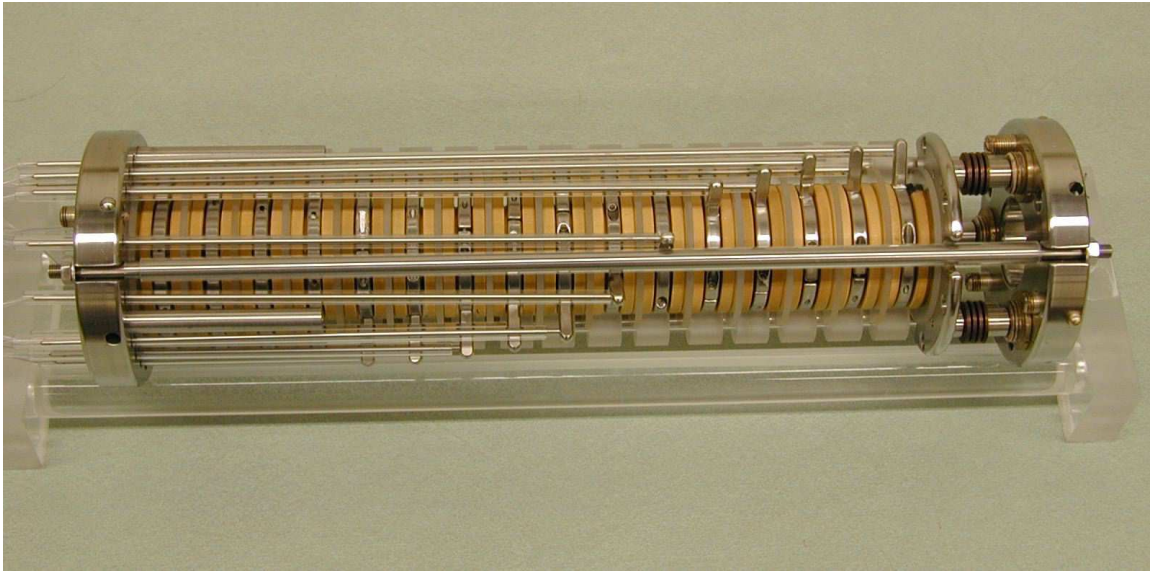}
\caption{\label{fig:trap} The proton trap.}
\end{figure}

The electrodes for the trap were fabricated from fused quartz to optical tolerances and coated with a thin conducting layer of gold.   Adjacent segments were separated by 2~mm-thick insulating spacers also made of fused quartz. The trap is shown in Fig.~\ref{fig:trap}. The length of each electrode and spacer was measured at room temperature using a coordinate measuring machine in the NIST Fabrication Technology Division. Relevant dimensions of the trap are given in Table~\ref{tab:metrol}; the precision is known to better than $\pm 5\;\mu m$.  Changes in the dimension due to thermal contraction are at the 0.01~\% level for quartz~\cite{CRC82}. Electrodes 1 through 3 are referred to as the ``door,'' and their voltages do not change. Following the door are a variable number (3 to 10) of grounded electrodes that comprise the central trap (see Fig.~\ref{fig:schematic}). This number determines the trap length. The three electrodes after the trap are called the ``mirror.'' The position of the mirror determines the trap length.

The shape of the electrostatic potential near the door and mirror is the same for all traps with 3 through 10 grounded electrodes in the trapping region, so the effective length of the end region, while unknown, is in principle constant.  The length of the trap can then be written
\begin{equation}
L = nl + L_{\rm end},
\label{eq:TrapLen}
\end{equation}
where $n$ is the number of grounded electrodes and $l$ is the physical length of one electrode plus an adjacent spacer (21.6~mm). $L_{\rm end}$ is an {\em effective} length of the 
two end regions; it is proportional to the physical length of the end regions {\em and} the 
probability that protons created there will be trapped. From Eq. (\ref{eq:mean_p}) and (\ref{eq:mean_alpha}), one sees that the ratio of proton counting rate to alpha counting rate is
\begin{equation}
\frac{\dot{N_p}}{\dot{N}_{\alpha+t}} = \tau_n^{-1}\left( \frac{\epsilon_p}{\epsilon_0 v_o}\right)
(nl + L_{\rm end}).
\label{eq:FitFunc}
\end{equation}

We fit $\dot{N_p} / \dot N_{\alpha+t}$ as a function of $n$ to a straight line and determine $\tau_n$ from the slope, so there is no need to know the value of $L_{\rm end}$, provided that
it is the same for all trap lengths. Because of the symmetry in the trap's design, $L_{\rm end}$ is approximately equal for all trap lengths that were used.

\begin{table}
\caption{\label{tab:metrol} The measured lengths of trap elements (spacers and electrodes), beginning with electrode 1 (upstream end).  The uncertainty in these measurements is $\pm 5\;\mu m$.}
\begin{ruledtabular}
\begin{tabular}{ld|ld}
 Trap Element 	& \multicolumn{1}{c}{Length (mm)}	&Trap Element 	& \multicolumn{1}{c}{Length (mm)}\\
\hline
electrode 1	&	18.600		&	electrode 9	&	18.600\\
spacer 1	&	3.000		&	spacer 9	&	3.019\\
electrode 2	&	18.479		&	electrode 10	&	18.599\\
spacer 2	&	3.005		&	spacer 10		&	3.002\\
electrode 3	&	18.646		&	electrode 11	&	18.576\\
spacer 3	&	2.951		&	spacer 11		&	3.001\\
electrode 4	&	18.591		&	electrode 12	&	18.562\\
spacer 4	&	3.014		&	spacer 12		&	2.998\\
electrode 5	&	18.759		&	electrode 13	&	18.593\\
spacer 5	&	2.997		&	spacer 13		&	3.005\\
electrode 6	&	18.573		&	electrode 14	&	18.558\\
spacer 6	&	3.002		&	spacer 14		&	2.929\\
electrode 7	&	18.481		&	electrode 15	&	18.586\\
spacer 7	&	3.008		&	spacer 15		&	3.002\\
electrode 8	&	18.550		&	electrode 16	&	18.600\\ 
spacer 8	&	3.003		&					&	\\
\end{tabular}
\end{ruledtabular}
\end{table} 

The electrodes and spacers were mounted in a stainless steel frame that was mounted rigidly inside the bore of the superconducting solenoid. Alignment jigs were constructed to allow the precise alignment of the trap axis with the neutron beam. For UHV compatibility and low thermal conductivity, stainless steel wire was used to connect each electrode to an electrical feedthrough at the vacuum interface. Outside the vacuum system, three high-voltage pulsers controlled by the data acquisition system provide DC voltage to the door, mirror, and central electrodes at the appropriate time in the trapping cycle.

\subsubsection{The Trapping Cycle}

The electrodes on the proton trap operate in three distinct modes: trapping protons, counting protons, or clearing the trap.  The typical proton trapping period was 10~ms in duration.  The period was selected primarily to avoid intermittent instability sometimes observed in the behavior of the trap.  Given our neutron fluence rate, this range of trapping times makes it unlikely to trap multiple protons, thus reducing the magnitude of the dead time corrections.

The first mode of operation during the measurement is the trapping mode. The door and mirror electrodes are held at $+800$~V, and the central electrodes are at ground.  The depth of the well is sufficient to confine axially protons that are created in the fiducial volume of the trap. The magnetic field confines them radially.

After approximately 10~ms, a signal is sent from the data acquisition system (DAQ) to acquire data from the proton detector.  Since the detector needs to be enabled only during extraction, the background is significantly reduced by the ratio of the extraction time to the trapping time (typically a factor of about $125$ in our experiment).  21~$\mu$s after the detector is enabled, the door electrodes are grounded and a graduated potential is imposed on the central electrodes to flush out protons that may have only a small amount of axial momentum. This is the counting mode and is also referred to as the ``ramp'' configuration.  The previously trapped protons now exit the trap and adiabatically follow the magnetic field lines.  These field lines bend by $9.5^{\circ}$ in the region beyond the trap and pass through the silicon detector, where protons are accelerated and detected.

The counting mode remains active for $76~\mu$s, a time that is sufficient to permit all protons to exit the trap.  The next signal establishes a clearing mode where the ramp voltages are maintained  and all other electrodes of the trap are held at ground.  This procedure prevents charged particles, which may contribute to instability, from being trapped in any portion of the trap.  After $33~\mu$s, both the door and mirror are raised, the trapping mode is reestablished, and another trapping cycle begins.  The acquisition is disabled, and no additional events are recorded until the cycle repeats.  Both energy and timing information are recorded for the proton events. Data from the timing spectra are used to determine the proton rate since both the background and dead time correction are less complicated than in the energy spectra. A detailed discussion of the analysis method that is employed to extract the proton rate from the timing spectrum is given in Section~\ref{sec:metofanal}

\subsubsection{The Proton Detector}\label{sec:TheProtonDetector}

To detect the protons ejected from the trap, we used silicon surface barrier detectors and passivated ion-implanted planar detectors.  They have good energy resolution and high detection efficiency for protons with incident energies greater than approximately 20~keV.  To minimize detector capacitance, one wants to use a detector with a large depletion depth and the smallest area that completely encompasses all the protons originating from the trap.  We used detectors with a depletion depth of 300~$\mu$m and 300~mm$^{2}$ active area.  The detector and preamplifier were radiatively cooled to approximately 150~K to minimize the detector leakage current and the noise contribution from the  preamplifier.

One must be certain that all the protons originating from the trap will be seen by the detector.  The neutron beam diameter was established in two ways: 1) using Monte Carlo calculations and modeling of the NG-6 beamline and the collimation of the lifetime apparatus (taking into account the finite divergence of the neutron beam and the maximum radius of the orbits of the trapped protons), and 2) making detailed measurements of the neutron beam profile (discussed in Section~\ref{sec:Halo}).  Great care was taken to align the detector precisely with the proton beam to ensure that protons would not be lost beyond the active area of the silicon.

One must accelerate the low-energy protons to a high negative potential in order to register the event in the detector.  All silicon detectors have a thin layer on the surface that is inactive, commonly referred to as a dead layer.  This layer consists of gold and/or silicon dioxide, depending on the type of silicon detector. The recoil energy of protons from neutrons decay is so low (751~eV maximum) that they cannot penetrate the dead layer without additional acceleration. Typical acceleration voltages for this experiment range between $-25$~kV and $-35$~kV. We chose to place the proton detector at high voltage rather than the trap at a high positive potential to avoid trapping electrons in the decay region.

\subsubsection{Proton Detector Alignment}

Given the importance of the correct positioning of the proton detector, three distinct approaches were used to verify that the detector was centered precisely on the center of the proton beam originating from the trap: surveying with a theodolite, measuring the centroid with source electrons, and measuring the centroid using protons from neutron decay.

\paragraph {Survey}  Initially, the proton trap was aligned to the neutron beam and rigidly secured to the inside bore of the magnet.  Crosshairs were inserted into the upstream and downstream ends of the trap to perform the alignment.  A survey target centered on the detector was mounted on its linear motion feedthrough with 1~m of travel and moved into its operating position in the bore of the magnet.  The alignment axis at $9.5^{\circ}$ with respect to the beam was defined using a mirror inserted into the magnet bore.

\paragraph{Electron Scan}  For the second approach, we used an electron source inserted into the downstream electrode of the trap.  The source was $^{210}$Pb-$^{210}$Bi-$^{210}$Po in equilibrium, which produces a beta-decay electron with a 1.16~MeV endpoint energy. The energy is such that when the magnet is at its nominal 4.6~T field strength, the highest-energy beta particle has a cyclotron radius comparable to that of protons from neutron decay.  Thus, the trajectory of the electron is similar to that of a proton, and one can use the electron source to scan the distribution of particles originating from the center of the trap. By comparison the cyclotron radius of the 5.3~MeV alpha particles from the $^{210}$Po decay is much larger, making their contribution negligible.

A plastic mask with a 3~mm diameter hole in the center was placed over the face of a typical surface barrier detector.  The detector is mounted on a three-axis manipulator, and its position can be rastered in the two directions perpendicular to the direction of the electron beam (x,y) as  well as along the beam axis (z). The scans give the beam profile in the transverse directions as well as along the beam axis. The x and y data were fit to a Gaussian function to obtain the centroid of the electron distribution, which coincided well with the geometric axis determined by the survey. 

\begin{figure}
\includegraphics[width=6in]{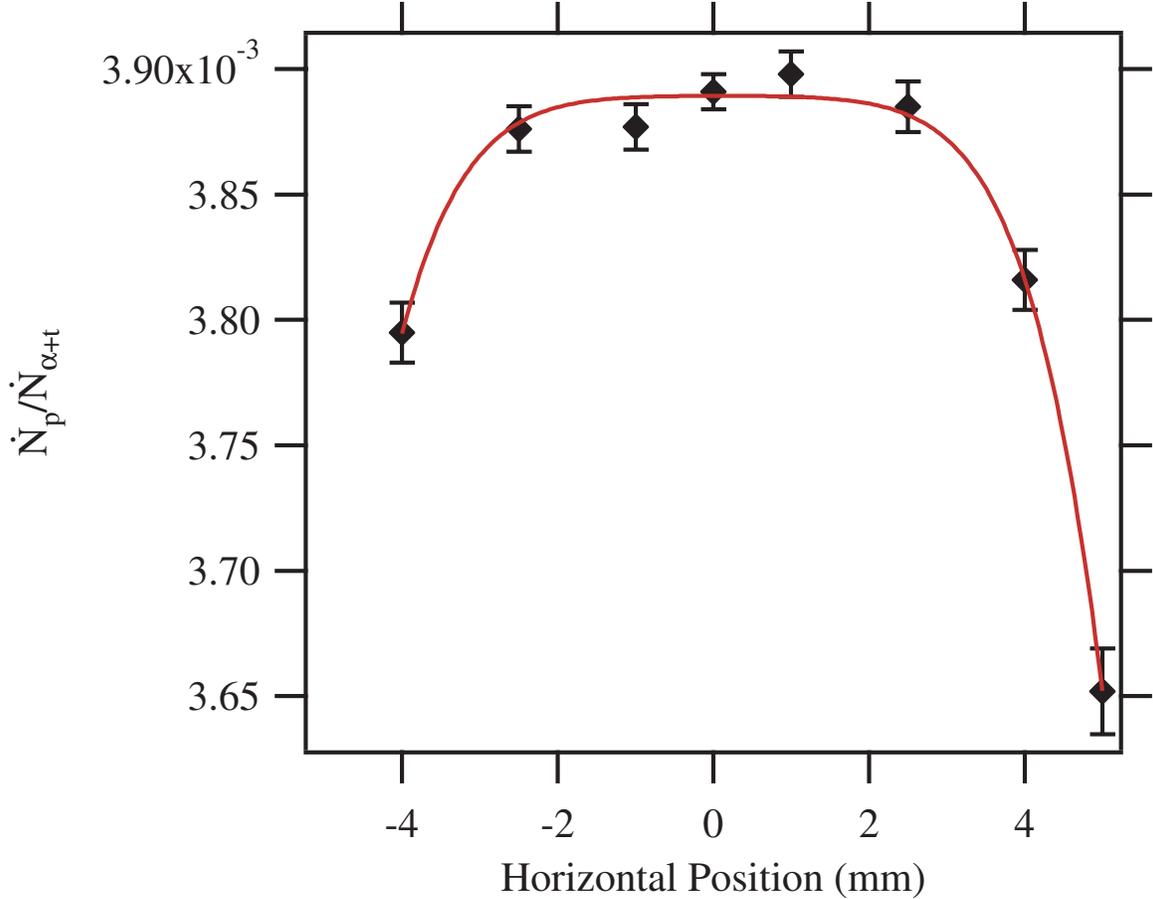}
\caption{\label{fig:ProtScan}Plot of the proton rate normalized to the neutron detector counting rate as a function of the horizontal position of the detector. The solid line is a fit to the data points.}
\end{figure}

\paragraph {Proton Scan}  Finally, the most accurate check on the positioning of the detector comes from decay protons themselves.  Although it is clearly the relevant measure of the alignment, the significantly lower event rate (in comparison with the electron source) makes it a more time consuming test.   The principle of the measurement is the same as with the electron source although no mask was used due to the lower count rate. One obtains the centroid by fitting the data to the convolution of a square distribution and a Gaussian function. Figure~\ref{fig:ProtScan} shows a scan for one direction of detector motion. In order to avoid proton rate drifts due to reactor power fluctuations, the relevant quantity to plot is the ratio of proton rate to neutron rate versus the position of the detector. 

For both the electron and proton scans, the position of the centroids in the two directions of the detector plane agreed with those obtained through alignment done by the theodolite to better than 1~mm.  Scans done in the z direction demonstrated the insensitivity in the rate as a function of the detector position along the magnet field axis. Additional measurements that were done to quantify the fraction of protons that still might miss the active area detector are discussed in Section~\ref{sec:Halo}.

\subsection{Neutron Counting}\label{sec:NeutronCounting}

The absolute number of neutrons passing through the proton trap is determined by measuring the  products from the $^{6}$Li(n,t)$^{4}$He reaction and relating that rate to the incident neutron fluence rate.  The total rate at which these reaction products are detected depends on the neutron fluence rate, the total detector solid angle, the neutron absorption cross section, and the deposit areal density.  The detector consists of a target surrounded by four silicon semiconductor detectors with a solid angle defined by precision-machined apertures and operates by counting the tritons and alpha particles produced by neutron capture on the $^{6}$Li.  It is shown schematically in Fig.~\ref{fig:schematic}. The geometry is chosen to make the solid angle subtended by the alpha detectors insensitive to first order in the source position. Two of these devices exist; the second device has been used in efforts to determine absolute neutron fluence at the level of 0.1~\%~\cite{CHO03}.  Additional details of the construction and operation may be found in Refs.~\cite{RIC93,CHO00}

The target in the neutron monitor consists of a thin ($\approx 0.4$~mm) 50~mm-diameter single-crystal wafer of silicon with an evaporated deposit of $^{6}$LiF. This deposit is thin enough that the
neutron fluence rate is only slightly attenuated, and the products from the $^{6}$Li(n,t)$^{4}$He reaction suffer negligible scattering or energy loss in passing through it.  The alpha particles and tritons produced by the neutron absorption reactions in $^{6}$Li are detected by four surface barrier detectors, each of which has a solid angle defined by a diamond-turned precision aperture, as shown in
Fig.~\ref{fig:schematic}.

\begin{figure}
\includegraphics[width=6in]{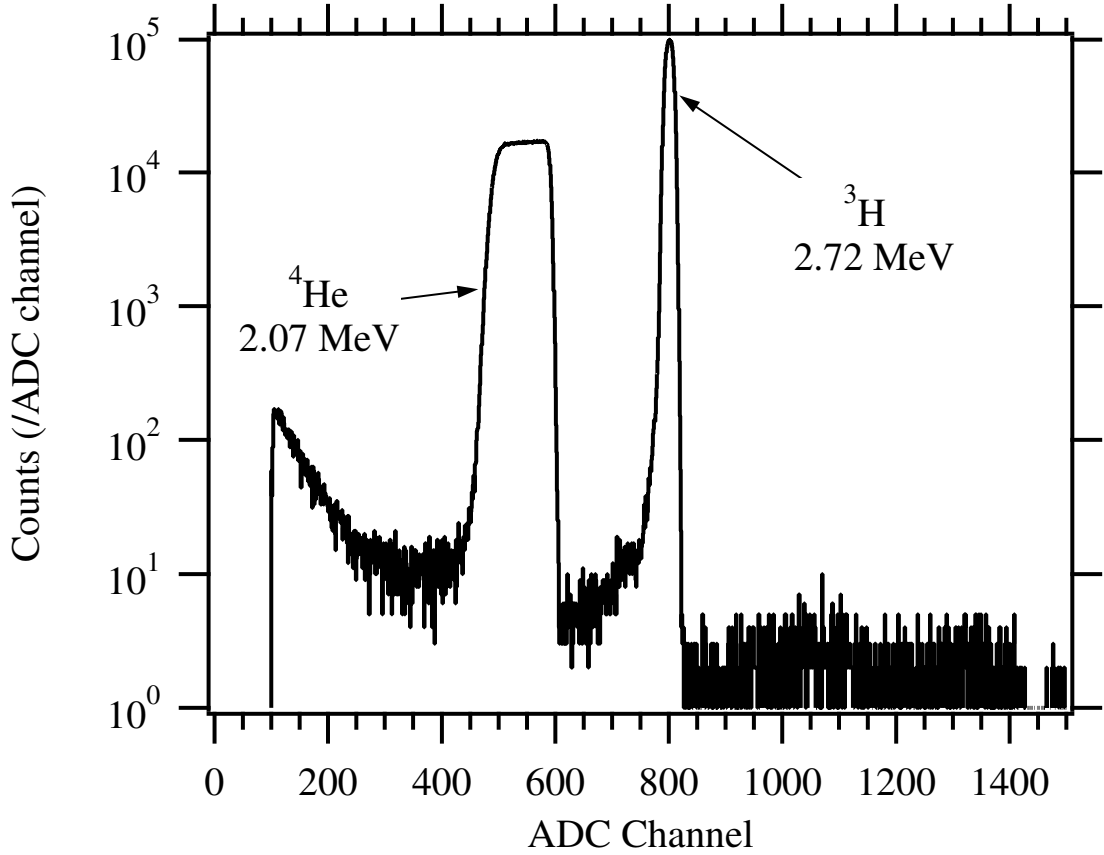}
\caption{\label{fig:1_vSpectrum} Plot of a typical pulse height spectrum from ${^6}$Li reaction products incident on a silicon detector in the neutron monitor.}
\end{figure}

${^6}$Li has a thermal (n,t) cross section that is large and known to 0.14~\%~\cite{CAR93}. Note that we use the evaluated nuclear data files (ENDF/B-VI) combined-analysis uncertainty from the R-matrix evaluation and not the expanded uncertainty. The energy dependence of the ${^6}$Li cross section is well known for cold and thermal neutrons and corresponds very closely to a strict $1/v$ dependence.  The deviation from pure $1/v$ behavior is less than 0.03~\% at thermal and subthermal energies~\cite{BER61}.  Fig.~\ref{fig:1_vSpectrum} shows a typical pulse height spectrum from one of the silicon detectors.  Both the triton and alpha particle peaks are well resolved from the electronic noise.  

\subsubsection{Neutron Counting Efficiency}\label{sec:NeutronCountingEff}

The neutron monitor is characterized by a parameter $\epsilon_{0}$
that denotes the ratio of detected alphas/tritons to incident neutrons. 
It is defined as
\begin{equation}
\epsilon_{0} =\frac{\sigma_{0}}{4\pi}
\int\int\Omega(x,y)\rho(x,y,)\phi(x,y) dx dy \label{eq:neutroneff},
\end{equation}
where $\sigma_{0}$ is the cross section at thermal ($v_{0} = 2200$~m/s) velocity, $\Omega(x,y)$ is the detector solid angle, $\rho(x,y)$ is the areal mass density distribution of the deposit, and
$\phi(x,y)$ is the areal distribution of the neutron intensity on the target. $(x,y)$ are the coordinates normal to the beam axis.  The $^{6}$Li thermal cross section is ($941.0 \pm 1.3$)~b ~\cite{CAR93}.   The neutron detector solid angle has been measured in two independent ways: mechanical contact metrology and calibration with a $^{239}$Pu alpha source of known
absolute activity.  The metrology was performed using a coordinate measuring machine to measure precisely the diameter of each aperture and the distance of each aperture center to the center of the target.  In the second method, we placed an alpha source in the location of the deposit and measured the alpha rate in the detectors.  The ratio of the measured alpha counts to the total alpha activity determines the solid angle.  The absolute source activity can be traced back to an interlaboratory comparison of the absolute activity of actinide targets by low solid-angle alpha particle counting~\cite{DEN99}.  The results of the two solid angle measurements, conducted several years apart, agree to better than 0.1~\%. The value of $\Omega /4 \pi = 0.004196 \pm 0.1$~\%.

The $^{6}$LiF (and $^{10}$B) targets were fabricated at the Institute for Reference Materials and Measurements in Geel, Belgium in two separate efforts.  The manufacture of deposits and characterization of the $^{6}$LiF areal density were exhaustively detailed in measurements performed over several years~\cite{TAG91,PAU91,SCO92,GIL93,PAU95,SCO95}. Target materials were deposited by evaporation in batches onto silicon wafers in mounts that simultaneously revolved around the source to improve spatial uniformity.  The areal densities were determined by measuring
the thermal neutron induced charged particle reaction rates in a thermal neutron beam at the BR1 reactor at SCK/CEN, MOL, Belgium. The masses of six of the deposits were measured at IRRM by isotope dilution mass spectrometry and correlated with their reaction rates so that the masses, and hence the areal densities, of the remaining deposits could be deduced. The result of these measurements gives $\rho = (39.30 \pm 0.10)~\mu\rm{g/cm^{2}}$.

One should note that $\rho$ is an average density of the deposit.  The density distribution of ${^6}$LiF is not a constant but is a function of the radius.  This is an inevitable consequence of the fabrication process.  Thus, for a given neutron density, the absorption is not constant over the deposit.  This necessitates a small correction for the detector efficiency that involves integrating the neutron beam profile $\phi(x,y)$ over the deposit areal density distribution, as indicated by Eq.~(\ref{eq:neutroneff}). 

In addition, there are several other small corrections for the detector efficiency that must be considered, such as the neutron absorption in lithium and the silicon substrate, and incoherent neutron scattering the silicon.  These topics are addressed in Section~\ref{sec:neudetlos}.

\subsection{Data Acquisition System}

The data acquisition system was managed by a PC running National Instruments LabWindows/CVI 5.5 under Microsoft Windows~\cite{DISCLM}.  The computer controlled CAMAC-based modules generated periodic pulses for the trap and acquisition timing, accumulated time-to-digital converter (TDC) and analog-to-digital converter (ADC) proton spectra, counted single-channel analyzer (SCA) pulses from neutron capture, and operated a local beam shutter.  The computer also operated the high voltage supply for the proton detection electronics, read a digital voltmeter to monitor the trap temperature via a resistor, and read the current being drawn by one of the two ion pumps.  NIM-based electronics were used to generate an analog signal from the proton detector, 16 SCA pulses from the neutron detector, and an SCA signal from a fission-based neutron fluence monitor positioned at C1.  Finally, home-built TTL-based units converted the periodic timing pulses into TTL levels that controlled the trap voltages and gated the data acquisition.

Prior to the beginning of a run, the superconducting magnet coils were energized to 110~amps, producing a 4.6~T field along the trap.  The high voltage detector potential was applied using a separate program to raise the voltage from zero slowly.  The high voltage supply was attached to the proton detection subsystem which was in turn attached to ground via a 9~G$\Omega$ resistor.  Thus, 3.3~$\mu$A of current was drawn with DC voltage of 30~kV.  During the voltage ramp, this current was monitored for signs of instability.  An isolation transformer was used to provide AC power to the proton electronics which, like the proton detector, were maintained at high voltage.

In a normal data sequence, a particular trap length was configured manually and the following acquisition sequence was executed: 30 minutes of acquisition with the local shutter closed followed by either 4 or 8 hours of acquisition with the local shutter open.  This cycle was repeated until the trap length was changed.  Typically one day (8 hours) or one night (16 hours) was devoted to each trap length.  For most of our precision runs, all 8 different trap lengths were sampled.  The beam-off runs were very important.  The local LiF shutter stops cold neutrons while allowing those fast neutrons and gamma-rays not attenuated in the beam filters to pass through the apparatus.  In most of the beam-off data sets the above background rate was essentially zero.  In a few sets, however, there was a measurable, trap length dependent rate that was subtracted from the beam-on data set.  This phenomenon is discussed in detail in Section~\ref{sec:metofanal}. 

\subsubsection{Timing Signals}

Two crystal oscillators generated all of the signals necessary for controlling the trap voltages and gating the data acquisition.  One oscillator controlled the trap time while the other generated signals necessary for managing trap voltages and gate signals at the end of each trapping cycle.  The precision lifetime data were taken with trapping times of 5~ms and 10~ms.  For those times and a typical observed proton rate of 4~s$^{-1}$, the probability of observing one proton in a single trapping cycle is 0.02 and 0.04, respectively.  The analysis algorithm, which corrects for the dead time of the TDC, takes into account those instances when more than one proton arrives during a single trapping cycle. Since the correction necessary for multiple hits increases with rate, it was important to minimize this correction by operating with short trapping times (see Section~\ref{sec:metofanal}).

The data were acquired in one minute intervals.  During this time neutrons were counted continuously and the proton trap was cycled continuously.  At the end of the interval, data acquisition was disabled and the data were read out.  The readout period was brief and a new interval began immediately following its conclusion.

At the end of each trapping cycle, the following sequence of events occurs:
\begin{description}
\item[$t = 0\;\mu${\rm s:}] The proton detector ADC is gated on, and a pulse is sent to the start input of a single channel TDC.
\item[$t = 21\;\mu${\rm s:}] A signal is sent to open (i.e., ground) the door electrodes; simultaneously the ramp is turned on. The ramp voltages,  which range from approximately $+20$~V downstream to 0~V upstream in the longest trap, force protons to leave the trap very quickly. A delay of 21~$\mu$s between detector-on and trap-open allows one to sample the background proton rate. The sampling is done at the end of this sequence.
\item[$t = 97\;\mu${\rm s:}] The ramp is maintained but the mirror electrodes are grounded to flush out any electrons that may have accumulated in the positive potential of mirror electrodes. Between $t = 21\;\mu$s and $t = 97\;\mu$s, any proton that may have been trapped will arrive at the detector. Any pulse above threshold stops the TDC and causes an ADC conversion in the proton signal channel.
\item[$t = 127\;\mu${\rm s:}] The mirror and door electrodes are raised, and the ramp is turned off.
\item[$t = 159\;\mu${\rm s:}] The proton detection channel is gated off.
\item[$t = 999\;\mu${\rm s:}] A second DC level-sensing ADC is requested to read a voltage. Typically, this voltage is proportional to the door voltage. It is done as a check on the stability of the trap voltages.
\end{description}

\subsubsection{Trap Voltages}

It is extremely important that the trap be operated with stable, reproducible voltages because the observed proton count rate is a strong function of the applied voltages.  Changing the height of the door or mirror causes the end effects to change thereby changing the observed rate.  We used a single, stable DC power supply to generate the mirror and door voltages.  Two TTL-controlled, two-state switches fed this voltage onto the trap electrodes at the appropriate times.  This ensured that the mirror and door voltages were identical.  The switches were fast (rise time $\approx 2~\mu$s) and applied essentially all of the input voltage to the electrodes.  As these switches operate by toggling their output between two user-supplied voltages, a second DC supply was employed to set the zero level precisely.  It was found that operating with a slightly negative voltage ($\approx -1$ V) rather than ground caused the trap to be more stable.

A home-built, TTL-controllable operational amplifier was used to provide voltages to the central electrodes.  Its two-state output (40~V maximum, 0~V minimum) was stepped down using a resistive divider before being supplied to each of the electrodes in the central region of the trap.  For example, in the case of the 10 electrode-long trap, the ramp voltages were typically 20, 18, 16, 13, 11, 9, 7, 4, 2, and 0~V; while in the case of the 3 electrode-long trap, the ramp voltages were 4, 2, and 0~V. 

Setting of the trap length was the only manual step in the acquisition process.  It involved removing and reattaching coaxial cables from the switches to the appropriate electrodes.  Those electrodes that were outside of the door/central/mirror system -- 0 (7) electrodes in the case of the 10 (3) electrode-long trap -- were grounded.

\subsubsection{Data Inflow}\label{sec:DataInflow}

This section briefly describes the different sources of data that were recorded.

\paragraph{Proton Signal} As described in Section~\ref{sec:TheProtonDetector}, the proton
detector was a silicon semiconductor detector.  Its output was attached to an Amptek A250 preamplifier.  Both the detector and preamp were radiatively cooled by a graphite-coated beryllia insulator thermally connected to the liquid nitrogen reservoir of the cryostat. The typical operating temperature was approximately 150~K, sufficient to minimize the detector leakage current and field effect transistor noise in the preamplifier.  The preamp signal was amplified and shaped with an ORTEC 472A spectroscopy amplifier with a shaping time of $1~\mu$s.  The resultant proton energy signal was transmitted to ordinary ground by means of a LeCroy 5612 analog fiber optics transmitter and LeCroy 5613 receiver/controller.  The threshold for acquiring an event was fixed in the discriminator level of a LeCroy 3512 ADC.  The busy-out of the ADC was directed to the stop input of a LeCroy 4204 TDC.  

\paragraph{Fission Chamber} A fission chamber was placed just upstream of the C1 collimator in order to sample a small fraction of the entire neutron beam.  Each output pulse was fed into an SCA whose windows were carefully set in order to generate countable pulses proportional to capture fluence.  This signal, which was counted in a CAMAC scaler, was frequently useful as a diagnostic.

\paragraph{Neutron Monitor} The output of each of the four silicon detectors from the neutron monitor was amplified, shaped, and fed into two SCAs that provided rates for four regions on the ${^6}$Li decay-product spectrum (see Fig.~\ref{fig:1_vSpectrum}).  Each region was based on a threshold above which everything was counted, producing a total of 16 channels that were input into CAMAC scalers. The neutron monitor is discussed in detail in Section~\ref{sec:NeutronCounting}.

\paragraph{Pressure Parameters}

As discussed in Section~\ref{sec:VacuumSystem}, the vacuum system consisted of three main sections: the proton detector, the bore of the superconducting solenoid, and the neutron detector.  Ion pumps were employed on the proton and neutron ends of the apparatus.  All of the precision data were acquired with the gate valve at the neutron (proton) end ion pump closed (open). The current drawn by one of the two ion pumps was logged by the computer and is related to the pressure in a straightforward manner.  This quantity was frequently used as a diagnostic, especially when the apparatus was cooled down or warmed up.

\paragraph{High Voltage}

An acceleration voltage was applied to the high voltage components of the apparatus by a Bertan Associates Series 225 power supply.  That section of the apparatus was connected to ground through a 9\,G$\Omega$ resistor.  With a voltage of 30\,kV, one observed a constant current of 3.3\,$\mu$A.  At the end of each minute-long counting cycle, the supply voltage, current, and operating status were read.  The voltage and current were included in the data stream, while the status was monitored to check for an anomalous situation requiring an emergency shutdown of the apparatus.

\subsubsection{Data File Contents}

Data files were written to disk every minute to allow analysis of trends that occur on time scale short in comparison to the overall run time. It was convenient to store the information in three files.

The first file contains parameters that are relevant to the run. The data are input by the operator and include the trap length, beam on/off status, thickness of gold layer on the proton detector, acceleration potential, trapping time, beam collimation parameters, door/ramp/mirror electrode voltages, and thickness of bismuth in the filter cryostat. 

The second file records scaler counts accumulated during one minute intervals from the sixteen SCA channels from the neutron detector, one SCA channel from the fission chamber, 2 (2) windowed integrals from the TDC (ADC) histograms, the number of TDC and ADC values written, the number of trap openings, and the number of ``bad'' events.  Bad events are defined to be those where either three or more ADC events are registered in a trap cycle or the ADC value is above a high threshold (both are exceedingly unlikely events).  The file also contains once per minute acquisitions of the high voltage and current, clock time, and the trap-closed voltage.

The third file contains a list of all ADC and TDC values written in each one minute interval.  With these values for each conversion, it is possible to generate different TDC/ADC cuts to focus on specific regions of interest in the time-energy domain. This is very useful in the beam-off runs where one can extract an accurate estimate of the beam-off trapped rate by placing precise cuts around the region where decay protons arrive. For the lifetime data, no cuts are used because they complicate the correction for dead time losses.

During acquisition, the computer monitored the number of TDC and ADC entries that accumulated every 50~ms. That rate would increase dramatically if there were problems in the system. If the rate exceeded an operator-set limit, the computer would lower the high voltage to zero and close a beam shutter, thereby putting the apparatus in a safe mode.

Prior to evaluating the neutron and proton rates, a determination is made of which minute intervals to use.  This is done by plotting all the running sums and looking for nonstatistical anomalies.  In most
cases the entire file is usable.  The most common reasons for choosing to exclude parts of a file are severe reactor power fluctuations or a loss of neutron flux due to a loss of helium gas flow in the neutron
flight path.  Once a range has been selected, mean alpha and proton rates are evaluated as discussed below.

\section{Data Reduction}\label{sec:analysis}

\subsection{\label{sec:metofanal}Method of analysis}

Extracting the neutron lifetime requires the accurate determination of all the parameters from Eq.~(\ref{eq:mean_tau}).  In this section we address the method of determining $\dot{N}_{\alpha+t}$ and ${\dot{N}_{p}}$ from the data stream. 

\subsubsection{Determination of $\dot{N}_{\alpha+t}$}

As mentioned, there are sixteen counts associated with the neutron monitor $N_{ij}$, where $i = (1,2,3,4)$ denotes the physical silicon detector (up, down, east, west relative to the beam direction) and
 $j = (a,b,c,d)$ denotes an SCA window of the Li decay-product spectrum. For the purpose of determining the lifetime, the two relevant regions are  a) alpha particles, tritons, and high energy events and d) high energy events. The other two regions are used for systematic checks. Thus, the expression for the total number of alpha particles is
\begin{equation}
  \label{eq:AlphaRate}
  N_{\text{tot}} = \frac{1}{2} \sum_{i=1}^{4} (N_{ia} - N_{id}),
\end{equation}
where the factor of one half is included because both alpha and triton particles are counted.  The relative size of the background is typically less than 0.05~\%.  One obtains the alpha particle rate by dividing the total number of alpha particles by the live time
\begin{equation}
  \label{eq:alphadot}
  \dot{N}_{\alpha+t} = \frac{N_{\text{tot}}}{t_{\text{trap}}
  N_{\text{open}}},
\end{equation}
where the live time is determined by the product of the
trap time $t_{\text{trap}}$ and the number of trap openings
$N_{\text{open}}$.

Average counting rates of alpha and triton particles in one of the four detectors for all of the runs is given in Table~\ref{tab:runs}. The statistical uncertainty from neutron counting is much less than from proton counting and amounts to 0.01\%, or 0.1~s on the measurement of the neutron lifetime. There is a small but non-negligible correction that must be applied for the dead time of the neutron counting. Using the nonparalyzable model of dead time and a value of 0.5~$\mu$s, the correction for the counting rate is $+0.01$\%, and we take the uncertainty to be equal to that value.

\subsubsection{Determination of $\dot{N}_{p}$}

For a particular run within a series, the trapped proton rate is obtained by subtracting the beam-off proton rate for that trap length from the beam-on proton rate.  Series-averaged beam-off rates are determined for each trap length for which there is data.  Where there is no data, the beam-off rate is set to zero.

\paragraph{Beam-on rate}

\begin{figure}
\includegraphics[width=6in]{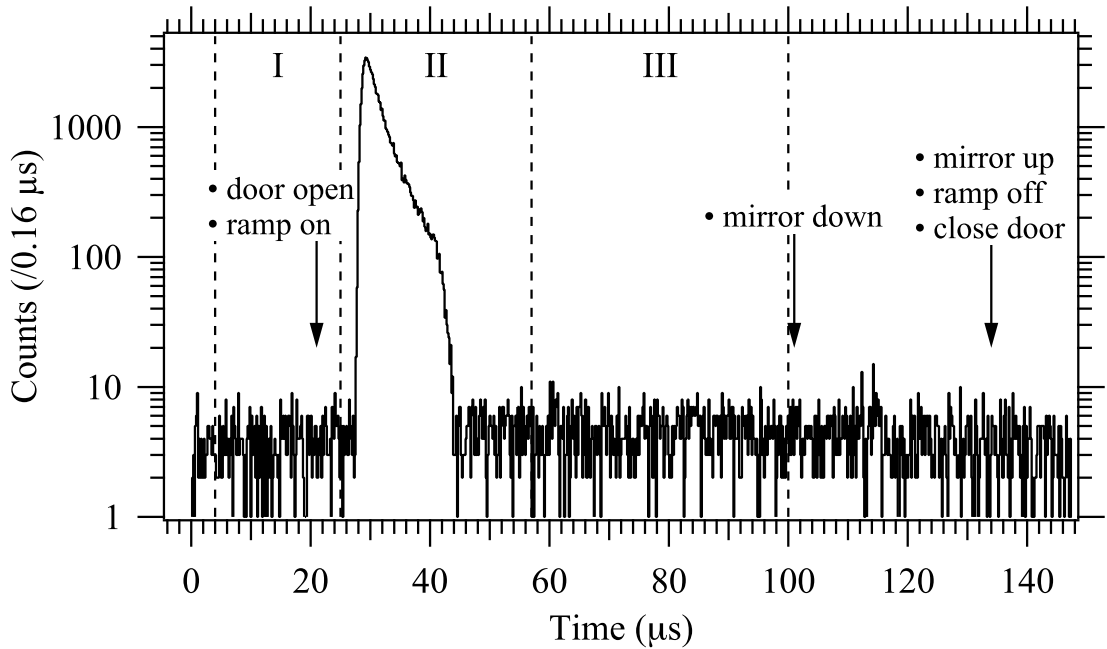}
\caption{\label{fig:TDCSpectrum} Measured timing spectrum showing trap  changes of state (arrows), and three regions of   interest (dashed vertical lines) used to extract proton and background rates.  As a first step in the analysis, this spectrum must be corrected for intrinsic dead time of the TDC.}
\end{figure}

Although the proton rates in this experiment are very low, it is essential that all trapped protons be counted. The beam-on proton rate is determined from the timing spectrum since the dead time correction and background subtraction, while containing important subtleties, are much simpler than for the energy spectrum (compare Figs.~\ref{fig:TDCSpectrum} and \ref{fig:ProtPk}).  Some protons will not appear in the timing spectrum due to the intrinsic dead time of the TDC. This dead time arises due to the fact that the TDC cannot accept multiple stop events.  Thus, one must correct for instances where more than one event could have stopped the TDC. This correction can be done analytically.

The experimentally measured time spectrum, $N_{i}^{\text{exp}}$, is corrected for the dead time of the TDC using the expression
\begin{equation}
  \label{eq:DeadTime}
  N_{i} =
  \frac{N_{i}^{\text{exp}}}{1-\frac{\sum_{j=1}^{i-1}
  N_{j}^{\text{exp}}}{N_{\text{open}}}}.
\end{equation}
In this equation the indices $i$ and $j$ refer to timing channel numbers. For purposes of analysis the timing spectrum is divided into three regions (Fig.~\ref{fig:TDCSpectrum}); regions I and III correspond to background while region II contains the proton peak. Region I occurs before any trapped protons have reached the detector, and region III occurs after the trapped protons have all reached the detector but while the trap is still open.

The backgrounds in the two regions are not equal because there is a small but non-negligible number of neutrons that decay in the trap while the trap is open.  We call this the in-flight contribution. Protons seen from these decays cause the background in region III to exceed that in region I by a small amount.  A correction is made for this in a two step process.  Initially one assumes that the backgrounds are equal on both sides and the mean background per channel multiplied by number of channels in region II is subtracted from the sum over region II.  The remaining sum in region II divided by the live time gives an estimate of the trapped proton rate $r_p^{\text{one}}$ for the second step.  The number of decays in-flight per channel
$N_{\text{inflt}}$ coming from the trapping region is given by $r_p^{\text{one}} N_{\text{open}} t_{\text{pc}}$ where $t_{\text{pc}}$ is the time per TDC channel (either $0.08~\mu$s or $0.16~\mu$s).
$N_{\text{inflt}}$ is subtracted from each count in region III, and the mean background per channel is again multiplied by number of channels in region II and subtracted from the sum over region II to yield a new estimate $r_p^{\text{two}}$.  This value must be adjusted up to take into account the fact that the trap is not trapping for approximately $100~\mu$s out of each cycle.  The number of counts missed as a consequence $N_{\text{missed}}$ is $( t_2 - t_1 ) r_p^{\text{one}} N_{\text{open}}$, where the trap door is opened (closed) at time $t_1$ ($t_2$), less those that are already included in the region II sum.
Those already counted in region II are $N_{\text{inflt}} C$, where $C$ is the number of channels in region II minus 10.  The 10 left-most channels in region II occur before any protons from the trap have
reached the detector.  Finally, the trapped proton rate is given by
\begin{equation}
  \label{eq:prate}
  \dot{N}_{p} = r_p^{\text{two}} + \frac{N_{\text{missed}}}{t_{\text{trap}}
  N_{\text{open}}}.
\end{equation}
The uncertainty on this quantity is obtained by combining the uncertainty on the number of counts in region II and the uncertainty of the background subtraction in quadrature.

\paragraph{Beam-off rate}

From the beam-on data, it is possible to determine tight timing and energy windows that can be used to place cuts on beam-off data.  This allows one to count the number of protons that come during the beam-off time with a very high signal-to-noise ratio.  At the conclusion of a series there are frequently several beam-off runs for each trap length.  The total number of protons divided by the total live time gives an estimate of the beam-off rate. Table~\ref{tab:beam-off} summarizes all of the observed beam-off rates.

The correction for the beam-off background is small but not negligible. If the measured beam-off rates are simply set to zero, the final neutron lifetime changes by $-0.72$~s, or roughly half of the statistical uncertainty of our result.

\subsubsection{Experimental Validation of Proton Counting}

Given the complexity of the analysis procedure and the necessity of counting all the protons at the 0.1~\% level, we devised an experimental test of the proton counting electronics. We generated random events electronically to verify that the number of events determined through the analysis agreed with a straightforward tally of the events in a scaler. The DAQ was modified to simulate a Poisson distribution of pulses by triggering on ``random'' noise from two amplifiers in series.  The door triggered a delayed gate of about $30~\mu$s width.  A logical AND was made using  the random pulses and gate. That output was counted in a scaler channel in the DAQ and also combined with the usual background noise from the surface barrier detector using a logical OR.  The OR output was sent to the TDC stop.  The data were analyzed using the standard procedure with two minor exceptions: the in-flight correction (Section~\ref{sec:metofanal}) was removed and the event window was increased to accommodate the gate.  The agreement between the total scaler counts and number of events produced by the analysis was better than 0.1~\%, indicating the reliability of the proton counting electronics and analysis method at the desired level of precision.

\subsection{Data  Summary}

The data files were written in groups labeled by a series and run number. A series number was incremented when the experimental conditions changed significantly. Within a series, a run number changed when either the trap length was changed or the beam was turned off or on. The number of runs varied among series. Many series were devoted to the study of systematic effects.  Some of these included raster scans of the detector using protons (for alignment), tests of the trapping efficiency as a function of the  door/mirror voltage, and validation of the Monte Carlo results.  In addition, some series were used for calibrations, cool-down periods, detector testing, and other routine measurements.  In the 8-month period from June 2000 through February 2001, we acquired the data used in determining the value of the neutron lifetime reported here. 

\subsubsection{Experimental Parameters}

Table~\ref{tab:runs} gives a listing of the 13 series used in lifetime analysis and some of the relevant experimental parameters. Although the trap is comprised of 16 separate electrodes, only 8 different lengths were used.  The door and mirror electrodes were consistently fixed at a length of three electrodes, and a 3-electrode trap was the shortest used.  That leaves 8 possible trapping schemes of lengths 3 to 10 electrodes.  A typical run consisted of 7 or 8 lengths, but we were able to obtain some measures of the lifetime from runs whose purpose was for systematic measurements; these runs may have only needed 2 or 3 lengths. The trap timing was typically operated at 10~ms although it was changed on occasion to search for systematic differences.

During all of the runs, the cold neutron beam had either 25.4~cm or 20.3~cm of single-crystal bismuth in the beam path. C1 gives the diameter of the first collimator in the beam; it was varied to study possible proton losses due to beam halo. The apertures were made out of either $^6$Li-loaded glass or plastic.  
The counting rate of the alpha particles and tritons, which is directly proportional to the neutron fluence rate, is given by $\langle \dot N_{\alpha+t}\rangle$. The value indicates the rate in one of the four detectors averaged over all the detectors and the entire series. The final three columns of the table give the voltage on the detector, the door/mirror combination, and the maximum voltage on the  ramp of the central electrodes.

\subsubsection{Lifetime Fit Results}

\begin{figure}
\includegraphics[width=6in]{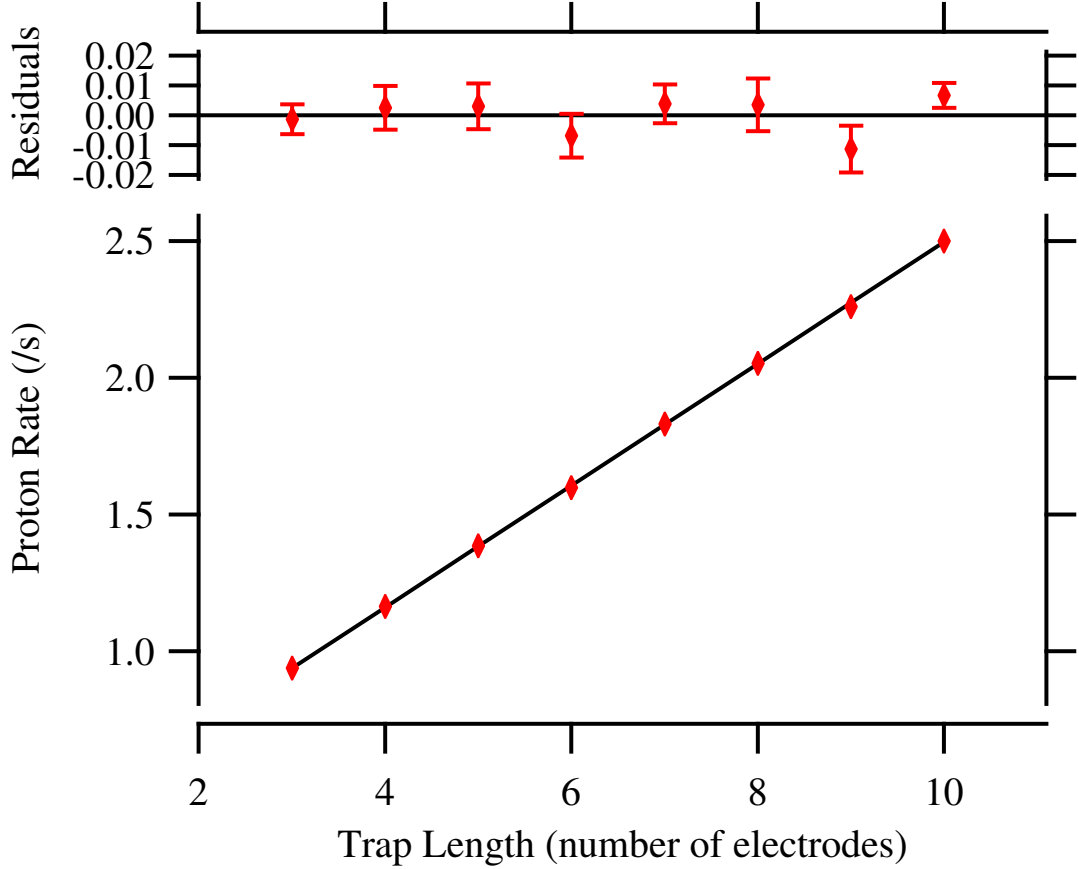}
\caption{\label{fig:ElectrodeFit} Linear fit of typical raw proton count rate ${\dot{N}_{p}}$ versus trap length data. These data have not yet been corrected for nonlinearities discussed in Section~\ref{sec:MC}.}
\end{figure}

In order to obtain the neutron lifetime from Eq.~(\ref{eq:FitFunc}), one performs a least-squares linear fit of ${\dot{N}_{p}}/{\dot{N}_{\alpha+t}}$ versus trap length $L$ for the beam-on runs within a given series.
The trap length is taken to be $21.6 \times n$~mm where $n$ is the number of electrodes and 21.6~mm is the average length of an electrode/spacer combination.  There are, however, subtle corrections to the trapping efficiency that affect the value of the fit. Section~\ref{sec:MC} discusses corrections that arise from the dimensions of each the electrode/spacer combinations differing slightly from the average value, as well as effects from nonuniformity of the magnetic field and beam divergence. Figure~\ref{fig:ElectrodeFit} shows an example of a typical fit before the corrections are applied.

Table~\ref{tab:taus} gives a summary of the fit parameters for each series used in determining the neutron lifetime. The value in the second column is the slope of the linear fit to the proton-to-neutron ratio for each trap length. $\rm Tau_{Lost}$ is the value of the lifetime after the correction for lost protons, which is discussed in Section~\ref{sec:protonlosses}. $\sigma_{Tau}$ is the statistical  uncertainty on the slope. The y-intercept of the fit and its uncertainty are given.  Note that the intercepts are not zero, as one might expect, due to the end effects of the trap. The intercepts, however, should be a constant irrespective of the series, but they differ by as much as 13~\%. This is attributed to rates of background events that are constant throughout a series but may differ among the series.  The final two columns are the chi-squared per degree-of-freedom, $\chi^{2}/dof$,  of the linear fit and the statistical probability for obtaining a worse fit. 

\begin{table*}
  \caption{\label{tab:beam-off}Series-averaged beam off rates (protons
    per second)  The beam-on rate for a 3-electrode long trap is given
    for comparison.}
  \begin{ruledtabular}
    \begin{tabular}{cccccccccc}
      &&\multicolumn{8}{c}{Trap length (electrodes)} \\
      Series & 3 (beam-on rate) & 3 & 4 & 5 & 6 & 7 & 8 & 9 & 10 \\
      \hline
      121 & 0.93 & 0.0000 & 0.0000 & 0.0001 & 0.0001 & 0.0001 & 0.0000 & 0.0001 & 0.0003 \\
      125 & 0.93 & 0.0000 & & & & & & & 0.0001 \\
      130 & 0.74 & 0.0000 & & & & & & & 0.0006 \\
      134 & 1.09 & & & & & & & & 0.0006 \\
      140 & 1.09 & 0.0000 & & 0.0000 & 0.0000 & 0.0000 & 0.0000 & 0.0006 & 0.0007 \\
      142 & 2.30 & 0.0000 & 0.0000 & 0.0000 & & 0.0002 & 0.0003 & 0.0003 & 0.0006 \\
      143 & 1.97 & 0.0002 & 0.0003 & 0.0006 & 0.0003 & 0.0000 & 0.0006 & 0.0017 & 0.0004 \\
      149 & 1.84 & 0.0000 & 0.0000 & 0.0000 & 0.0000 & 0.0000 & 0.0003 & 0.0003 & 0.0000 \\
      151 & 1.88 & 0.0000 & 0.0000 & 0.0006 & 0.0000 & 0.0000 & 0.0000 & 0.0000 & 0.0011 \\
      154 & 1.98 & 0.0069 & 0.0008 & & 0.0017 & 0.0004 & 0.0011 & 0.0061 & 0.0109 \\
      155 & 2.01 & 0.0050 & 0.0006 & 0.0011 & 0.0011 & 0.0008 & 0.0015 & 0.0028 & 0.0296 \\
      166 & 1.92 & 0.0003 & & & & 0.0003 & & & \\
      170 & 1.86 & 0.0031 & 0.0033 & 0.0057 & 0.0071 & 0.0065 & 0.0049 & 0.0040 & 0.0051 \\
    \end{tabular}
  \end{ruledtabular}
\end{table*}

\begin{table*}
\caption{\label{tab:runs} Some of the relevant run parameters for the series used in the determination of the neutron lifetime.  A discussion of the parameters is in the text.}
\begin{ruledtabular}
\begin{tabular}{lcccccccccc}
Series 	& 	Date & 	\# Runs & 	\# Trap & 	Timing & 	Bi & 	C1 &$\langle \dot N_{\alpha+t}\rangle$&  ${\rm 
V_{detector}}$ & ${\rm V_{door}}$ & ${\rm V_{ramp}}$ \\
 &   & & Lengths & (ms) & (cm) & (cm) &(/s)& (kV) & (V) & (V)\\
\hline
121 & 06/27/00	&	38 & 8 & 10 & 25.4 & 3.17		&156&	27.5 & 825 & 45 \\
125 & 08/25/00	&	7 & 2 & 10 & 25.4 & 3.17 		&153&	27.5 & 950 & 45 \\
130 & 09/09/00	&	7 & 2 & 10 & 20.3 & 3.17 		&123&	30.0 & 850 & 45 \\
134 & 09/15/00 &	3 & 2 & 10 & 20.3 & 3.17 		&179&	30.0 & 950 & 20 \\
140 & 09/28/00 &	15 & 8 & 10 & 20.3 & 3.17 	&181&	32.5 & 850 & 20 \\
142 & 10/04/00 &	18 & 8 & 10 & 20.3 & 5.08 	&383&	32.5 & 800 & 20 \\
143 & 10/10/00 &	18 & 8 & 5 & 20.3 & 4.45 		&323&	32.5 & 800 & 20 \\
149 & 10/26/00 &	17 & 8 & 10 & 20.3 & 4.45 	&317&	27.5 & 800 & 20 \\
151 & 11/01/00 &	11 & 7 & 10 & 20.3 & 4.45 	&318&	32.5 & 800 & 20 \\
154 & 11/13/00 &	21 & 8 & 10 & 20.3 & 4.45 	&326&	30.0 & 800 & 20 \\
155 & 11/19/00 &	29 & 8 & 10 & 20.3 & 4.45 	&329&	32.5 & 800 & 20 \\
166 & 12/14/00 &	13 & 3 & 10 & 20.3 & 4.45 	&322&	27.5 & 800 & 20 \\
170 & 02/22/01 &	21& 8  & 10 & 20.3 & 4.45 	&314&	27.5 & 800 & 20 \\
\end{tabular}
\end{ruledtabular}
\end{table*}

\section{Systematic Corrections and Uncertainties}\label{sec:systematics}

This section describes the systematic corrections that modify the measured neutron lifetime. The corrections are organized into four subsections that discuss the systematics related to neutron counting, beam halo, the proton trap, and proton counting. Table~\ref{tab:Sys} summarizes all the systematic corrections and their associated uncertainty and directs the reader to the specific section where the correction is determined.

\begin{table*}
\caption{\label{tab:Sys} Summary of the systematic corrections and  uncertainties for the measured neutron lifetime. Several of these terms also appear in Table~\ref{tab:corrections} where it is seen that their magnitude depends weakly on the running configuration.  In those cases, the values given in this table are the configuration average. The origin of each quantity is discussed in the section noted in the table. }
\begin{ruledtabular}
\begin{tabular}{lddc}
Source of Correction & \multicolumn{1}{c}{Correction (s)} & \multicolumn{1}{c}{Uncertainty (s)} &Section\\
\hline
$^{6}$LiF deposit areal density 				& 		& 2.2 	&  \ref{sec:neudetlos}\\
$^{6}$Li cross section 						& 		& 1.2 	& \ref{sec:NeutronCounting}\\
Neutron detector solid angle 					& 		& 1.0 	&\ref{sec:NeutronCountingEff}\\
Absorption of neutrons by ${}^{6}$Li	& +5.4	 	& 0.8 	&\ref{sec:NeutronAbsorption}\\
Neutron beam profile and detector solid angle 		& +1.3 	& 0.1 	&\ref{sec:NeutronAbsorption}\\
Neutron beam profile and ${}^{6}$Li deposit shape	&-1.7 	&0.1 		&\ref{sec:NeutronAbsorption}\\
Neutron beam halo 							& -1.0 	& 1.0  	&\ref{sec:HaloUnc}\\
Absorption of neutrons by Si substrate 			&+1.3 	&0.1 		&\ref{sec:NeutronAbsorption}\\
Scattering of neutrons by Si substrate			&-0.2 	&0.5 		&\ref{sec:NeutronScat}\\
Trap nonlinearity 							&-5.3 	&0.8 		&\ref{sec:MC}\\
Proton backscatter calculation 					& 		&0.4 		&\ref{sec:BkScUnc}\\
Neutron counting dead time					&+0.1	&0.1		&\ref{sec:NeutronCounting}\\
\hline
Proton counting statistics 						& 		& 1.2 	&\ref{sec:ExtraptoZero}\\
Neutron counting statistics 					& 		& 0.1 	&\ref{sec:NeutronCounting}\\
\hline
Total &-0.1 &3.4\\
\end{tabular}
\end{ruledtabular}
\end{table*}

\subsection{Determination of Neutron Detector Losses}\label{sec:neudetlos}

To first order the observed $\text{alpha}$ and $\text{triton}$ particle rate $\dot{N}_{\alpha+t}$ is given by
\begin{equation}
  \label{eq:idealn}
  \dot{N}_{\alpha+t} = 2 \frac{\Omega(0,0)}{4\pi}
  \frac{N_\text{A}\bar{\rho}\sigma_{0}}{A} \dot{N}_{n},
\end{equation}
where $\dot{N}_{n} = \int_{A} da \int_{v} dv \, \frac{v_{o}}{v}\, I(v) \phi(x,y)$ is the 2200 m/s equivalent neutron rate, $I(v)$ is the fluence rate per unit velocity, $\phi(x,y)$ is the areal distribution of the neutron intensity on the target, $A = 6.01512$~g/mol is the atomic weight of ${}^{6}$Li, $N_\text{A} = 6.0221415 \times 10^{23}$~mol$^{-1}$ is the Avogadro constant, $\bar{\rho} = (39.30 \pm 0.10)~\mu$g/cm$^{2}$ is the average areal density of ${}^{6}$Li of the deposit, $\sigma_{0} = (941.0 \pm 1.3$)~b is the $^{6}$Li absorption cross section at 2200\,m/s, and $\Omega(0,0) / 4\pi = 0.004196 \pm 0.1$~\% is the fractional solid angle subtended at the center of the deposit by the detector.  The factor of two is required because both alphas and tritons are detected.

Building on Eqs.~(\ref{eq:mean_alpha}) and (\ref{eq:neutroneff}) and taking into account neutron attenuation in the deposit, a more accurate expression for $\dot{N}_{\alpha+t}$ is

\begin{eqnarray}\label{eq:cor1}
\dot{N}_{\alpha+t} =\int_{A}da\int_{v} dv ( 1-&e&^{-\rho(x,y)\sigma_{0}\frac{v_{o}}{\mathstrut v}\frac{N_\text{A}}{\mathstrut A}})\nonumber\\&&I(v) \phi(x,y)\frac{\Omega(x,y)}{2\pi},
\end{eqnarray}
where $\Omega(x,y)$ is the detector solid angle and $\rho(x,y)$ is the areal density of the $^{6}$Li deposit.  By design, $\Omega(x,y)$ is nearly constant over the beam distribution and $\rho(x,y)$ is both small and nearly constant, making the ratio of Eqs.~(\ref{eq:cor1}) and (\ref{eq:idealn}) nearly one.  Nevertheless, the precise ratio is a correction that must be included in each of our 13 series-based lifetime values (Table~\ref{tab:taus}).  The solid angle $\Omega(x,y)$ is easily calculated given the positions of the deposit and the apertures; $\rho(x,y)$ was measured during the manufacture of the deposits; and $\phi(x,y)$ was measured using the dysprosium image method (Section~\ref{DyScan}). 

Rather than calculate the ratio of Eqs.~(\ref{eq:cor1}) and (\ref{eq:idealn}) directly, we decompose the calculation into three terms with corresponding numerical corrections $c_1$, $c_2$, and $c_3$
(see Table~\ref{tab:corrections}), each of which leads to an additive correction to the lifetime.  The first takes into account the exponential attenuation of neutrons in the ${}^{6}$Li deposit ($c_1$). The second accounts for the neutron beam profile and position dependent detector solid angle ($c_2$), and finally, the third accounts for the neutron beam profile and the position dependent ${}^{6}$Li areal density ($c_3$).  Because each of these effects is small, nothing of consequence is lost by separately calculating them. In addition to these effects, two additional neutron loss mechanisms are included in this section ($c_4$ and $c_5$).  Before the neutrons pass through the ${}^{6}$Li deposit, they pass through two perfect crystal Si wafers: one serves as the backing that holds the ${}^{6}$Li deposit and the other is situated between the neutron detector and the proton trap in order to prevent charged particles from streaming into the trapping region.  There will be neutron absorption ($c_4$) and scattering ($c_5$) from these wafers.  A final correction ($c_6$) accounting for a neutron beam halo is discussed in Section~\ref{sec:Halo}.

\begin{table*}
\caption{\label{tab:taus} Results from the fit of proton-to-neutron  counts versus trap length for the series used in the determination of  the neutron lifetime.  Column 2 contains the measured lifetime; in column 3 a small correction has been made for lost protons (see Section~\ref{sec:protonlosses}); column 4 contains the 1-$\sigma$   statistical uncertainty; column 5 contains the proton-neutron ratio
  at zero trap length (the intercept); column 6 contains its   uncertainty; column 7 gives the reduced chi-squared for the fit;  and column 8 gives the probability of getting a larger reduced chi-squared.
}
\begin{ruledtabular}
\begin{tabular}{lccccccc}
Series & Tau & ${\rm Tau_{Lost}}$ & $\sigma_{{\rm Tau}}$ & y-intercept &
$\sigma_{{\rm y-intercept}}$
& $\chi^{2}/\text{dof}$ & Probability\\
  & (s)  & (s) & (s) & $\times 10^{8}$ & $\times 10^{8}$ & & \\
\hline

121, 122, 124	& 892.4 & 892.3 & 3.1 & 1.348 & 0.027 & 1.061 & 0.370 \\
125 			& 884.1 & 884.1 & 4.0 & 1.397 & 0.031 & 0.425 & 0.832 \\
130 			& 885.9 & 885.8 & 4.8 & 1.362 & 0.034 & 2.235 & 0.048 \\
134 			& 889.1 & 889.0 & 5.0 & 1.464 & 0.043 & 0.269 & 0.604 \\
140 			& 889.3 & 889.3 & 3.0 & 1.387 & 0.025 & 0.651 & 0.812 \\
142 			& 891.8 & 891.8 & 2.4 & 1.375 & 0.022 & 0.928 & 0.536 \\
143 			& 892.3 & 892.2 & 2.3 & 1.375 & 0.018 & 1.604 & 0.059 \\
149 			& 909.9 & 902.7 & 2.6 & 1.285 & 0.020 & 1.155 & 0.300 \\
150, 151 		& 901.1 & 897.4 & 2.6 & 1.326 & 0.021 & 0.763 & 0.651 \\
154 			& 888.0 & 886.2 & 2.2 & 1.402 & 0.018 & 0.913 & 0.566 \\
155 			& 890.7 & 889.4 & 2.5 & 1.426 & 0.019 & 0.940 & 0.553 \\
165, 166 		& 899.0 & 897.7 & 4.4 & 1.376 & 0.024 & 1.375 & 0.177 \\
170 			& 888.5 & 886.2 & 3.5 & 1.267 & 0.030 & 1.304 & 0.168 \\

\end{tabular}
\end{ruledtabular}
\end{table*}

Table~\ref{tab:configurations} lists the four configurations of beam collimation and thickness of Bi filter material, and Table~\ref{tab:corrections} provides values and uncertainties (in seconds) for the six corrections discussed in this section for each of these configurations.  Multiple configurations were employed in order to check for unknown systematic effects.  At our level of accuracy, none were seen.  The sum of the configuration-appropriate column in Table~\ref{tab:corrections} has been added to each of the thirteen values of $\tau_n$ that appear in Table~\ref{tab:corrections}. The uncertainties are provided in the last column of Table~\ref{tab:corrections}.  These uncertainties are added in quadrature to those associated with other systematic effects in order to obtain a final systematic uncertainty.  The remainder of this section discusses corrections 1 through 5 and their uncertainties in detail.

\begin{table*}
  \caption{The 13 runs enumerated in Table~\ref{tab:taus} were carried     out under four different configurations of beam collimation and thickness of Bi filter material.  These configurations are     enumerated here along with the number of runs belonging to each,  the average measured neutron equivalent rate, and the estimated  mean beam wavelength. The resulting differences in beam size and wavelength distribution lead to four different sets of systematic adjustments for the 13 lifetimes.}
  \begin{ruledtabular}
  \begin{tabular}{ddccdd}
    \multicolumn{2}{c}{Configuration} & Label & Number &
    \multicolumn{1}{c}{Measured neutron equivalent} &
    \multicolumn{1}{c}{Estimated mean} \\
    \multicolumn{1}{c}{C1 (cm)} & \multicolumn{1}{c}{Bi thickness (cm)} & & of Runs& \multicolumn{1}{c}{rate ($\times10^{7}$ s${}^{-1}$)} &
    \multicolumn{1}{c}{wavelength (nm)} \\
    \hline
    4.4 & 20.3 & A & 7 & 4.14 & 0.457 \\
    3.2 & 20.3 & B & 3 & 2.30 & 0.427 \\
    3.2 & 25.4 & C & 2 & 1.99 & 0.447 \\
    5.1 & 20.3 & D & 1 & 4.93 & 0.474 \\
  \end{tabular}
  \end{ruledtabular}
  \label{tab:configurations}
\end{table*}

\begin{table*}
  \caption{Systematic effects and uncertainties associated with neutron counting.  Each correction is discussed in detail in the text.}
  \begin{ruledtabular}
    \begin{tabular}{lcddddd}
 & Variable & \multicolumn{4}{c}{Adjustment to lifetime (s)} & \multicolumn{1}{c}{Uncertainty} \\
Configuration & name & \multicolumn{1}{c}{A} & \multicolumn{1}{c}{B} & \multicolumn{1}{c}{C} & \multicolumn{1}{c}{D} &\multicolumn{1}{c}{(s)}\\
      \hline
      Absorption of neutrons by ${}^{6}$Li & $c_1$ & 5.5 &  5.2 & 5.4 & 5.6 & 0.8 \\
      Neutron beam profile and detector solid angle & $c_2$ & 1.5 & 1.0 & 1.0 & 1.7 & 0.1 \\
      Neutron beam profile and ${}^{6}$Li deposit shape & $c_3$ & -1.7 & -1.9 &  -1.9 & -1.6 & 0.1 \\
      Absorption of neutrons by Si substrates & $c_4$ & 1.3 & 1.2 & 1.3 & 1.3 & 0.1  \\
      Scattering of neutrons by Si substrates & $c_5$ & -0.2 & -0.2 & -0.2 & -0.2 & 0.5 \\
    \end{tabular}
  \end{ruledtabular}
  \label{tab:corrections}
\end{table*}


\subsubsection{A Model of $I(v)$}\label{sec:VelocModel}

Corrections $c_1$ and $c_4$ in Table~\ref{tab:corrections} depend on $I(v)$ which in turn depends upon the chosen running configuration (Table~\ref{tab:configurations}).  To address these corrections, a detailed model for the detected neutron fluence was developed.  Conceptually, the model consists of three factors:
\begin{enumerate}
\item A function for the cold source brightness (neutrons per second   per square centimeter per steradian per angstrom).  This function   was constructed by starting from an accurate wavelength distribution measurement made at the end of one of the cold-neutron guide tubes and dividing it by the material and guide transmission functions. After making minor adjustments to guide tube reflectivities, the final
function reproduces the total flux measured at the ends of each  guide quite well~\cite{COO04}.
\item Transmission functions for each of the materials through which the beam passes, including Al (1.1~cm), CO${}_2$ (68.9~cm), Mg (0.15~cm), He gas (127~cm), air (3.7~cm), pyrolytic graphite (0.4~cm), and Bi at liquid nitrogen temperature (20.3~cm or 25.4~cm as shown in Table~\ref{tab:configurations}).  Cross section evaluations are included for each of these materials.  In the case of Bi and pyrolytic graphite, these calculations are problematic because these are perfect or nearly perfect crystals and the scattering through them will be dependent upon their imperfectly known state.
\item A transmission function for the $^{58}$Ni-coated neutron guide tubes and experimental collimation.  A Monte Carlo program using a simple elastic collision model was used for this purpose.  The reflectivity of the guide tubes is a critical parameter in this calculation.  In the course of matching the cold source brightness to neutron fluence rates measured at the end of each guide, a reflectivity for each guide was obtained.  The value for our guide is 0.963.
\end{enumerate}
Terms 1 and 2 involve analytic functions whereas the third calculation results in transmission versus wavelength pairs at a specified set of wavelengths.  Fortunately, these points fit very well to the equation
\begin{equation}
  \label{eq:erf}
  T(x) = a \, \text{erf}(bx+c) + d,
\end{equation}
where ``erf'' stands for the error function and $a$, $b$, $c$, and $d$ are fit parameters.  $I(v)$ is proportional to the product of the source brightness function, the filter transmission function, and $T(x)$.

The average neutron equivalent rate measured in each of the four configurations is given in column 5 of Table~\ref{tab:configurations}. This same quantity can be predicted by numerically integrating
\begin{equation}
  \label{eq:numint}
  \int_{v} dv \, \frac{v_{o}}{v}\, I(v)\,.
\end{equation}
When this is done, the observed neutron equivalent rate is roughly 65~\% of the predicted value for all four configurations.  Another  issue is the rate ratio of rows 2 and 3, which differ from one by roughly twice as much as the predicted value and where the only difference is the amount of Bi present.  Chilled, perfect crystal Bi was used to absorb unwanted photons coming from the core because it weakly scatters those neutrons whose wavelengths are between 0.2\,nm and 0.7\,nm, the Bragg cutoff.  This band encompasses the bulk of the neutrons.  In order to account for the ratio and absolute value
discrepancies, we include in our transmission function one additional term that corresponds to additional neutron scattering between 0.2\,nm and 0.7\,nm.  An essentially wavelength-independent
transmission factor is tuned so as to obtain the best possible agreement with observed rates and the row 2 and 3 rate ratio. When this is done, the all-around agreement is excellent.  This additional term changes the implied shape of the neutron wavelength spectrum and consequently the correction factor $c_1$.

\subsubsection{Neutron Absorption Corrections}\label{sec:NeutronAbsorption}

The relative correction accounting for absorption of
neutrons by the Li foil is given by
\begin{equation}
  \label{eq:cor_1}
  c_1 = \frac{\displaystyle \bar{\rho}\sigma_{0}
  \frac{N_\text{A}}{A}v_{o}\int_{v}\! dv \, \frac{1}{v} \, I(v)}{\displaystyle
  \int_{v}\! dv \, ( 1 - e^{\scriptstyle -\bar{\rho}\sigma_{0}\frac{
  v_{o}}{\mathstrut v}\frac{\mathstrut
  N_\text{A}}{\mathstrut A}} )\, I(v)},
\end{equation}
where the integration over $x$ and $y$ cancels in this estimate.  This correction is sensitive to the wavelength distribution.  The uncertainty of this calculation is taken to be 100~\% of the difference between the lifetime resulting from analysis that includes the additional neutron scattering term in the filter transmission function and the lifetime resulting from analysis without it, or 0.8\,s.

\begin{figure}
  \includegraphics[width=6in]{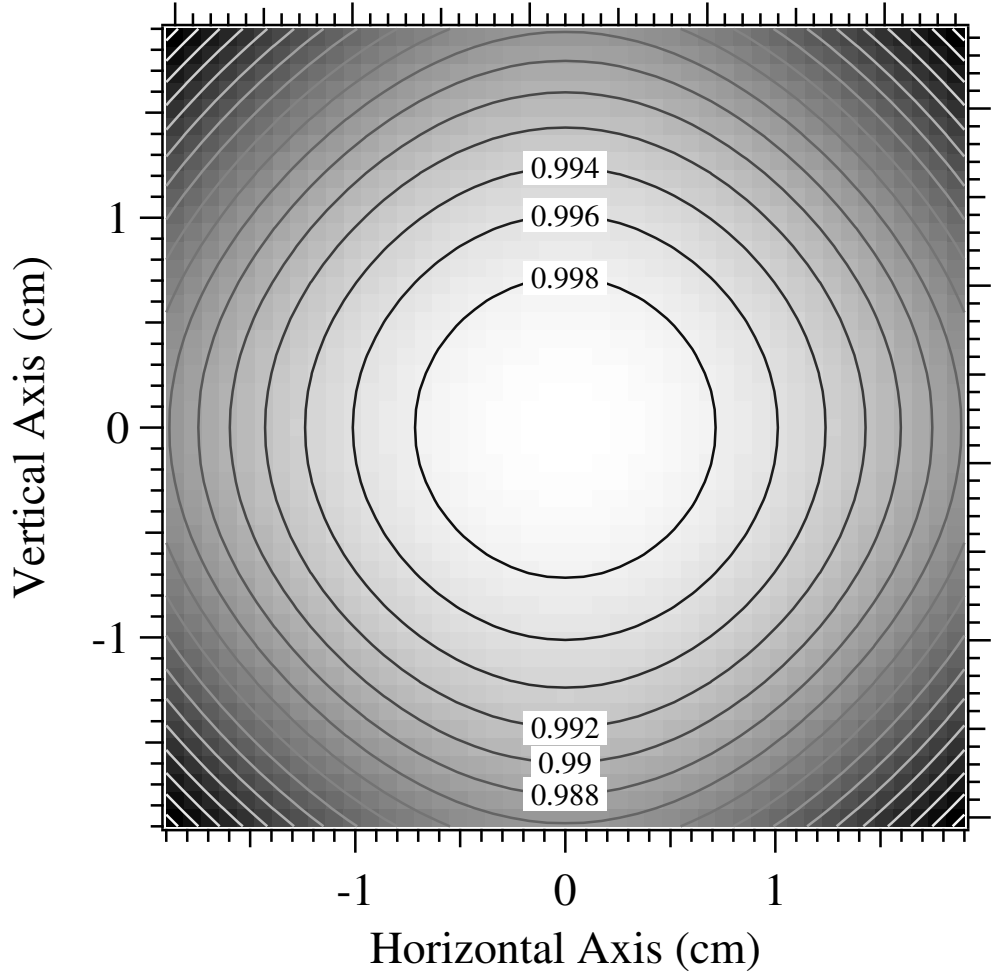}
  \caption{Relative solid angle efficiency of the neutron detector as a function of position on the Li deposit. The value is normalized to unity at the center of the deposit.  The deviation from circular contour lines reflects the existence of the four symmetrically placed detectors.}
  \label{fig:alphadeteff}
\end{figure}

\begin{figure}
  \includegraphics[width=6in]{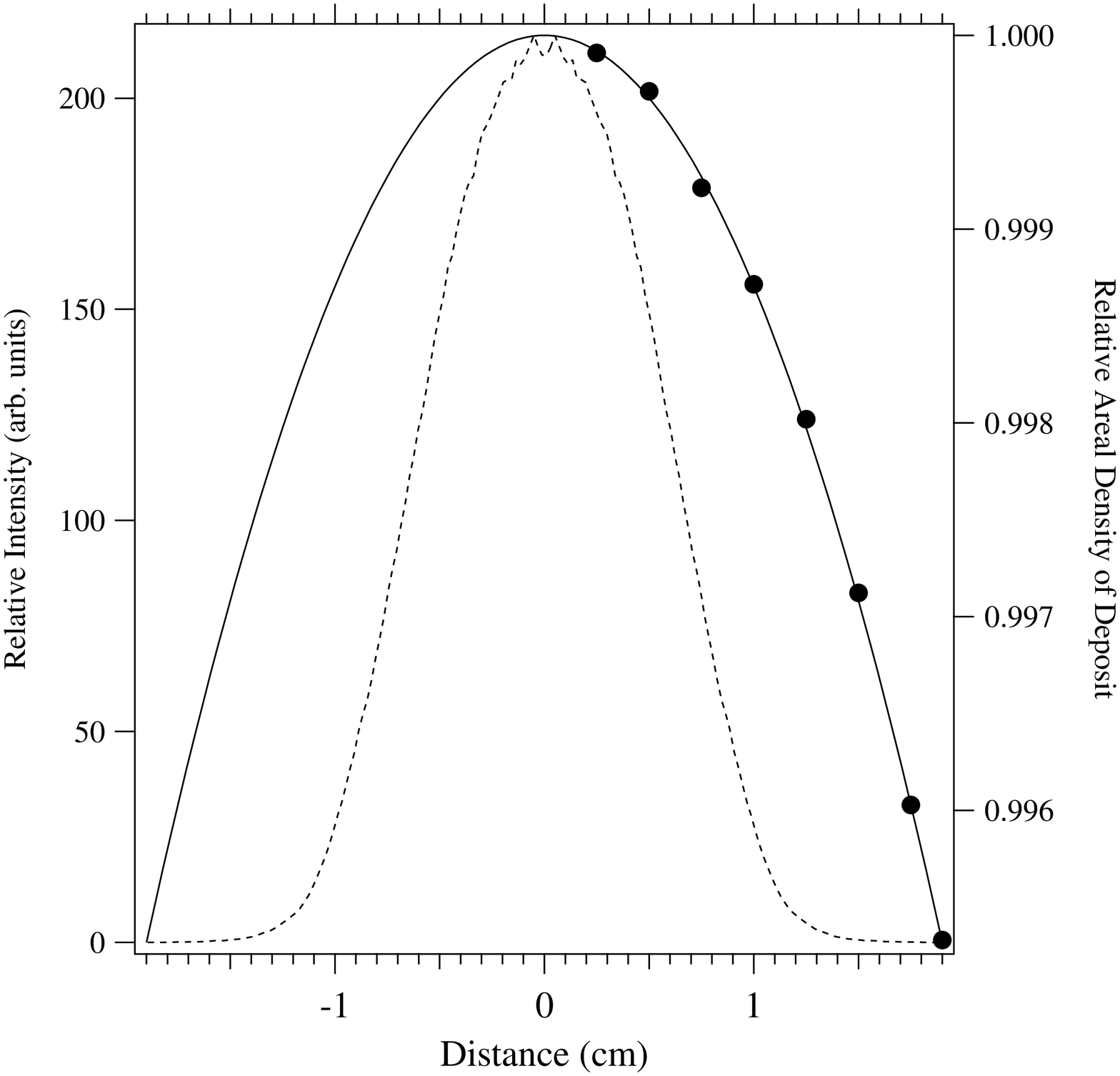}
  \caption{Measured neutron beam intensity as a function of radius on the deposit (dashed line) as well as data and fit for the Li deposit areal density versus radius (solid circles and line).}
  \label{fig:foilandbeam}
\end{figure}

Corrections $c_2$ and $c_3$ in Table~\ref{tab:corrections} require knowledge of the neutron beam profile at the Li deposit.  This was measured for configuration A using the dysprosium image method (see Table~\ref{tab:configurations}).  The radial distribution is shown as the gray curve in Fig.~\ref{fig:foilandbeam}.  The curve fits well to the function
\begin{equation}
  \label{eq:fofr}
  \phi(x,y) = \phi(r) = a \, e^{-\frac{r^{b}}{2 c^{\mathstrut 2}}},
\end{equation}
where parameters $a$, $b$, and $c$ are determined from the fit. Typical values of $b$ and $c$ are 2.59 and 9.73 when the radius $r$ is measured in millimeters.  The dysprosium image provides a picture of the neutron beam intensity distribution, but since the neutron absorption probability in the deposit is neither unity nor small, the neutron velocity weighting of the intensity distribution necessarily lies between these two limits.  Monte Carlo calculations reveal that the difference is negligible for the lifetime experiment.  $c_2$ and $c_3$ will, of course, depend on the values in column 1 of Table~\ref{tab:configurations}.  In order to estimate the corrections for configurations B, C, and D, Eq.~(\ref{eq:fofr}) was used with the radius multiplicatively scaled by the ratio of the predicted beam hard radii at the deposit
\begin{equation}
  \label{eq:rscaling}
  \phi_{j}(r) = \phi_{\text{A}}(r
  \frac{r_{\text{A}}}{r_{j}}),
\end{equation}
where $r_{}$ is the calculated hard radii for configuration $j \in \{A, B, C, D\}$.  The straightforward formula for the hard radius is
\begin{equation}
  \label{eq:hardradius}
  r = r_{\text{qtz}} + \frac{r_{\text{c1}} + r_{\text{c2}}}{z_{12}}
  z_{\text{dep-qtz}},
\end{equation}
where $r_{\text{qtz}} = 3.7$\,mm is the radius of the quartz tube, $z_{\text{dep-qtz}} = 1447$\,mm is the distance from the end of the quartz tube to the Li deposit, $r_{\text{c1}}$ and $r_{\text{c2}} = 4.2$\,mm are the radii of the two defining beam apertures, and $z_{12} = 4880$\,mm is the distance between the two apertures.  For the three values of $r_{\text{c1}}$ appearing in Table~\ref{tab:configurations}, the three hard radii at the Li deposit are 9.7\,mm, 11.6\,mm, and 12.5\,mm giving scaling factors of 1.19, 1.00, and 0.92, respectively. The uncertainties for $c_2$ and $c_3$ (0.1\,s) are taken to be 100~\% of the lifetime difference between scaling and not scaling $r$.

The solid angle subtended by the four surface barrier detectors as a function of position on the deposit $\Omega(x,y)$ can be calculated from the known geometry.  This function is shown graphically in
Fig.~\ref{fig:alphadeteff}.  The relative correction for the position dependent detector solid angle is given by
\begin{equation}
  \label{eq:cor_2}
  c_2 = \frac{\displaystyle \int_{-19}^{19} dy
  \int_{-\sqrt{19^2-y^2}}^{\sqrt{19^2-y^2}} dx \, \phi(\sqrt{x^2
  + y^2})}{\displaystyle \int_{-19}^{19} dy
  \int_{-\sqrt{19^2-y^2}}^{\sqrt{19^2-y^2}} dx \, \Omega(x,y) \phi(\sqrt{x^2
  + y^2})},
\end{equation}
where $\phi(r)$ is given by Eq.~(\ref{eq:fofr}) and 19\,mm is the radius of the deposit.

The radial thickness profile of the deposit has been measured to be
\begin{equation}
  \label{eq:foilprofile}
  \rho(r) = \bar{\rho} \frac{\displaystyle
  1-(1-0.995)\left(\frac{r}{19}\right)^2}{\displaystyle
  1-\frac{\mathstrut 0.005}{2}},
\end{equation}
where $r$ is measured in millimeters.  Measured relative thicknesses are shown in Fig.~\ref{fig:foilandbeam}.  The thickness falls off by 0.5~\% at the edge of the deposit ($r = 19$\,mm).  The denominator ensures that the average areal density is $\bar{\rho}$.

The relative correction for the position dependent ${}^{6}$Li areal density is given by
\begin{equation}
  \label{eq:cor_3}
  c_3 = \frac{\displaystyle \int_{0}^{\mathstrut 19} \phi(r) \,
  dr}{\displaystyle \int_{0}^{19} \rho(r) \phi(r) \, dr},
\end{equation}
where $\phi(r)$ is given by Eq.~(\ref{eq:fofr}) and 19\,mm is the radius of the deposit.

To an excellent approximation the relative correction for neutron absorption in Si is given by
\begin{equation}
  \label{eq:Siabs}
  c_4 = 2 \sigma_{\text{Si}} \frac{\bar{\lambda}}{\lambda_0}
  \rho_{\text{Si}} t_{\text{Si}},
\end{equation}
where the factor of two is included because the beam passes through two wafers, $\sigma_{\text{Si}} = 0.171$\,b is the Si absorption cross section at 2200\,m/s, $\bar{\lambda}$ appears in column 6 of
Table~\ref{tab:configurations}, $\lambda_0 = 0.1798$\,nm is the neutron wavelength corresponding to 2200\,m/s, $\rho_{\text{Si}} = 4.996 \times 10^{22}$\,/cm${}^{3}$ is the atom density of Si, and
$t_{\text{Si}} = 0.34$\,mm is the thickness of one wafer.  The uncertainty is taken to be the spread in the four values of $c_4$ given in Table~\ref{tab:corrections}.

\subsubsection{Neutron Scattering}\label{sec:NeutronScat}

\begin{figure}
  \includegraphics[width=6in]{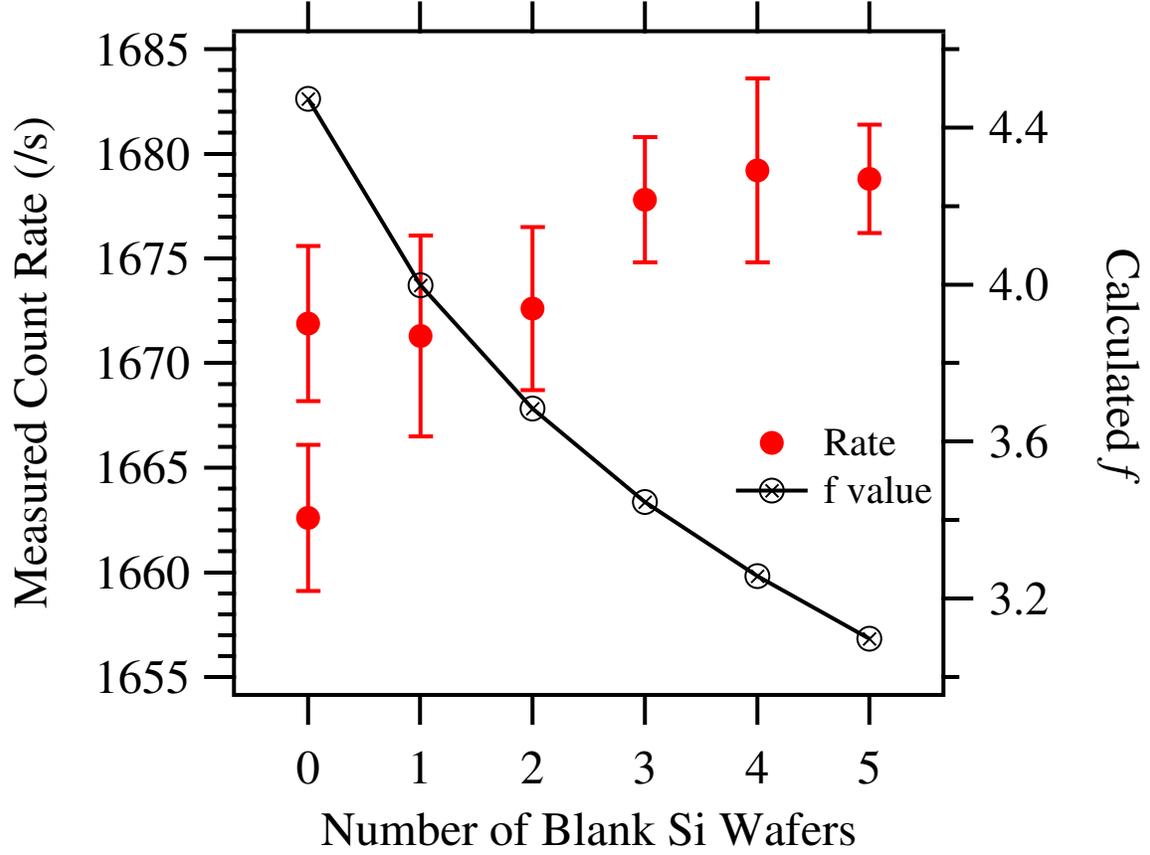}
  \caption{Measured count rate and scattering-induced enhancement factor $f$ versus the number of Si blanks placed behind the detection deposit.}
  \label{fig:nscattering}
\end{figure}

To investigate the correction for neutron loss due to scattering in the Si wafers, a separate experiment was carried out wherein the neutron detector was placed in a thermal neutron beam of average
wavelength 0.17\,nm and the output rate was measured as identical Si wafers were stacked one after another behind the neutron absorbing deposit.  The measured count rate as a function of number of additional wafers can be seen in Fig.~\ref{fig:nscattering}.  In the geometry that was used, the neutron absorber faced the incoming neutron beam while the blank wafers were added downstream.  When neutrons scatter in the Si, some are returned back through the neutron absorber (${}^{10}$B in this measurement) where they will again contribute to the observed signal. If they come back at an angle relative to their incoming directions, they can see a greater thickness of absorber and thus contribute more to the signal.  There are two quantities of interest: the probability of scattering in a single wafer $\epsilon_{\text{Si}}$ and the enhancement $f(i)$ experienced due to the neutrons passing through the absorber at varying angles. The expression that gives the observed rate is
\begin{equation}
  \label{eq:roger}
  R(i) = a ( 1 + \frac{f(i)}{2} \, i \, \epsilon_{\text{Si}} ),
\end{equation}
where $a$ is an overall constant and the index $i$ is the number of Si wafers behind the absorbing deposit starting with $i=1$. The factor of two appears because half of the neutrons scatter into the forward direction thereby escaping detection.

The function $f(i)$ was calculated in a simple Monte Carlo program.  It turns out that $\epsilon_{\text{Si}}$ is much larger than the incoherent scattering cross section prediction for perfect crystal Si wafers.  We conclude that this scattering above the perfect crystal prediction is occurring at the damaged surfaces of the wafers.  In the Monte Carlo program, a 50~\% probability is given to the neutron for scattering at either face.  Using this model, values of $f(i)$ were tabulated and shown in Fig.~\ref{fig:nscattering}). Finally, with these values and the measured rates, a value of $\epsilon_{\text{Si}} = 0.00105$ is obtained.  This value agrees reasonably well with a previous result obtained through a different procedure, suggesting that our assumption of isotropic scattering is correct.

In the lifetime experiment, there are two Si wafers.  The one that holds the absorbing $^{6}$LiF deposit is oriented with the Si facing upstream.  A second wafer is located further upstream to prevent charged particles from streaming between the two regions.  The expression that accounts for neutron scattering is
\begin{equation}
  \label{eq:nscat}
  c_5 = \epsilon_{\text{Si}} \left((1 - \frac{f}{2}) +
  (1-\frac{\Omega_{\text{dep}}}{4\pi})\right),
\end{equation}
where $\Omega_{\text{dep}} = 7.85 \times 10^{-3}$\,sr is the solid angle subtended by the absorbing deposit at the location of the upstream Si wafer.  The second term in Eq.~(\ref{eq:nscat}) accounts for the neutrons lost due to scattering in the distant wafer, while the first term accounts for scattering from the wafer holding the absorbing Li. The occurrence of $1-\frac{f}{2}$ rather than $\frac{f}{2}$ which appears in Eq.~(\ref{eq:roger}) accounts for the opposite orientation of the absorbing deposit relative to the neutron beam.  Again the Monte Carlo program provides a value for the enhancement $f = 4.46$. Combined with $\epsilon_{\text{Si}}$ from the independent experiment, Eq.~(\ref{eq:nscat}) can be evaluated.  The two terms in Eq.~(\ref{eq:nscat}) each contribute about 1\,s to the lifetime, \emph{in the opposite sense}.  The uncertainty on this quantity is taken as 0.5~s.  This allows for the possibility that the scattering occurs throughout the Si wafer or at either surface.

\subsection{Neutron Beam Halo}\label{sec:Halo}

The neutron beam was tightly collimated to ensure that less than 0.1~\% of the neutrons would lie in the halo of the beam.  In other words, greater than 99.9~\% of the neutrons exiting the final guide tube and passing through the proton trap are incident upon the 38-mm diameter LiF deposit.  Similarly for protons, the largest extent of the neutron beam in the trap must be such that greater than 99.9~\% of the protons created in the trap will follow magnetic field lines that terminates on the active area of the silicon detector. One must include the cyclotron  motion of the protons as they travel toward the detector.

\subsubsection{Dysprosium Image Method}\label{DyScan}

\begin{figure}
\includegraphics[width=6in]{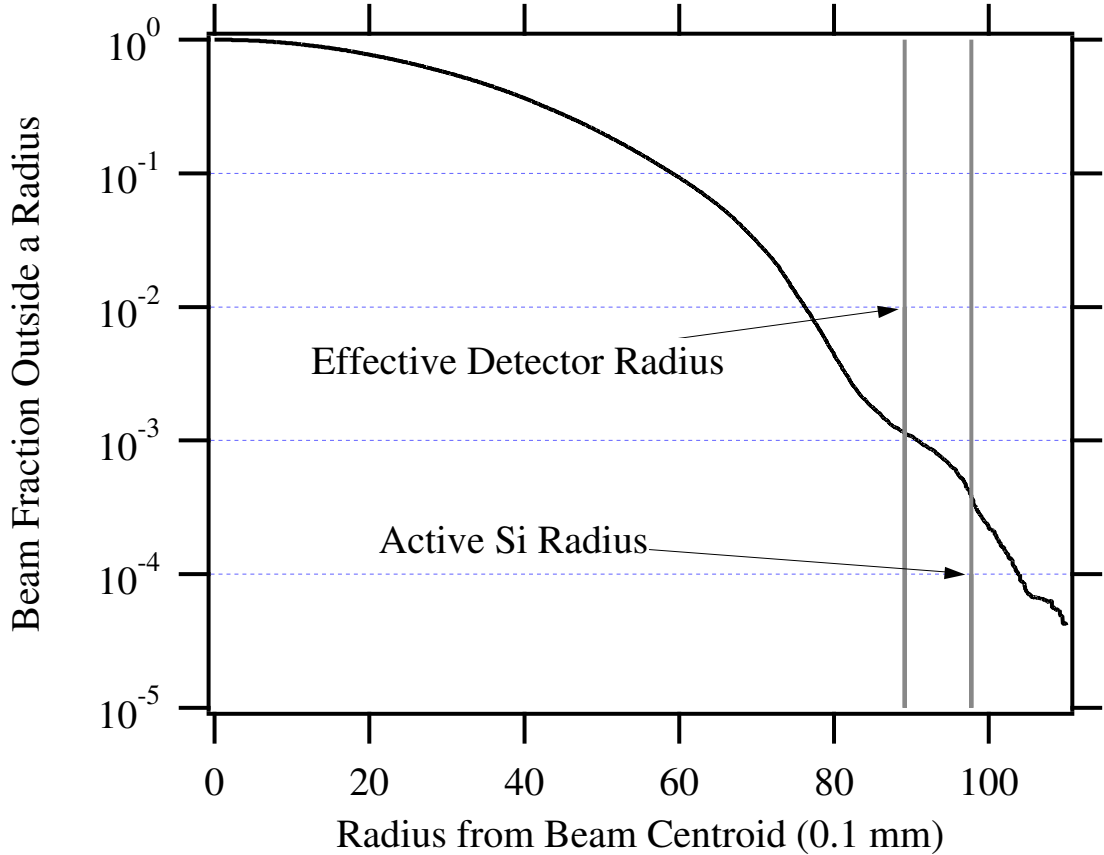}
\caption{\label{fig:fractions} The plot shows the measured fraction of neutrons outside a given radius from the centroid of the beam.  The image was taken at the downstream end of the trap.  }
\end{figure}

A neutron imaging technique was employed to profile the beam at three locations along the beamline. Neutrons are incident on an absorbing foil with a high thermal neutron absorption cross-section, a decay branch into beta particles, and few competing decay modes.  After irradiation the decay electrons from the foil expose a film that can be read out by an image reader.  The intrinsic pixel resolution of the image is $100~\mu$m, but the actual resolution is worse due to the electron range and other systematic effects related to obtaining the image and performing the irradiation.  The estimated resolution is less than 0.5~mm.  Although there is a number of suitable metals for use as the transfer foil, we used natural dysprosium, the relevant isotope being $^{164}$Dy with its large neutron absorption cross-section, convenient half-life, and lack of competing decays.  Other applications of Dy foil activation for neutron imaging are found in Refs.~\cite{CHE96,CHE00,CHO00}.

The film is read out as a logarithmic scale and covers almost five decades of dynamic range, making it ideal for sensitive neutron intensity profile measurements.  The logarithmic scale can be
converted to a linear scale through a function supplied by the manufacturer of the film reader.  The linear scale is referred to as a photostimulable luminescence. Although the capture of neutrons is not strictly proportional to $1/v$ due to the non-negligible thickness of the dysprosium foil, the linear scale is a good indicator of the beam intensity.

Beam images were obtained at three positions along the neutron beamline: just upstream of the trap, just downstream of the trap, and at the position of the neutron detector.  The images at the trap determine the envelope of protons that will be incident on the active area of the silicon detector; the image at the neutron detector gives the fraction of the neutron beam covered by the detector deposit.  

To obtain the fraction of the beam inside a given beam radius, one must first subtract the background from the film that is unrelated to the beam image.  The outline of the Dy foil on the film is clear and has a known area, so the background value is obtained by averaging a large number of pixel values outside that area of the foil.  This value, which was typically three orders of magnitude smaller than the maximum intensity in the peak, is subtracted from the area of the Dy foil.  One obtains the fraction at a given radius by taking the ratio of the sum of all pixels outside a radius to the total sum of all the pixels over the area of the Dy foil.  Figure~\ref{fig:fractions} gives an example of a beam fraction displayed as a function of radius.

\subsubsection{Uncertainty in the Neutron Beam Halo}\label{sec:HaloUnc}

There are several systematic effects that can affect the value of the beam halo fraction: background subtraction, alignment of the image with the beam axis, technique of exposing the film, and blooming of the image.  We have determined the magnitude of these effects to be small at the
0.1~\% level with the exception of image blooming.  We found that high values of the beam halo fraction are correlated with high values of the maximum intensity in the image. This was shown by comparing the fraction for an exposure with a large maximum intensity value and subsequently re-measuring in two distinct circumstances: letting the activity of the foil decay away and reducing the film exposure time.  Both cases have the same effect of lowering the fraction even though the nothing has changed from the initial beam irradiation. 

In order to measure the beam halo fraction in the presence of intensity-dependent image blooming, we varied the intensity and performed an extrapolation. We made two intensity measurements at the same position and exposed them for 6~s, 180~s, and 600~s, thus varying the maximum intensity.  The six fractions at the effective detector radius were plotted versus the maximum intensity value
and a linear extrapolation performed to obtain the fraction at zero intensity.  The intercept value is $1.1\times10^{-3}$.  We take it to be the correction for lost protons and use the value itself as the uncertainty in the correction.  We consider it to be a conservative estimate since most of the systematics cause the fraction to increase.  In addition, the trap images could not be taken at the true position of the trap electrodes due to practical considerations.  Instead, they were done approximately 10~cm beyond the ends of the trap, which means that the beam has expanded significantly from the trap volume where the lifetime was measured.  Trying to interpolate a beam shape between two images taken at the trap ends and then correcting each trap length used in the experiment is feasible, but the small size of the effect does not justify the effort. Instead, we assign a larger uncertainty.

To measure the extent of the neutron beam at position of the neutron detector,  we put a Dy foil in the same mount that holds the LiF deposit. The fraction at 19~mm was consistently around 0.03~\%, which is a small enough value that we did not perform a similar series of measurements  to improve on the accuracy of that number.

\subsection{Trap Corrections}\label{sec:MC}

If the proton trap and magnetic field are perfectly uniform, the effective length of the end region $L_{\rm end}$ will be the same for all trap lengths, and Eq.~(\ref{eq:FitFunc}) will yield a straight line. In reality there are some effects that introduce nonlinearities that must be corrected for. These corrections are discussed in detail in this section.

\subsubsection{Determination of Trap Corrections}

\begin{figure}
\includegraphics[width=3in]{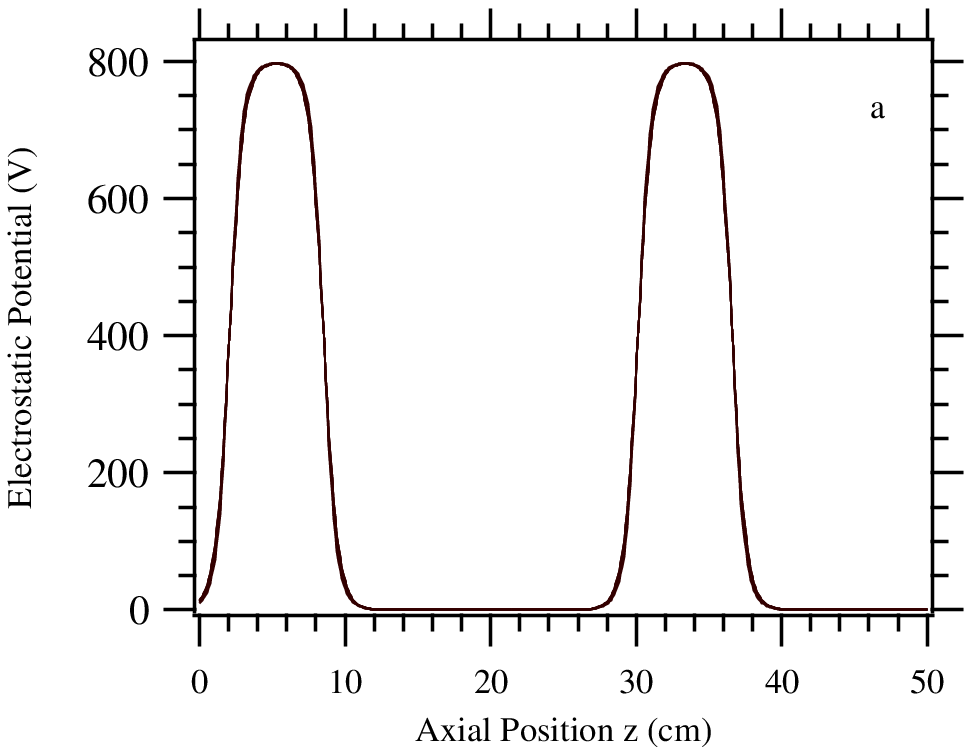}\hspace{0.3 in}
\includegraphics[width=3in]{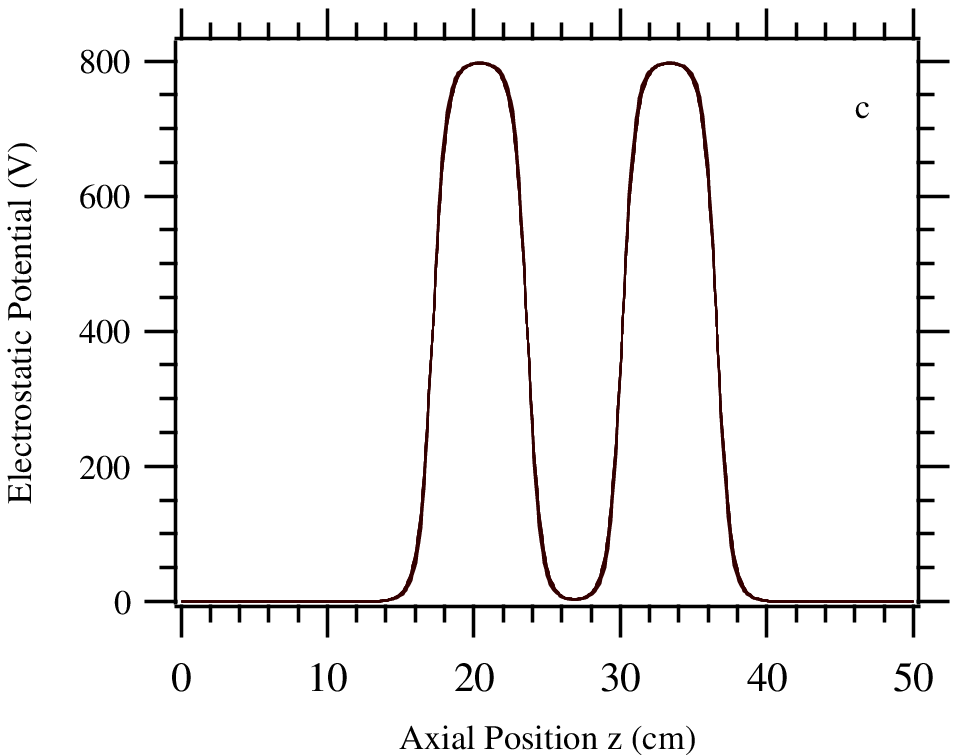}
\includegraphics[width=3in]{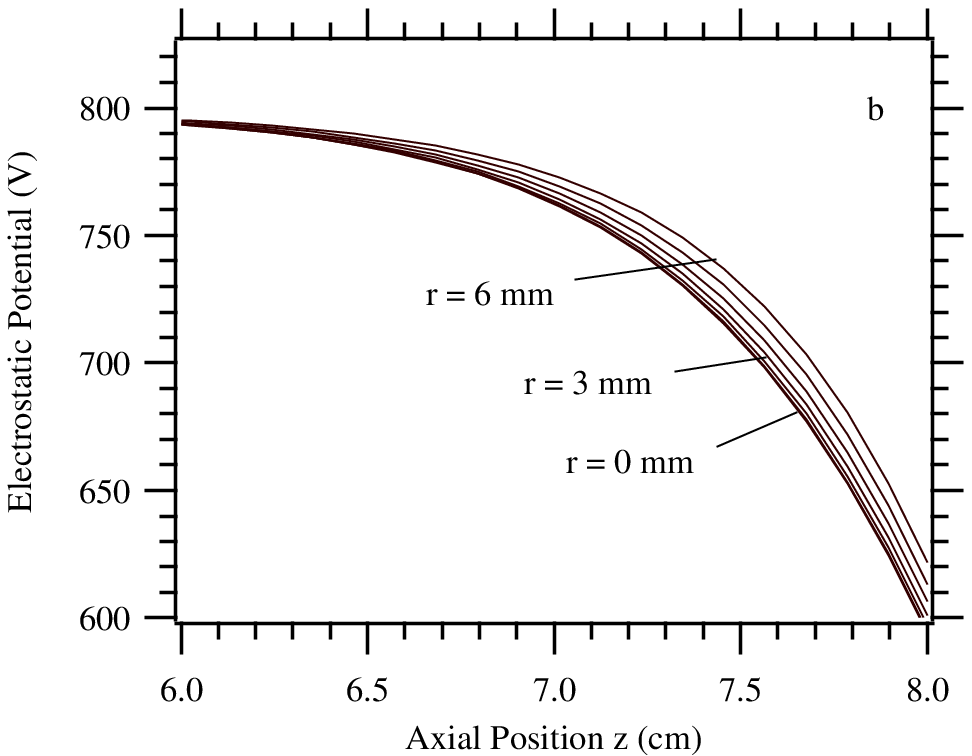}\hspace{0.3 in}
\includegraphics[width=3in]{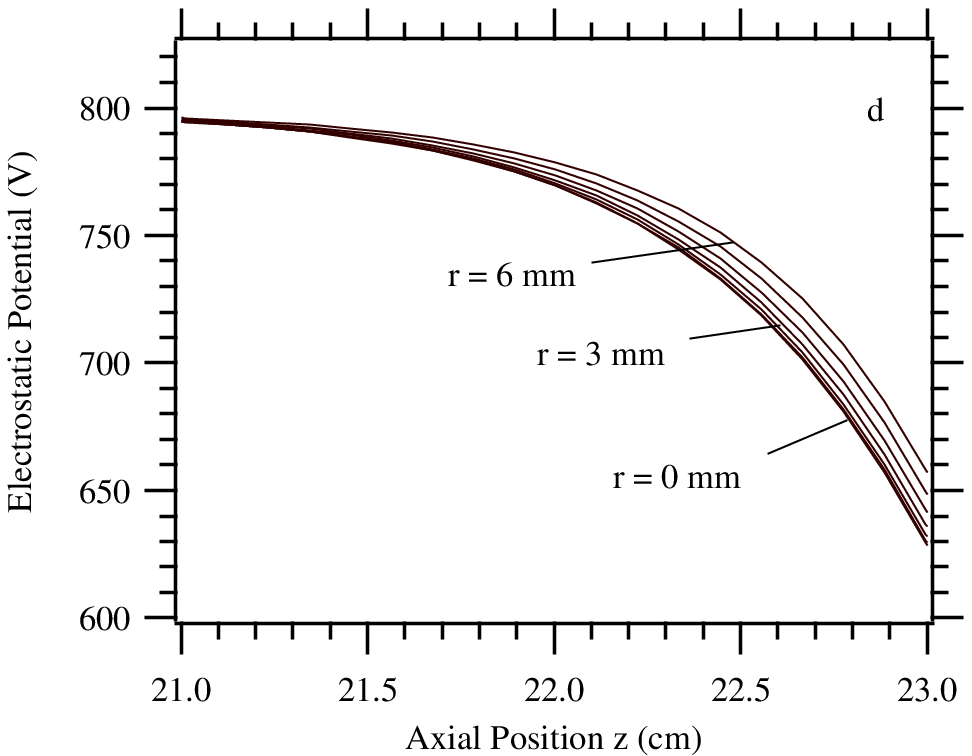}
\caption{\label{fig:trapPot} The calculated electrostatic potential of the trap corresponding to
10 grounded electrodes in the center region (a,b) and 3 grounded electrodes in the center region (c,d).}
\end{figure}

The electrostatic potential produced by the trap electrodes can be described to an accuracy of better than 0.02~\% by an approximate solution to the Laplace equation for a set of axially symmetric lenses, as discussed in Ref.~\cite{BYR93}. This calculated potential is shown in Fig.~\ref{fig:trapPot}a and b (c and d) as a function of the  axial trap coordinate $z$ for the 10-electrode (3-electrode) trap, {\em i.e.} the configuration with a three-electrode mirror and three-electrode door, each at a potential of $+800$~V, and ten (three) grounded electrodes in the central trapping region. The radial dependence of the potential is evident in Fig.~\ref{fig:trapPot}b and \ref{fig:trapPot}d. Near the trap axis, the maximum end potential is slightly lower than the mirror/door electrode potential of $+800$~V. 

In the regions near the door and mirror (the ``end regions''), a neutron decay proton can be created at an elevated potential and still be trapped. In this case, a proton is trapped if its initial (at birth) sum of  electrostatic potential energy and axial kinetic energy is less than the maximum end potential. 
Some protons created in the end regions are trapped and some are not. This complication
makes the effective length $L$ of the trap difficult to determine precisely.  Because of the symmetry in the trap's design, $L_{\rm end}$ is approximately equal for all trap lengths that were used. There are three small but important effects, however, that spoil this equality.
\paragraph {Nonuniformity of the magnetic field:} The motion of a charged particle in a Penning trap has been discussed elsewhere~\cite{BYR65, BYR72, KRE93}. The trap in this experiment is not a {\em true} Penning trap in that it lacks an axial quadrupole electrostatic field. However the basic ideas of charged particle motion are the same. There are three motional modes:

\begin{enumerate}
\item An axial ``bounce'' or back-and-forth motion with period $T_z$.
\item A cyclotron (circular) motion about the magnetic field lines with period $T_c = 2 \pi m_p / e B$, where $m_p$ and $e$ are the proton's mass and charge.
\item A magnetron drift motion perpendicular to the magnetic field lines with period $T_m$. The magnetic and electric fields inside the trap are primarily axial. They have small radial components and their azimuthal components are negligible, so the  $\vec{E} \times \vec{B}$ force is purely azimuthal and causes a slow azimuthal (magnetron) drift of the trapped particle's helical path in a circle about the trap's axis. The fraction of the particle's transverse energy in the magnetron mode is negligibly small, and the trap has excellent azimuthal symmetry, so the magnetron motion is of no consequence to us and will henceforth be ignored.
\end{enumerate}

It can be shown that for a charged particle moving in a magnetic field, the quantity $p_{\perp}^2 / \vec{B}\cdot \vec{l}$ is an adiabatic invariant (for example, see Ref.~\cite{JAC75}) . Here $\vec{l}$ is a unit vector along the {\em guiding center path}, {\em i.e.} the path of the center of cyclotron motion and $p_{\perp}$ is the particle momentum perpendicular to $\vec{l}$. The adiabatic condition requires that the magnetic field vary slowly over one cycle of the particle's motion, or
\begin{equation}
\frac{d(\vec{B}\cdot \vec{l})}{dl}\left(\frac{l_{cyc}}{\vec{B}\cdot \vec{l}}\right) \ll 1,
\label{eq:adCond}
\end{equation}
where $l_{cyc} = 2\pi v_z / \omega_{cyc}$ is the pitch of cyclotron motion (the length along $\vec{l}$ of one cycle). In our proton trap the quantity in Eq.~(\ref{eq:adCond}) is everywhere less than $10^{-3}$, so the adiabatic condition is satisfied. If one defines the longitudinal kinetic energy (kinetic energy along the guiding center path) to be
\begin{equation}
K_l = \frac{\left(\vec{p}\cdot\vec{l}\:\right)^2}{2m_p},
\end{equation}
the adiabatic invariant causes $K_l$ to vary as
\begin{equation}
dK_l = -\left(\frac{p_{\perp}^2}{\vec{B}\cdot \vec{l}}\right) \frac{d(\vec{B}\cdot \vec{l})}{2m_p}.
\end{equation}

A variation in the magnetic field along $\vec{l}$ will cause a corresponding change in $K_l$. The 
dot product $\vec{B}\cdot \vec{l}$ acts, in effect, as a one-dimensional scalar potential that is
proportional to the transverse energy of the proton. This quantity is treated as a  {\em magnetic pseudopotential}. By dividing by the proton charge, we can express the magnetic pseudopotential for a particular proton trajectory as a voltage that is a function of position along the guiding center path, with zero defined to be the initial position at birth. It is added to the electrostatic
potential, also a function of position, to get the total potential associated with every point along the path. In this picture, a proton will be trapped if its initial sum of longitudinal kinetic energy plus electrostatic potential energy is less than  the maximum total potential energy (electrostatic plus the magnetic pseudopotential) along its trajectory. The maximum end potential is large enough (about 800~V) and the magnitude of the magnetic pseudopotential is small enough (less than 30~V for all proton trajectories in our trap) that all  protons created at ground potential in the central region will be trapped.

For protons created in the end region, at an elevated potential,  the trapping probability will depend on the magnetic pseudopotential along its trajectory.  As we vary the trap length, the door electrodes remain fixed, but the position of the mirror is moved with respect to the magnet; therefore the shape of the magnetic field, and hence the size and shape of the magnetic pseudopotential in the mirror region, is slightly different for different trap lengths.  So while $nl$ remains proportional to the number of trap electrodes, $L_{\rm end}$ will vary with trap length.

\paragraph {Divergence of the neutron beam passing through the trap:} The neutron beam diverges slightly as it passes through the trap. Because the mirror is moved as we change trap length, the radial distribution of proton birth locations will vary with trap length. The electrostatic potential has a slight radial dependence as seen in Figs.~\ref{fig:trapPot}b and \ref{fig:trapPot}d. This causes the trapping probability for protons created near the ends to change slightly with trap length, which causes $L_{\rm end}$ to vary slightly with trap length.

\paragraph {Variation in trap electrode and spacer lengths:} Each of the electrodes is nominally identical with a length of 18.6~mm and an inner radius of 13.0~mm. The electrodes in the trap are separated by spacers of nominal length 3.0~mm. Slight variations in these lengths (see Table~\ref{tab:metrol}) will cause the total trap length to deviate from strict proportionality to the number of electrodes, and also cause  $L_{\rm end}$ to vary slightly with trap length.

The variation of $L_{\rm end}$ with trap length as described above will cause the data points of 
$\dot{N_p} / \dot{N_{\alpha+t}}$ versus $n$ to deviate from a straight line. A Monte Carlo simulation of  the experiment was developed in order to correct for these effects.  

\begin{figure}
\includegraphics[width=6in]{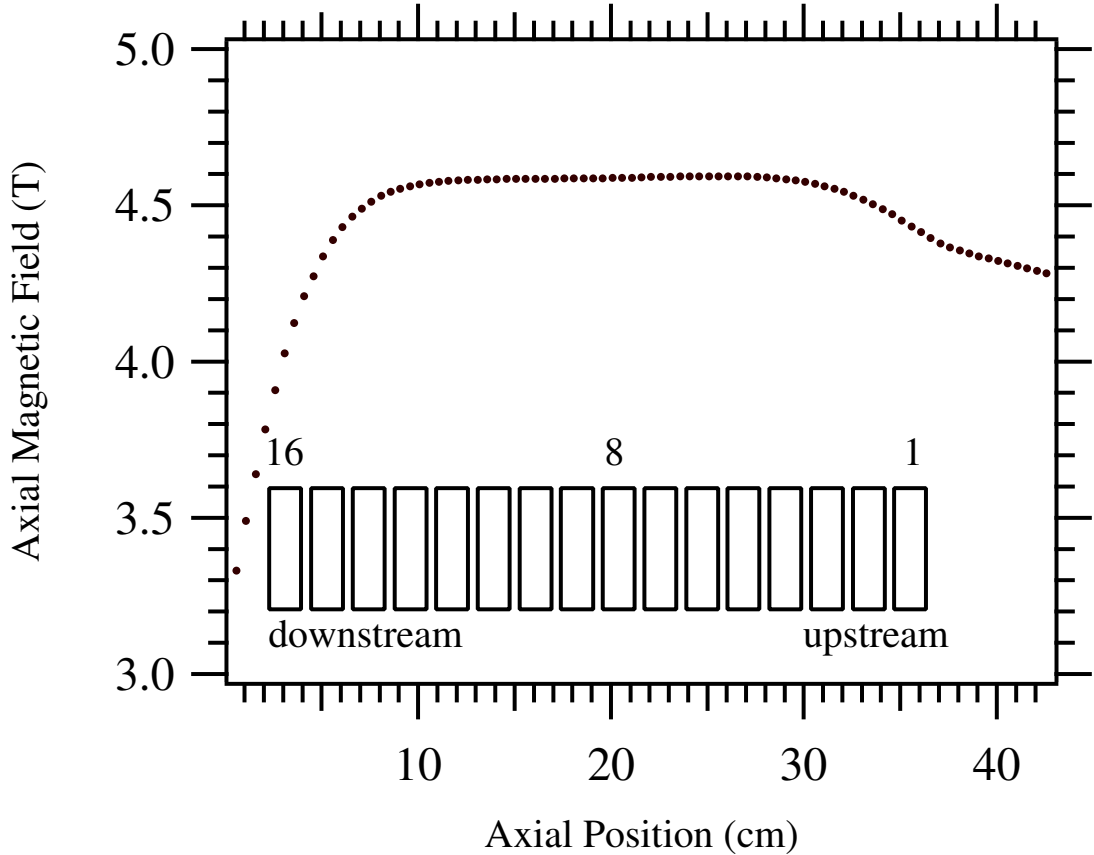}
\caption{\label{magTrap} The axial magnetic field inside the trap, measured along the axis using an axial Hall probe. Also shown are the positions of the 16 trap electrodes.}
\end{figure}

For the simulation we calculated the electrostatic potential for the electrode geometry of our trap (see Fig.~\ref{fig:trapPot}) by the method in Ref.~\cite{BYR93}. The measured lengths of the electrodes and spacers are as given in Table~\ref{tab:metrol}. To calculate $\vec{B}(r,z)$, we first measured $B_z$ along the axis of the trap ($r=0$) using an axial Hall probe. The measured points of $B_z$ versus $z$ are shown in Fig.~\ref{magTrap}. Noting that $\nabla \cdot \vec{B} = 0$ and $\nabla \times \vec{B} = 0$, we expand $\vec{B}(r,z)$ about $B_z(r=0,z)$:
\begin{eqnarray}
B_r(r,z) & \approx & -\frac{1}{2} r \frac{\partial B_z(r=0,z)}{\partial z} \\
B_z(r,z) & \approx & B_z(r=0,z) - \frac{1}{4} r^2 \frac{\partial^2 B_z(r=0,z)}{\partial z^2}.
\end{eqnarray}
Derivatives of $B_z$ higher than the second derivative can be neglected. 

In the Monte Carlo simulation, neutron decay recoil protons were generated according to the energy distribution of Nachtmann~\cite{NAC68} with the Fermi function $F(Z,E)$ evaluated using the expansion of Wilkinson~\cite{WIL89}. For each proton, a neutron decay vertex was chosen at random according to the measured radial neutron intensity distribution. Each proton was given a random direction, and the trajectory of its guiding center path was calculated using the calculated $\vec{B}(r,z)$. The electrostatic potential and magnetic pseudopotential at each point along the path determined whether the proton was trapped. A proton whose trajectory was reversed by {\em both} the door and mirror was considered trapped. 

Five different cases of varying complexity were simulated by the Monte Carlo in order to develop an understanding of the different contributions to the trap nonlinearities. For each case eight different trap lengths, from three to ten grounded electrodes in the central region (in correspondence to the experimental trap lengths), were calculated. 

\begin{description}

\item[{\rm Case A:}] The most realistic case. The magnetic field inside the trap was calculated from the 
measured axial magnetic field as described above.  The measured values of the trap electrode and spacer lengths from Table~\ref{tab:metrol} were used. We used the neutron beam radial intensity distribution that was measured at the end of the trap (see Section~\ref{sec:Halo}). The radius of this distribution was scaled by trap position $z$ to form a cone with half-angle 4.15~mrad, determined by the geometry of the beam collimators, so that the neutron beam distribution and divergence were both accurately modeled in the simulation. 

\item[{\rm Case B:}] The same as Case A, except a uniform 4.5~T axial magnetic field was used instead of the calculated field.

\item[{\rm Case C:}] The same as Case B, except the neutron beam was nondivergent. The radial intensity distribution measured at the end of the trap was used throughout the trap.

\item[{\rm Case D:}] The same as Case B, except the neutron beam had zero radius (line source).

\item[{\rm Case E:}] The same as Case D, except we used electrodes of uniform length 18.6~mm, and spacers of uniform length 3.0~mm.

\end{description}

A total of $4 \times 10^7$ protons were generated for each trap length, except for Case E, which had $2.1\times 10^8$ decays per trap length. A nominal neutron lifetime of $\tau_n = 885$~s was used in all runs to establish the neutron lifetime. For each Monte Carlo case we determined the ratio of proton trapping rate to thermal-neutron fluence rate $\dot{N_p} / \dot{N_0}$. The thermal-neutron fluence rate $\dot{N}_{0}$ is related to the alpha counting rate $\dot{N}_{\alpha+t}$ in the experiment  by the factor $\epsilon_0$, the overall efficiency of counting the reaction products for a thermal neutron that passes through the trap. Also implicit here is an assumption that the proton counting efficiency $\epsilon_p$ equals unity in the Monte Carlo.

\begin{figure}
\includegraphics[width=6in]{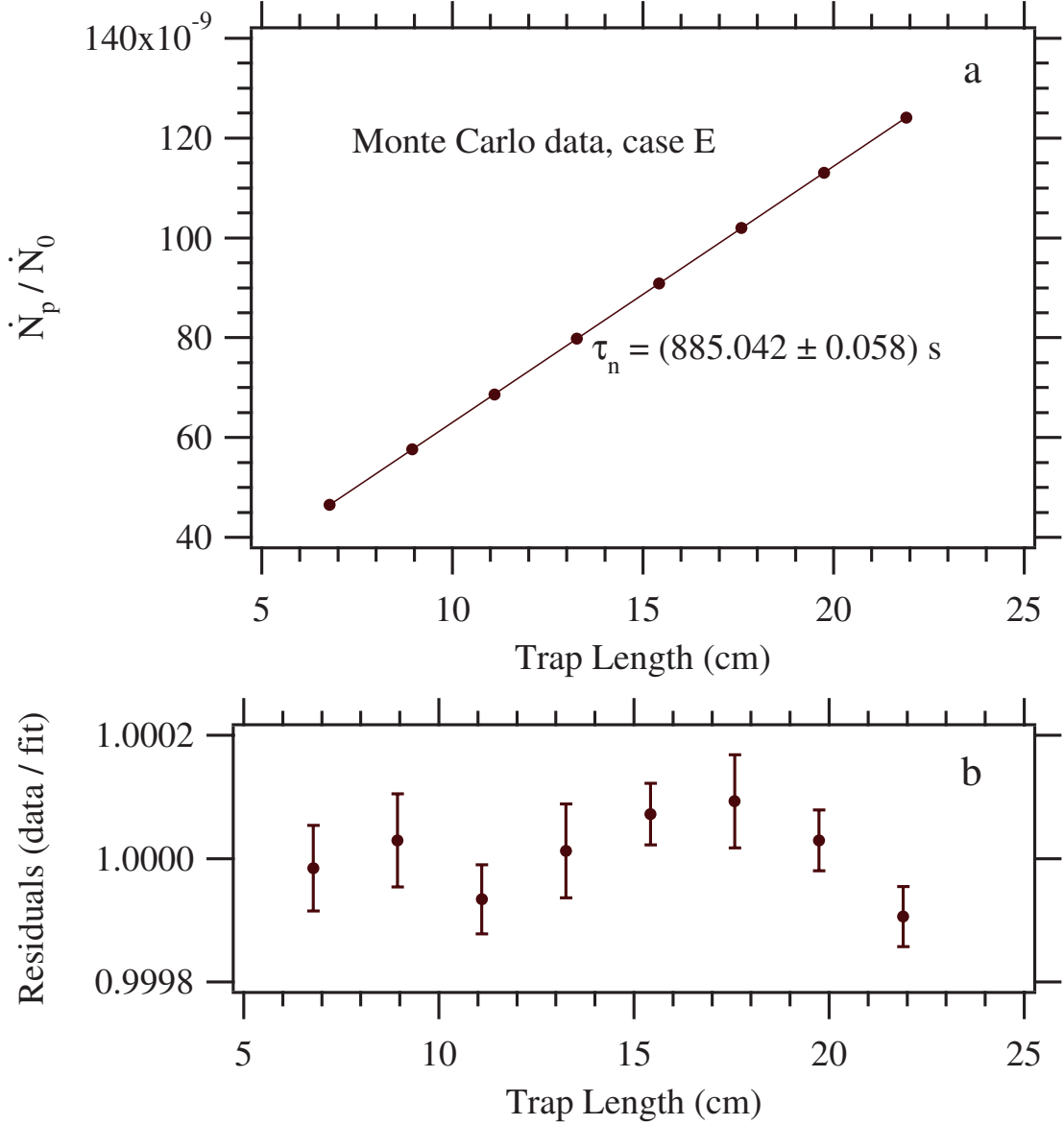} 
\caption{\label{caseE} Data from Monte Carlo case E: a) the proton/neutron ratio $\dot{N_p} / \dot{N_0}$ versus trap length; b) the data/fit residuals}
\end{figure}

In Case E all of the effects described earlier that cause $L_{\rm end}$ to vary with trap length are avoided,  so $\dot{N_p} / \dot{N_0}$ versus $n$ should yield a perfectly straight line. The Monte Carlo data for Case E are shown in Fig.~\ref{caseE}.  A linear fit to the trapped proton/neutron ratio: $\dot{N_p} / \dot{N_0}$, gives a slope that corresponds to $\tau_n = (885.042 \pm 0.058$)~s, with a chi-squared of 9.3 for 6 degrees of freedom. The input neutron lifetime of 885~s was precisely recovered in this case. 

\begin{figure}
\includegraphics[width=6in]{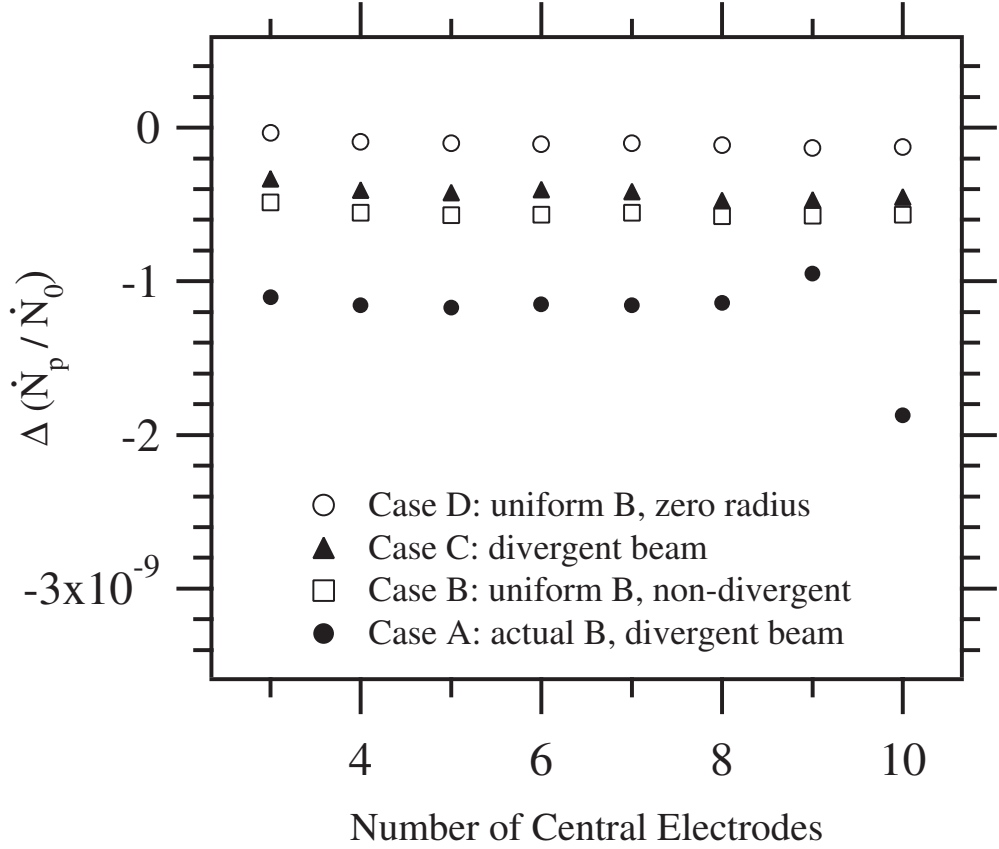}
\caption{\label{MCdiff} The difference in $\dot{N_p} / \dot{N_0}$ between Monte Carlo Cases A and E (solid circles), B and E (squares), C and E (triangles), and D and E (open circles). The statistical uncertainties are smaller than the plot symbols.}
\end{figure}

Figure~\ref{MCdiff} shows the differences in $\dot{N_p} / \dot{N_0}$ between Cases A, B, C, D and Case E ({\em e.g.} the open circle points are $\dot{N_p} / \dot{N_0}$ [Case D] minus $\dot{N_p} / \dot{N_0}$ [Case E]). The effects on the slope of $\dot{N_p} / \dot{N_0}$ due to differences in the electrode and spacer lengths and beam divergence are seen to be very small. The main differences in $\dot{N_p} / \dot{N_0}$ between Cases B, C, and E are in the vertical offsets, caused by the different neutron radial distributions between these cases. When a proton is created at a large radius from the trap axis, the shape of the electrostatic potential in the end region is more square (see Fig.~\ref{fig:trapPot}), so fewer protons created in the end region are trapped. As a result, $L_{\rm end}$ decreases with increasing beam radius. The divergent beam of Case B has a slightly smaller average radius than the nondivergent beam of Case C, so its $\dot{N_p} / \dot{N_0}$ data are slightly higher. 

The largest effect on both the offset and the slope comes from the actual magnetic field shape in the trap. Looking at Fig.~\ref{magTrap}, one can see that the axial magnetic field is slightly lower at the door electrodes (electrodes 1 through 3), so the magnetic pseudopotential in the door is negative relative to both the central region of the trap and the mirror for trap lengths 3 through 8. For these trap lengths, some protons created at an elevated potential in the mirror have enough longitudinal kinetic energy to escape through the door, while they would not have escaped with a perfectly uniform magnetic field. Therefore $L_{\rm end}$ is reduced, and we see a negative vertical offset of the solid circle points in Fig.~\ref{MCdiff}. For trap length 9 the mirror is on electrodes 13 through 15, where the magnetic field begins to drop off at the downstream end. The magnetic pseudopotential is slightly negative at the mirror, about equal to that in the door, so fewer protons created there can escape through the door, causing a small increase in $L_{end}$. Finally, for trap length 10, the mirror is on electrodes 14 through 16, where the magnetic field is lower. In this case the magnetic pseudopotential has dropped well below that of the door, and the effect is reversed. Some protons created near the door can escape through the mirror, so $L_{end}$ is significantly lower.

\par
To correct the experimental data for these nonlinear effects we divide the Monte Carlo $\dot{N_p} / \dot{N_0}$ data for case A by those of case E. This yields a set of correction factors, listed in Table \ref{MCcorrect}. The factor $\epsilon_p / \epsilon_0$ cancels when the Monte Carlo data are divided, as does the assumed neutron lifetime of 885~s, so the correction factors can be applied directly to the experimental data. We multiply the experimental $\dot{N_p} / \dot N_{\alpha+t}$ for each trap 
length by the corresponding correction factor to restore the linearity of the experimental data points. This results in a correction of $-5.3$~s to our neutron lifetime, dominated by the magnetic field gradient at the end of the trap in the 10-electrode configuration. For comparison, if we exclude all 10-electrode data from our complete analysis, the extrapolated neutron lifetime (equivalent to Fig.~\ref{fig:Tau}) is $(888.4 \pm 1.5)$~s, a consistent result, and the correction becomes only $-1.0$~s.

\begin{table}
\caption{\label{MCcorrect} Correction factors for the experimental data:  $\dot{N_p} / \dot{N_0}$[Case A] divided by $\dot{N_p} / \dot{N_0}$[Case E]. The statistical uncertainty is $\pm 0.0002$ on each point.}
\begin{ruledtabular}
\begin{tabular}{cc}
Trap length 			& Correction factor\\
(number of central electrodes)	&	\\
\hline
3		&	1.0243\\	
4		&	1.0205\\
5		&	1.0174\\	
6		&	1.0146\\	
7		&	1.0129\\	
8		&	1.0113\\
9		&	1.0085\\	
10		&	1.0153\\
\end{tabular}
\end{ruledtabular}
\end{table}

\subsubsection{Determination of Trap Correction Uncertainties}

The uncertainty in the Monte Carlo correction factors is dominated by the uncertainty in the magnetic field, which derives from two primary  sources: the uncertainty in the axial magnetic field map and the uncertainty in the trap position in the magnetic field coordinate system. These two independent sources of uncertainty were estimated separately.

\begin{figure}
\includegraphics[width=6in]{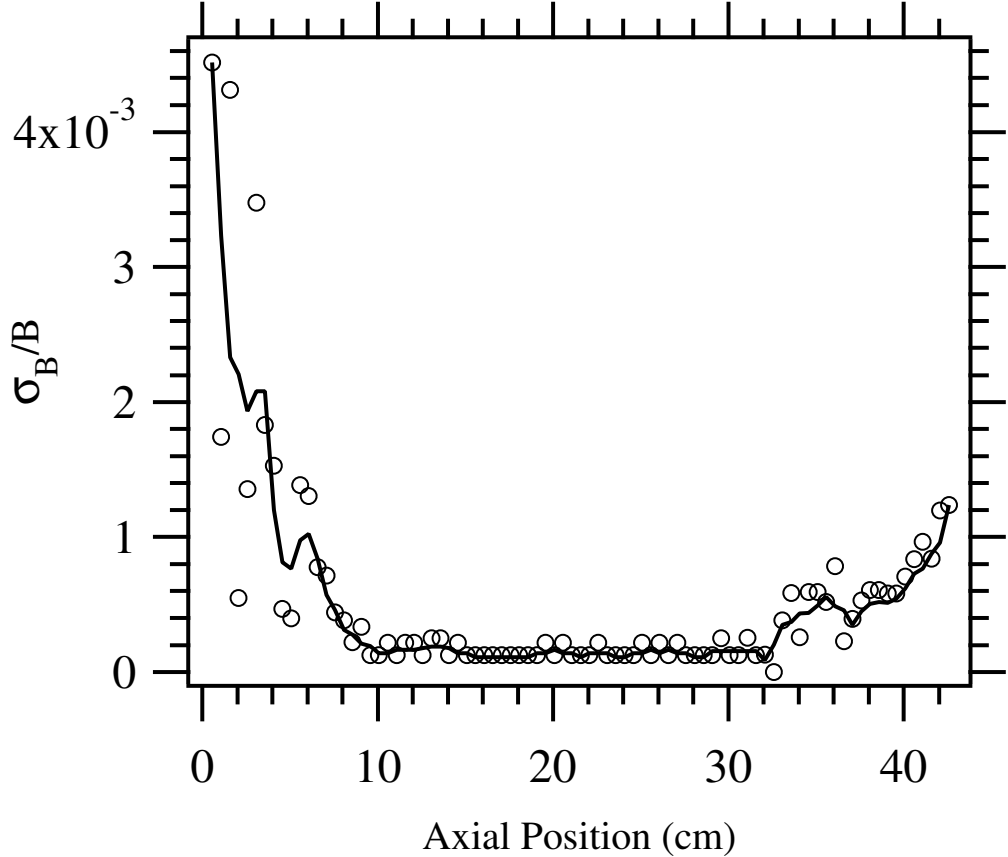} 
\caption{\label{BErr} The fractional standard deviation $\sigma_B / B$ of the axial magnetic field $B_z$ versus axial position $z$. The open circles used only three measurements of $B_z$ to find $\sigma_B / B$ and the solid line includes the measurements from adjacent points for a total of nine.}
\end{figure}

The magnetic field map was performed three times, each with measurements at 1~cm intervals  
along the axis of the bore. It is only the magnetic field gradient that affects the lifetime determination, so the absolute calibration of the Hall probe is unimportant. We need to know only the relative uncertainty due to zero drifts and random fluctuations. We can use the variance in the three 
measurements made at each point to estimate this uncertainty. An improved estimate can be made by combining the three {\em fractional} deviations in $B_z$ with the three {\em fractional} deviations from adjacent points, for a total of nine fractional deviations from which the variance
and standard deviation are estimated. This procedure is valid because the true standard deviation should depend only on $B$, which is quite close for adjacent points in $z$. The effect smooths out the estimated standard deviation function versus $B_z$, taking advantage of the 
fact that while only three measurements were taken at each $z$. This was done at many different $z$, so in effect we made a lot of measurements at each value of $B$. Figure~\ref{BErr} shows the estimated fractional standard deviation in $B_z$ as a function of axial position $z$.

Ten sets of dithered axial magnetic field data were generated by randomly varying the average measured $B_z$ by a Gaussian distribution at each point $z$ using the estimated $\sigma_B / B$ shown in Fig.~\ref{BErr} (solid line). For each magnetic field set the full Monte Carlo
simulation was performed, which yielded ten sets of correction factors analogous to those in Table~\ref{MCcorrect}. For each set of correction factors, the complete data analysis was performed to extract the neutron lifetime from our full data set, producing ten different values of the neutron lifetime. The standard deviation of these lifetimes is $\sigma_{MC,1} = 0.35$ s, which we take to be the 1 $\sigma$ uncertainty in the  lifetime due to the axial magnetic field map.
\par
The second important source of uncertainty is the axial position of the trap in the magnet bore. The transverse and angular alignments of the trap to the bore were done to high precision using a theodolite, and those uncertainties are negligible. The axial position was established to the nearest 0.8~mm (1/32 inch) using a scale. The uncertainty in the trap's axial position was conservatively estimated  to be $\pm 1$~mm. To estimate the resulting error in the lifetime, the trap position was shifted by $\pm 1$~mm in the Monte Carlo simulation. This small shift has a relatively large effect on $\dot{N_p} / \dot{N_0}$ for the 10-electrode trap, because the axial magnetic field is falling quite rapidly there. When the corresponding Monte Carlo correction factors were applied to the full data analysis, a shift in the lifetime of $\pm 0.71$ s was obtained, which we take to be the 1 $\sigma$ uncertainty $\sigma_{MC,2}$ in the lifetime due to the uncertainty in the trap position.
\par
The net systematic uncertainty in the lifetime due to the Monte Carlo correction is then the quadrature sum of $\sigma_{MC,1}$ and  $\sigma_{MC,2}$ which is $\sigma_{MC}$ = 0.79~s.

\subsection{Determination of Proton Detector Losses}\label{sec:protonlosses}

The efficiency of proton detection is less than unity due to several   well-known effects.  Some protons lose all of their energy before  traversing the dead layer of the detector and never reach the active  
layer.  One must also impose a discriminator threshold on the proton  pulse-height spectrum due to the detector and preamplifier noise.  This  results in the loss of protons that do not deposit their full energy  
into the detector and fall below the discriminator threshold. In  addition, some fraction of protons will Rutherford backscatter from the  inactive layer of the detector and will not be detected.  There is some  
probability, however, that those protons will be reflected back to the  detector and have another chance at being detected.  This quantity is  difficult to calculate but can be determined by measuring the lifetime  
at different calculated backscattering fractions and then fitting for  the slope and intercept.  The slope gives a measure of the effective  backscattering fraction, which will be reduced by those backscattered  
protons that are returned to the detector, and the intercept gives the  free neutron lifetime.

An accurate determination of the proton detection efficiency is one of the more time-consuming aspects of this work because it requires an extrapolation to remove the dependency on backscattering.  Thus, one must measure the neutron lifetime at many values of the calculated backscatter fraction.  This value is varied experimentally by acquiring data at several acceleration voltages and by using detectors with different dead layer thickness and composition.  The calculated value for the fractions was determined using two independent methods, a Monte Carlo calculation using SRIM 2003~\cite{ZIE03} and an analytical calculation based on the Rutherford backscattering formula. The SRIM calculations were used to obtain the fraction since they consider energy loss mechanisms that are not included in analytical  calculations. The calculations, however, provide considerable insight into the loss processes and are presented for completeness. 

\begin{figure}
\includegraphics[width=6in]{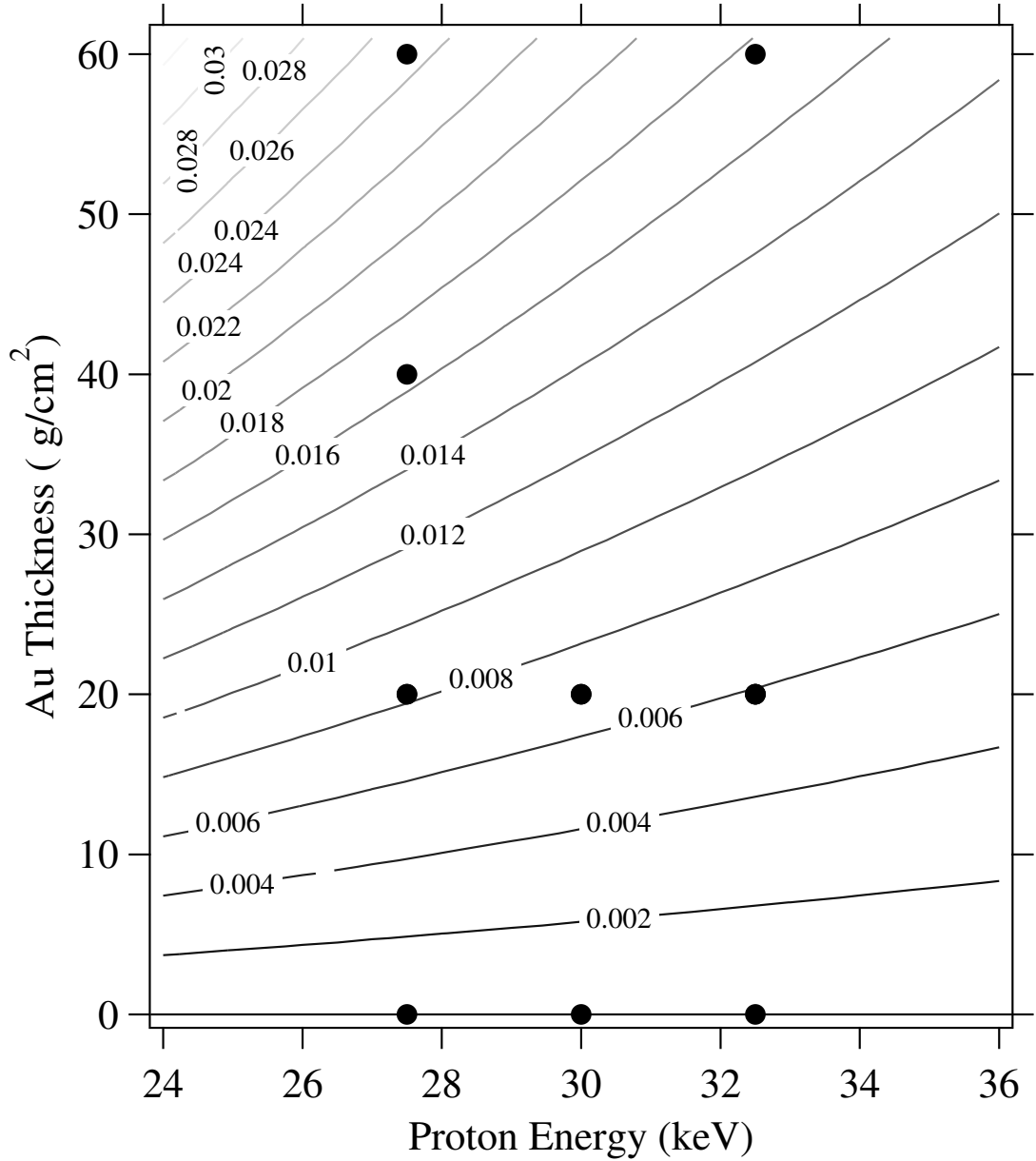}
 \caption{Backscattering probability versus dead  layer thickness and proton energy (Eq.~(\ref{eq:back})).  Each contour represents a line of constant Rutherford backscattering probability.  The contour labels give the Rutherford backscattering probability for protons impinging on a layer of gold.  The black circles indicate where lifetime measurements have been carried out.  Those entries with zero thickness correspond to the runs where so-called ``windowless'' detectors were employed. In fact these detectors have a non-negligible dead layer consisting of silicon and silicon dioxide.}
\label{fig:bckmap}
\end{figure}

The main source of energy loss for protons is Rutherford scattering from the atomic nuclei.  As the starting point, one considers protons of kinetic energy $E$ impinging on a material of atomic number Z.  The Rutherford backscattering cross section is given by
\begin{eqnarray}
  \label{eq:rbs}
  \sigma & = & 2 \pi \int_{\frac{\pi}{2}}^{\pi} \left(
  \frac{m_{e}c^2}{E}\frac{Zr_{e}}{4\sin^2 \frac{\theta}{2}}
  \right)^{2}\sin \theta \, d\theta\nonumber\\
  & = & \frac{\pi}{4} Z^2 r_{e}^{2} \left(\frac{m_{e}c^2}{E}\right)^2,
\end{eqnarray}
where $\theta$, the proton scattering angle, is integrated over all backward angles and $r_e$ is the classical radius of the electron ($\approx 2.8 \times 10^{-15}$\,m).  The use of a screened nucleus,
necessary to keep Eq.~(\ref{eq:rbs}) finite when the integral includes $\theta = 0$\,rad, makes a negligible difference over this angular range.  It follows that the backscattering probability $p$ is given by
\begin{equation}
  \label{eq:back}
  p = \frac{N_A}{A} \sigma \rho,
\end{equation}
where $A$ is the atomic weight of the material, $N_\text{A}$ is the Avogadro constant, and $\rho$ is the areal density of the material. This function is plotted in Fig.~\ref{fig:bckmap} as a function of $\rho$ and $E$.  Also shown in the same figure are black circles indicating the configuration space sampled in this work.

A second important effect is their energy loss as the protons travel through the detector material.  In our regime, the approximate formula for energy loss is
\begin{equation}
  \label{eq:eloss}
  \frac{dE}{d\rho} = -k \sqrt{E},
\end{equation}
where $E$ is the kinetic energy, $\rho$ is the areal density, and $k$ is a constant of proportionality.  We have for the two materials of interest in our detectors (Au and Si)
\begin{eqnarray}
  \label{eq:ks}
  k_{\text{Au}} & = & 11 \, \sqrt{\text{keV}} \,
  \frac{\text{cm}^2}{\text{mg}} \nonumber\\
  k_{\text{Si}} & = & 65 \, \sqrt{\text{keV}} \,
  \frac{\text{cm}^2}{\text{mg}} \, .
\end{eqnarray}
Note that $k$ is proportional to $1/A$ in contrast to $p$ which is proportional to $A$.  Thus, higher $Z$ materials have greater backscattering probability and less energy loss than do lower $Z$ materials.

Eqs.~(\ref{eq:back}) and (\ref{eq:eloss}) can be used to construct an analytic model of the proton loss process.  The proton energy as a function of distance into the detector is calculated by integrating
Eq.~(\ref{eq:eloss}) over $\rho$ (essentially distance).  Then Eq.~(\ref{eq:back}) is used to tally those protons that backscatter as they travel through the detector losing energy.

\subsubsection{Backscattering Calculation}

\begin{figure}
\includegraphics[width=6in]{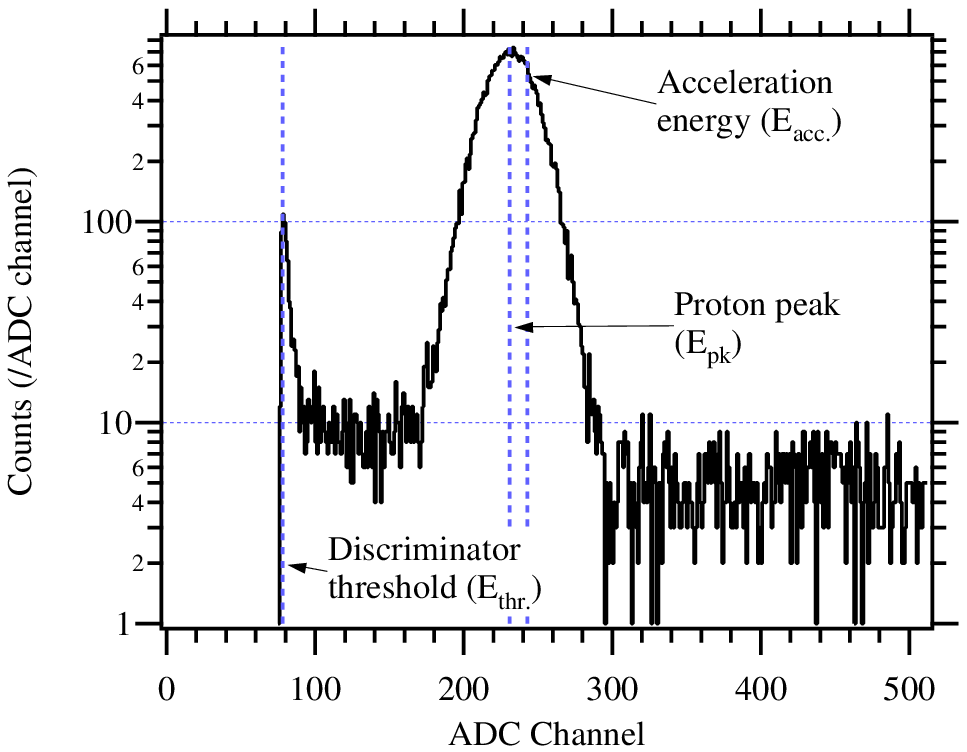}
\caption{\label{fig:ProtPk}A proton pulse-height spectrum for a typical  run.  The acceleration energy of the protons was 32.5 keV, and the detector was a surface barrier detector with $40~\mu$g/cm$^{2}$ of
  gold.  The energy loss, E$_{\text{loss}}$, is the difference between   the acceleration energy and the energy of the peak, or 1.64 keV.}
\end{figure}

Calculations using SRIM were used to determine the backscattering fractions since they take into consideration mechanisms (such as multiple scattering and energy loss upon scattering) that the analytical model does not. First, one must determine the dead layer thickness of each  detector.  We determined the dead layer thickness by measuring the difference between the acceleration energy of the proton  and the deposited energy. Figure~\ref{fig:ProtPk} shows a typical proton pulse-height spectrum. Detectors were calibrated {\it in situ}  using either an $^{241}$Am or $^{109}$Cd source.  The detector was mounted on the end of manipulator with one meter of travel.  It could be  retracted from its position for proton detection to a position where one could mount a source outside the vacuum can approximately 10~cm from  the detector.  The detector was still cold and under vacuum, so possible systematic problems due to gain shifts were minimized.  The relevant  gamma lines are 59.5~keV and 88.0~keV for the $^{241}$Am and $^{109}$Cd sources, respectively.  The energy resolution for a typical detector was 10~\% at the 59.5~keV line of $^{241}$Am; the full-width at half-maximum is 6.0~keV.  

\begin{figure}
\includegraphics[width=6in]{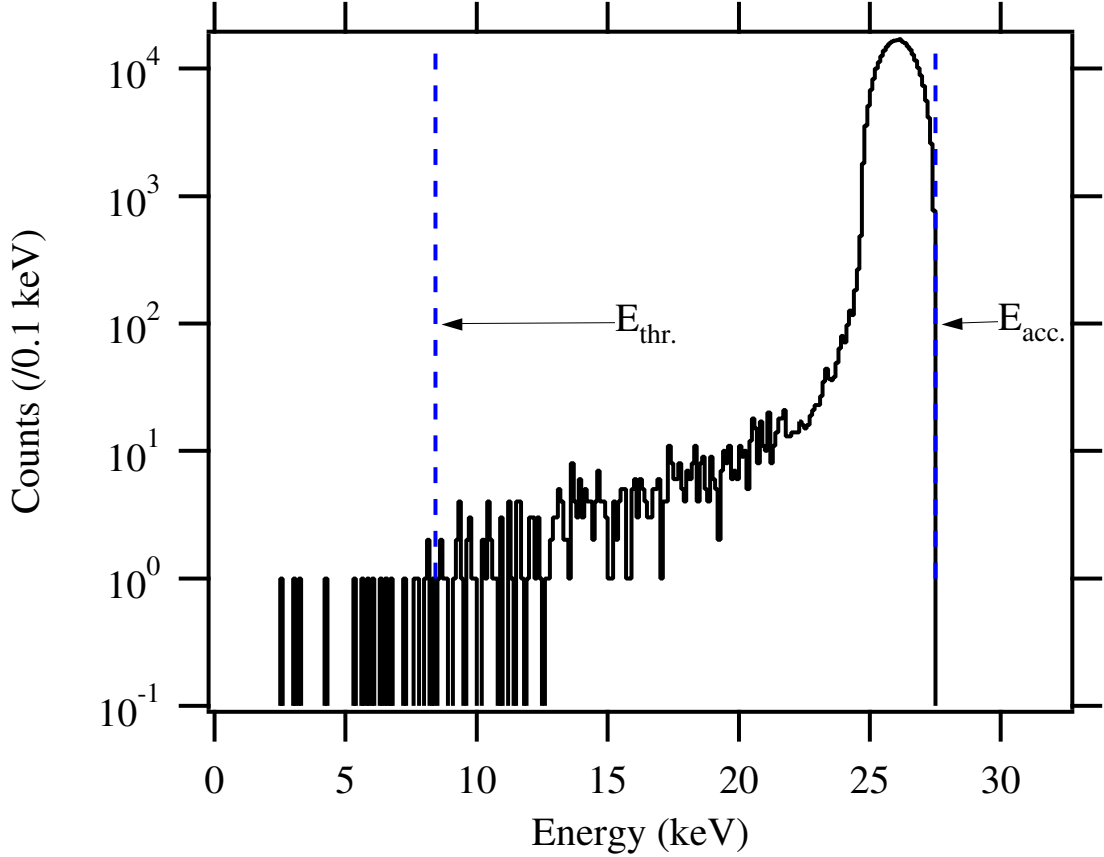}
\caption{\label{fig:SRIM} SRIM calculation of the energy spectrum of protons with incident energy of 27.5 keV transmitted through a $20~\mu$g/cm$^{2}$ gold dead layer. Of the initial $3\times10^{5}$ events for this example, 298,019  were transmitted with an average energy of 26.0~keV, 1980 backscattered  
and exited the detector, 18 entered the active silicon with an energy  below the registration threshold, and 1 stopped in the dead layer.}
\end{figure}

\begin{table*}
\caption{\label{tab:scat} Input parameters for determining the proton backscattering fraction and results from SRIM Monte Carlo calculations.}
\begin{ruledtabular}
\begin{tabular}{lcccrcccccc}
Series & Au & E$_{\text{acc.}}$ & E$_{\text{loss}}$ & E$_{\text{thr.}}$ & $f_{\text{Ruth}}$ & $f_{\text{AT}}$& $f_{\text{BT}}$ & $f_{\text{Stp}}$ & $f_{\text{Bsc}}$ & $f_{\text{Lost}}$\\
  & ($\mu$g/cm$^{2}$) & (keV) & (keV) & \text{(keV)} & (\%) & (\%) & (\%) &  (\%) & (\%) & (\%)\\
  \hline
121& 20 & 27.5& 1.47& 8.43& 0.0660 & 0.004 & 0.006 & 0.000 & 0.664 & 0.006\\
125& 20 & 27.5& 1.47& 8.43& 0.0660 & 0.004 & 0.006 & 0.000 & 0.664 & 0.006\\
130& 20 & 30.0& 1.30& 9.91& 0.0575 & 0.003 & 0.006 & 0.000 & 0.578 & 0.006\\
134& 20 & 30.0& 1.42& 9.91& 0.0575 & 0.003 & 0.006 & 0.000 & 0.578 & 0.006\\
140& 20 & 32.5& 1.64& 10.31& 0.0474 & 0.003 & 0.004 & 0.000 & 0.477 & 0.004\\
142& 20 & 32.5& 1.66& 10.31& 0.0474 & 0.003 & 0.004 & 0.000 & 0.477 & 0.004\\
143& 20 & 32.5& 1.69& 10.31& 0.0474 & 0.003 & 0.004 & 0.000 & 0.477 & 0.004\\
149& 60 & 27.5& 5.44& 10.66& 0.2354 & 0.020 & 0.656 & 0.143 & 2.374 & 0.799\\
151& 60 & 32.5& 5.87& 10.66& 0.1811 & 0.008 & 0.344 & 0.070 & 1.819 & 0.414\\
154&  0 & 30.0& 9.17& 11.22& 0.0158 & 0.036 & 0.117 & 0.084 & 0.194 & 0.201\\
155&  0 & 32.5& 9.70& 11.93& 0.0156 & 0.027 & 0.090 & 0.061 & 0.183 & 0.151\\
166& 40 & 27.5& 3.14& 9.89& 0.1479 & 0.011 & 0.129 & 0.009 & 1.490 & 0.138\\
170&  0 & 27.5& 8.60& 10.40& 0.0201 & 0.042 & 0.159 & 0.106 & 0.242 & 0.266\\
\end{tabular}
\end{ruledtabular}
\end{table*}

After one has measured the energy loss, one can input the appropriate parameters in the SRIM code to determine the two most relevant values: the fraction of protons that backscatter ($f_{\text{Bsc}}$), and thus have additional chances for detection, and the fraction of protons that cannot be detected ($f_{\text{Lost}}$) because they either stopped in the dead layer or deposited an amount of energy below the detection threshold.  When running the Monte Carlo, one tallies the classes of events that are the predominant contributors to those fractions.  These quantities may be written simply as
\begin{subequations}
\label{eq:Bksc}
\begin{equation}
f_{\text{Bsc}} = f_{\text{Ruth}}+f_{\text{AT}}
\label{eq:Bsc}
\end{equation}
\begin{equation}
f_{\text{Lost}} = f_{\text{Stp}}+f_{\text{BT}},
\label{eq:Lost}
\end{equation}
\end{subequations}
where $f_{\text{Ruth}}$ is the fraction of protons that Rutherford backscatter from the dead layer and exit the surface of the detector, $f_{\text{AT}}$ is the fraction that backscatter and exit from the active layer but did not deposit enough energy to be detected (i.e., their energy is still above the threshold for detection),
$f_{\text{Stp}}$ is the fraction that stop in the dead layer, and $f_{\text{BT}}$ is the fraction that arrives in the active layer but with an energy below the discriminator threshold.


In order to obtain the quantities in Eqs.~(\ref{eq:Bsc}) and (\ref{eq:Lost}), the SRIM calculation was done in two steps.  The first step inputs a proton of the appropriate energy incident on a dead layer of silicon dioxide and/or gold and an active silicon substrate, where the overall thickness of both is determined by a combination of the calibration and the manufacturer specification.  The average energy of the transmitted beam must be equal to the incident energy minus the energy loss in the dead layer.  The calculation is done iteratively by changing the dead layer thickness until those two quantities are equal.  From this calculation, one tallies $f_{\text{Ruth}}$, $f_{\text{BT}}$, and $f_{\text{Stp}}$. Figure~\ref{fig:SRIM} is a plot of the energy spectrum of transmitted protons from SRIM for 27.5~keV incident protons on a surface barrier detector with a 20~$\mu$g/cm$^{2}$ gold layer and 5.0~nm of silicon. The second step inputs the transmitted proton beam into active silicon, and one tallies the fraction of events that are backscattered and leave the active layer with a sufficient amount of energy that they could be detected it they were to return to the detector ($f_{\text{AT}}$).

For each series, 300,000 events are tallied in the Monte Carlo.  Table~\ref{tab:scat} gives all of the input parameters for determining  the proton detector backscattering fractions along with  the results from the SRIM calculations.

\subsubsection{Extrapolation to Zero Backscattering Fraction}\label{sec:ExtraptoZero}

\begin{figure}
\includegraphics[width=6in]{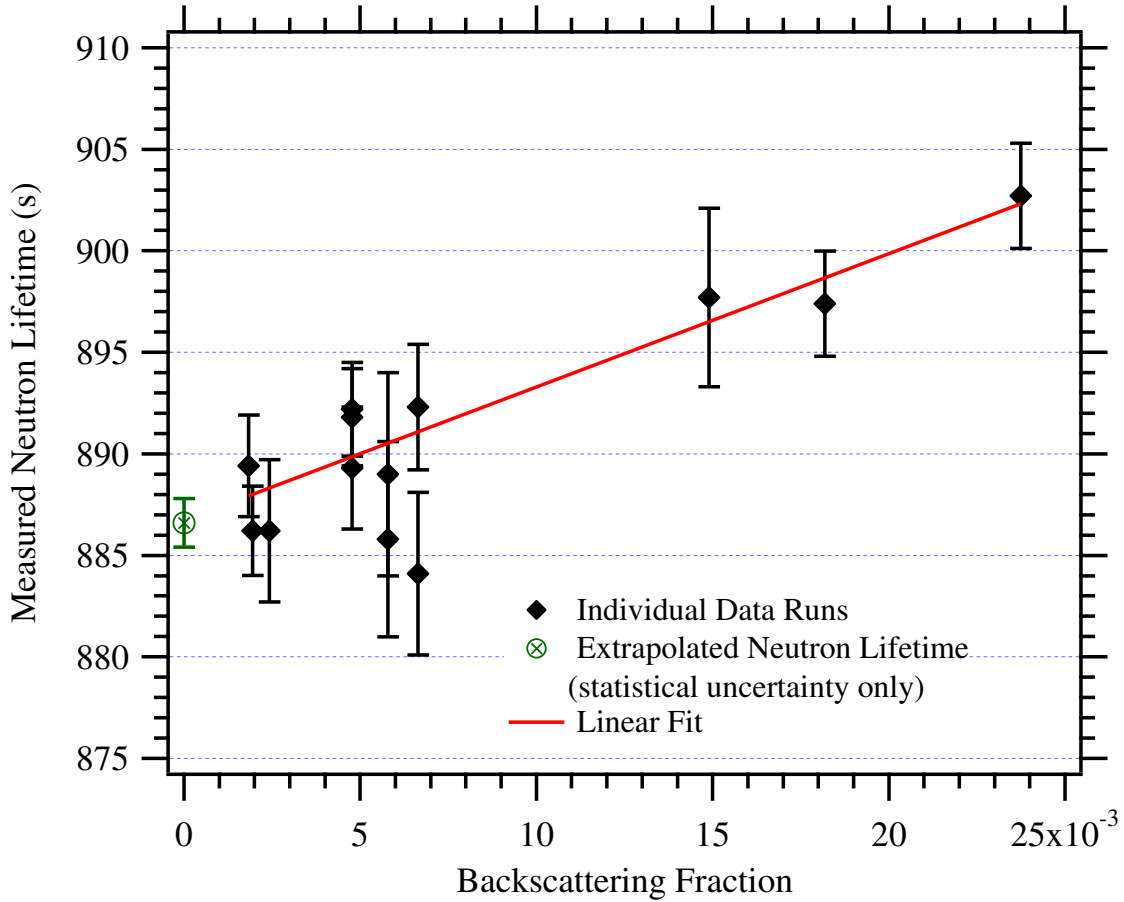}
\caption{\label{fig:Tau} A linear fit of the measured neutron lifetime versus the detector backscattering fraction $f_{Bsc}$.  The   extrapolation to zero backscattering gives the free neutron lifetime.   The measured lifetime values plotted here have already been adjusted for the known neutron and proton counting loss mechanisms.}
\end{figure}

With the values of $f_{\text{Bsc}}$ and $f_{\text{Lost}}$, one can determine the free neutron lifetime $\tau_n$.  Each measured lifetime $\tau_{\text{measured}}$ must be reduced by the factor $1+f_{\text{Lost}}$ since this fraction of decay protons was missed. These corrected values of the measured lifetime are plotted versus $f_{\text{Bsc}}$ and are fit to the linear form
\begin{equation}
\frac{\tau_{\text{measured,$i$}}}{1+f_{\text{Lost,$i$}}} = \tau_n + X
f_{\text{Bsc,$i$}}
\label{eq:magic}
\end{equation}
as shown in Fig.~\ref{fig:Tau}, where the index $i$ refers to a particular series.  The slope of the line $X$ may vary between 0 and 1.  The physical significance of $X = 0$ is that every backscattered proton returned to the detector and was registered; the physical significance of $X = 1$ is that no returning proton was registered in the detector.  Both of these extreme values are unlikely.  The electrostatic potentials are such that protons will be returned to the detector face, but the spectrum of those backscattered protons is lower and they encounter the same loss mechanisms as the incident protons.

The fitted value of the slope gives the fraction of returned protons that are detected.  For the data in Fig.~\ref{fig:Tau}, the slope is $0.74 \pm 0.13$.  The extrapolation to zero backscattering gives the
free neutron lifetime $\tau_n = (886.6 \pm 1.2)$~s, where the uncertainty is statistical.  Note that the intercept of Eq.~(\ref{eq:magic}) is insensitive to an overall multiplicative factor in the backscattering values.  Such a factor could change the slope, and thus the interpretation of how many returned protons are detected, but not the value of the neutron lifetime.

\subsubsection{Backscattering Calculation Uncertainties}\label{sec:BkScUnc}

We determined the fraction backscattered and the fraction lost by two independent methods to serve as a check on the values.  We use the SRIM results as the more accurate values due to the fact that
additional physics is included in the Monte Carlo code which is difficult to implement in the analytical calculation.  The two predominant differences are the treatment of multiple scattering and energy loss.

If one compares the results for the two methods, the $f_{\text{Bsc}}$ values are systematically lower for SRIM. The reason is that SRIM allows protons that have been backscattered to scatter again, and they may have sufficient energy to enter the active layer.  At first consideration, one might expect this fraction to be a negligible correction since the initial fraction is already a small number.  When the effect of energy loss in included, however, the backscattering probably increases significantly.  The energy spectrum of singly-scattered protons is broadened and shifted to lower energies, and hence, their probability for subsequent scattering is increased.  This effect produces the slightly smaller values for $f_{\text{Bsc}}$ from SRIM in comparison with the analytical calculation.  We checked this assertion by comparing the number of backscattered events from the analytical calculation with the comparable value from SRIM, that is, the number of single-backscattered events.

The main contributions to the systematic uncertainty of the backscattering values come from the Monte Carlo statistics and the measurements of the dead layer thickness.  The uncertainty in the latter value is dominated by the calibration of the detector.  The method of performing the {\it in situ} calibrations does not allow much time to collect data because one does not want to produce gain shifts due to warming of the detector.  The uncertainty in the peak of the calibration is typically $\pm 7$~\%.  We estimate the uncertainty on the gold thickness of surface barrier detectors to be $\pm 7$~\% of the nominal value.  This is a conservative value based on past measurements performed on similar detectors (from the same vendor) used in a previous in-beam measurement of the neutron lifetime~\cite{BYR90}; the agreement with the nominal values was better than $\pm 7$~\%.  We estimate the statistical uncertainty of each SRIM calculation to be 5~\%.  As a consequence of this, there is a 5~\% series-dependent uncertainty and a 7~\% detector-dependent uncertainty in each $f_{\text{Lost,$i$}}$ and $f_{\text{Bsc,$i$}}$ and these will add in quadrature.  In a simple Monte Carlo, which was repeated many times, each of the fractions was randomly varied by an appropriate normally distributed amount after which $\tau_n$ was determined via Eq.~(\ref{eq:magic}).  The standard deviation of derived $\tau_n$'s was 0.4\,s, making this our estimate for the uncertainty due to proton scattering in the detector.

\section{Results}\label{sec:results}

The result of the lifetime measurement is $\tau_{n} = (886.6\pm1.2{\rm [stat]}\pm3.2{\rm [sys]})$~s, which is the most precise measurement of the lifetime using an in-beam method. This result is in good agreement with the current world average~\cite{EID04}. The systematic uncertainty is dominated by neutron counting, in particular the areal density of the $^{6}$LiF deposit and the $^{6}$Li({\it{n,t}}) cross section.  A summary of all corrections and uncertainties was given in Table~\ref{tab:Sys}.

One notes that the lifetime produced by this measurement technique is inversely proportional to the value of the ${^6}$Li cross section, which is obtained from the current ENDF evaluation. The value could be made independent of the cross section by an absolute calibration of the neutron counter. Furthermore, such a calibration would improve the uncertainty on the lifetime significantly by eliminating the two largest systematic uncertainties. A cryogenic neutron radiometer that promises to be capable of such a calibration at the 0.1~\% level has recently been demonstrated~\cite{RIC98,CHO03}, and we are pursuing this method further. We expect that this experiment will ultimately achieve an uncertainty of approximately 2~s.

\section{Acknowledgments}

We would like to thank  J.~Adams, M.~Arif, A.~Carlson, C.~Evans, and G.~Lamaze of NIST, Z.~Chowdhuri of the University of Maryland/NIST, and P.~Dawber and J.~Byrne of the University of Sussex for very informative discussions and their continued interest in this experiment. We also thank P.~Robouch of the IRMM for assistance in the fabrication and characterization of the $^{6}$LiF and $^{10}$B targets used in this  work.  We acknowledge the support of the National Institute of Standards and Technology, U.S. Department of Commerce, in providing the neutron facilities used in this work.  This research was made possible in part by a grant from the U.S. Department of Energy interagency agreement No.\ DE--AI02--93ER40784  and the National Science Foundation PHY-0100348.


\begin{thebibliography}{99}
\bibitem{MAM89} W.~Mampe, P.~Ageron, C. Bates, J.~M.~Pendlebury, and A.~Steyerl, Phys.  Rev.  Lett.  {\bf 63}, 593 (1989).

\bibitem{MAM93} W.~Mampe, L.~N.~Bondarenko, V.~I.~Morozov, Yu.~N.~Panin, and A.~I.~Fomin, JETP Lett.  {\bf 57}, 82 (1993).

\bibitem{ARZ00} S.~Arzumanov, L.~Bondarenko, S.~Chernyavsky, W.~Drexel, A.~Fomin, P.~Geltenbort, V.~Morozov, Yu.~Panin, J.~Pendlebury, and K.~Schreckenbach, Phys.  Lett.  B {\bf 483}, 15 (2000).

\bibitem{PAU89} W.~Paul, F.~Anton, L.~Paul, S.~Paul, and W.~Mampe, Z. Phys.  {\bf 45}, 25 (1989).

\bibitem{SPI88} P.~E.~Spivak, Zh. Eksp. Fiz. {\bf 94}, 1 (1988).

\bibitem{HUF01} P.~R.~Huffman, C.~R.~Brome, J.~S.~Butterworth, K.~J.~Coakley, M.~S.~Dewey, S.~N.~Dzhosyuk, R.~Golub, G.~L.~Greene, K.~Habicht, S.~K.~Lamoreaux, C.~E.~H.~Mattoni, D.~N.~McKinsey, F.~E.~Wietfeldt, and J.~M.~Doyle, Nature {\bf 403}, 62 (2001).

\bibitem{NEZ92} V.~V.~Nezvizhevskii, A.~P.~Serebrov, R.~R.~Tal'daev, A.~G.~Kharitonov, V.~P.~Alfimenkov, A.~V.~Streikov, and  V.~N.~Shvetsov, JETP.  {\bf 75}, 405 (1992).

\bibitem{BYR96} J.~Byrne, P.~G.~Dawber, C.~G.~Habeck, S.~J.~Smidt, J.~A.~Spain, and A.~P.~Williams, Europhys.  Lett.  {\bf 33}, 187 (1996).

\bibitem{DEW03} M.~S.~Dewey, D.~M.~Gilliam, J.~S.~Nico, F.~E.~Wietfeldt, X.~Fei, W.~M.~Snow,   G.~L.~Greene,  J.~Pauwels, R.~Eykens, A.~Lamberty, and  J. Van Gestel, Phys. Rev. Lett. {\bf 91}, 152302 (2003).

\bibitem{EID04} S.~Eidelman et al. (Particle Data Group), Phys. Lett. B {\bf 592}, 1 (2004) (URL: http://pdg.lbl.gov).

\bibitem{HER01} P.~Herczeg, Prog. Nucl. Part. Phys.  {\bf 46}, 413 (2001).

\bibitem{JAC57} J.~Jackson, S.~Treiman, and H.~Wyld, Phys.  Rev.  {\bf 106}, 517 (1957).

\bibitem{TOW95} I.~S.~Towner and J.~C.~Hardy, in {\it Symmetries and Fundamental Interactions in Nuclei}, edited by W.~C.~Haxton and E.~M.~Henley (World Scientific, Singapore, 1995), pp. 183-249.

\bibitem{SHI03} I.~Shipsey in {\it Quark Mixing, CKM Unitarity}, edited by D.~Mund and H.~Abele (Mattes Verlag, Heidelberg, 2003) and hep-ex/0203033.

\bibitem{HOL77} B.~R.~Holstein and S.~B.~Treiman, Phys.  Rev. D {\bf 16}, 2369 (1977).

\bibitem{KUR02} A.~Kurylov and M.~J.~Ramsey-Musolf, Phys.  Rev. Lett {\bf 88}, 071804 (2002).

\bibitem{LAN88} P.~Langacker and D.~London, Phys.  Rev. D {\bf 38}, 886 (1988).

\bibitem{MAA90} J.~Maalampi and M.~Roos, Phys.  Rep.  {\bf 186}, 53 (1990).

\bibitem{LAN92} P.~Langacker and M.~Luo, Phys.  Rev. D {\bf 45}, 278 (1992).

\bibitem{MAR87} W.~J.~Marciano and A.~Sirlin, Phys.  Rev. D {\bf 35}, 1672 (1987).

\bibitem{FRA04} P.~Franzini, hep-ex/0203033. 

\bibitem{BAL03} P.~Ball,  J.~Flynn, P.~Kluit, and A.~Stocchi, eds. {\it Proceedings of the Second Workshop on the CKM Unitarity Triangle}, IPPP, Durham (2003) and Electronic Proceedings Archive eConf C0304052, 2003.

\bibitem{CAB03} N.~Cabibbo, E.~Swallow, and R.~Winston, hep-ph/0307214 (2003).

\bibitem{LOP99} R.~E.~Lopez and M.~S.~Turner, Phys.  Rev.  D  {\bf 59},  103502-1 (1999).

\bibitem{BUR99} S.~Burles, K.~M.~Nollett, J.~W.~Truran, and M.~S.~Turner, Phys.  Rev. Lett. {\bf 82}, 4176 (1999).

\bibitem{SPE03} D.~N.~Spergel et al. (WMAP collaboration), Astrophys.  J.  Suppl. {\bf 148},  175 (2003).

\bibitem{CYB03} R.~H.~Cyburt, B.~D.~Fields, and K.~A.~Olive Phys. Lett. B {\bf 567},  227 (2003).

\bibitem{WIL89} A.~P.~Williams,  Ph.D. thesis, University of Sussex (1989).

\bibitem{BYR90} J.~Byrne, P.~G.~Dawber, J.~A.~Spain, A.~P.~Williams, M.~S.~Dewey, D.~M.~Gilliam, G.~L.~Greene, G.~P.~Lamaze, R.~D.~Scott, J.~Pauwels, R.~Eykens, and A.~Lamberty, Phys. Rev. Lett {\bf 65}, 289 (1990).

\bibitem{SNO00} W.~M.~Snow, Z.~Chowdhuri, M.~S.~Dewey, X.~Fei, D.~M.~Gilliam, G.~L.~Greene, J.~S.~Nico, and F.~E.~Wietfeldt, Nucl.  Instrum.  Meth.  A {\bf 440}, 528 (2000).

\bibitem{BYR89} J.~Byrne, P.~G.~Dawber, J.~A.~Spain, M.~S.~Dewey, D.~M.~Gilliam, G.~L.~Greene, G.~P.~Lamaze, A.~P.~Williams, J.~Pauwels, R.~Eykens, J.~VanGestel, A.~Lamberty, and R.~D.~Scott,  Nucl. Instrum. Meth.  A {\bf 284}, 116 (1989).

\bibitem{BER61} A.~A.~Bergman and F.~L.~Shapiro, JETP {\bf 13}, 5 (1961).

\bibitem{HUF03} P.~R.~Huffman,  M.~Arif, M.~S.~Dewey, T.~R.~Gentile, D.~M.~Gilliam, D.~L.~Jacobson, J.~S.~Nico, and A.~K.~Thompson, {\it Application of Accelerators in Research and Industry: Seventeenth International Conference on the Application of Accelerators in Research and Industry}, ed. J.~L.~Duggan, AIP Conference Proceedings 680 (2003).

\bibitem{CRC82} {\it Handbook of Chemistry and Physics}, eds. R.~C.~Weast and M.~J.~Astle (CRC Press, Boca Raton, FL 1982).

\bibitem{CHO03} Z.~Chowdhuri, G.~L.~Hansen, V.~Jane, C.~D.~Keith, W.~M.~Lozowski, W.~M.~Snow,
M.~S.~Dewey, D.~M.~Gilliam, G.~L.~Greene, J.~S.~Nico, A.~K.~Thompson, and F.~E.~Wietfeldt,
Rev.  Sci. Instrum.  {\bf 74} (10), 4280 (2003).

\bibitem{RIC93} J.~M.~Richardson, Ph.D. thesis, Harvard University, 1993.

\bibitem{CHO00} Z.~Chowdhuri, Ph.D. thesis, Indiana University, 2000.

\bibitem{LAM02} G.~P.~Lamaze (private communication, 2003).

\bibitem{CAR93} A.~D.Carlson, W.~P.~Poenitz, G.~M.~Hale, R.~W.~Peelle, D.~C.~Doddler, C.~Y.~Fu, and W.~Mannhart {\it The ENDF/B-VI Neutron Cross Section Measurement Standards}, NISTIR 5177 (1993).
\bibitem{DEN99} B.~Denecke, R.~Eykens, J.~Pauwels, P.~Robouch, D.~M.~Gilliam, P.~Hodge, J.~M.~R.~Hutchinson, and J.~S.~Nico,  Nucl.  Instrum.  Meth.  A {\bf 438}, 124 (1999).

\bibitem{TAG91} H.~Tagziria, J.~Pauwels, J.~Verdonk, J.~Van~Gestel, R.~Eykens, D.~M.~Gilliam,  R.~D.~Scott, J.~Byrne, and P.~Dawber, Nucl. Instrum. Meth. A {\bf 303}, 123 (1991).

\bibitem{PAU91} J.~Pauwels, R.~Eykens, A.~Lamberty, J.~Van~Gestel, H.~Tagziria, R.~D.~Scott, J.~Byrne, P.~Dawber, and D.~M.~Gilliam, Nucl. Instrum. Meth. A {\bf 303}, 133 (1991).

\bibitem{SCO92} R.~D.~Scott, J.~Pauwels, R.~Eykens, J.~Byrne, P.~G.~Dawber, and D.~M.~Gilliam, Nucl. Instrum. Meth.  A {\bf 314}, 163 (1992).

\bibitem{GIL93}  D.~M.~Gilliam, G.~P.~Lamaze, M.~S.~Dewey, and G.~L.~Greene, Nucl. Instrum. Meth.  A {\bf 334}, 149 (1993).

\bibitem{PAU95} J.~Pauwels, R.~D.~Scott, R.~Eykens, P.~Robouch, J.~Van~Gestel, J.~Verdonck, D.~M.~Gilliam, and G.~Greene, Nucl.  Instrum.  Meth.  A {\bf 362}, 104 (1995).

\bibitem{SCO95} R.~D.~Scott, P.~D'hondt, R.~Eykens, P.~Robouch, D.~F.~G.~Reher, G.~Sibbens, J.~Pauwels, D.~M.~Gilliam, Nucl. Instrum.  Meth. A {\bf 362}, 151 (1995).

\bibitem{DISCLM} Certain trade names and company products are mentioned in the text or identified in illustrations in order to adequately specify the experimental procedure and equipment used. In no case does such identification imply recommendation or endorsement by the National Institute of Standards and Technology, nor does it imply that the products are necessarily the best available for the purpose.

\bibitem{COO04} J.~Cook (private communication, 2004).

\bibitem{BYR93} J.~Byrne and P.~G.~Dawber, Nucl.  Instrum.  Meth. A {\bf 332}, 363 (1993).

\bibitem{BYR65} J.~Byrne and P.~S.~Farago, Proc.  Phys.  Soc.  London {\bf 88}, 801 (1965).

\bibitem{BYR72} J.~Byrne, Proc.  Roy.  Soc.  Edinburgh, Sect.  A {\bf 70}, 47 (1972).

\bibitem{KRE93} M.~Kretzschmar, Phys.  Scripta RS {\bf 20}, 7 (1993).

\bibitem{JAC75} J.~D.~Jackson, {\em Classical Electrodynamics, Second Ed.}, (John Wiley and Sons, 1975), p.589--590.

\bibitem{NAC68} O.~Nachtmann, Zeit.  f\"{u}r Phys.  {\bf 215}, 505 (1968).

\bibitem{WILK89} D.~H.~Wilkinson, Nucl. Instrum.  Meth. A {\bf 275}, 378 (1989).

\bibitem{ZIE03} J.~F.~Ziegler, Stopping and Range of Ions in Matter (SRIM-2003). Available at www.srim.org.

\bibitem{PAR89} N.~R.~Parikh, The University of North Carolina at Chapel Hill (private communication, 1989).

\bibitem{CHE96} Y.~T.~Cheng, T. Soodprasert, and J.~M.~R.~Hutchinson, Appl. Radiat. Isot. {\bf 47}, 1023 (1996).

\bibitem{CHE00} Y.~T.~Cheng and D.~F.~R.~Mildner, Nucl.  Instrum.  Meth.  A {\bf 454}, 452 (2000).

\bibitem{RIC98} J.~M.~Richardson, W.~M.~Snow, Z.~Chowdhuri, and G.~L.~Greene, IEEE Trans. Nucl. Sci. {\bf 45}, 550 (1998).


\end{thebibliography}
\end{document}